\documentclass[10pt,twocolumn,letterpaper]{article}

 \usepackage{cvpr}              

\usepackage{multirow}
\usepackage{wrapfig}
\usepackage{dblfloatfix}
\usepackage{cuted}

\newcommand{\vtheta}{\boldsymbol{\theta}}

\definecolor{cvprblue}{rgb}{0.21,0.49,0.74}
\usepackage[pagebackref,breaklinks,colorlinks,allcolors=cvprblue]{hyperref}

\newcommand{\dhr}{\textcolor{orange}}

\def\paperID{8354} 
\def\confName{CVPR}
\def\confYear{2026}

\title{Learning to Translate Noise for Robust Image Denoising}

\author{
Inju Ha$^{*1}$ \qquad 
Donghun Ryou$^{*2}$ \qquad 
Seonguk Seo$^{3}$ \qquad 
Bohyung Han$^{1,2}$ \\
$^1$ECE \& $^2$IPAI, Seoul National University \quad $^3$Meta \\
{\tt\small \{hij1112, dhryou, bhhan\}@snu.ac.kr} \quad {\tt\small seonguk@meta.com}
\\
\href{https://hij1112.github.io/learning-to-translate-noise/}{\tt\small https://hij1112.github.io/learning-to-translate-noise/}
\vspace{-2.5mm}
}
\begin{document}
\maketitle

\begingroup
\renewcommand\thefootnote{} 
\footnotetext{\protect\hspace{-0.5em}${}^*$indicates equal contribution.}
\endgroup

\begin{strip}
\centering
\includegraphics[width=0.75\textwidth]{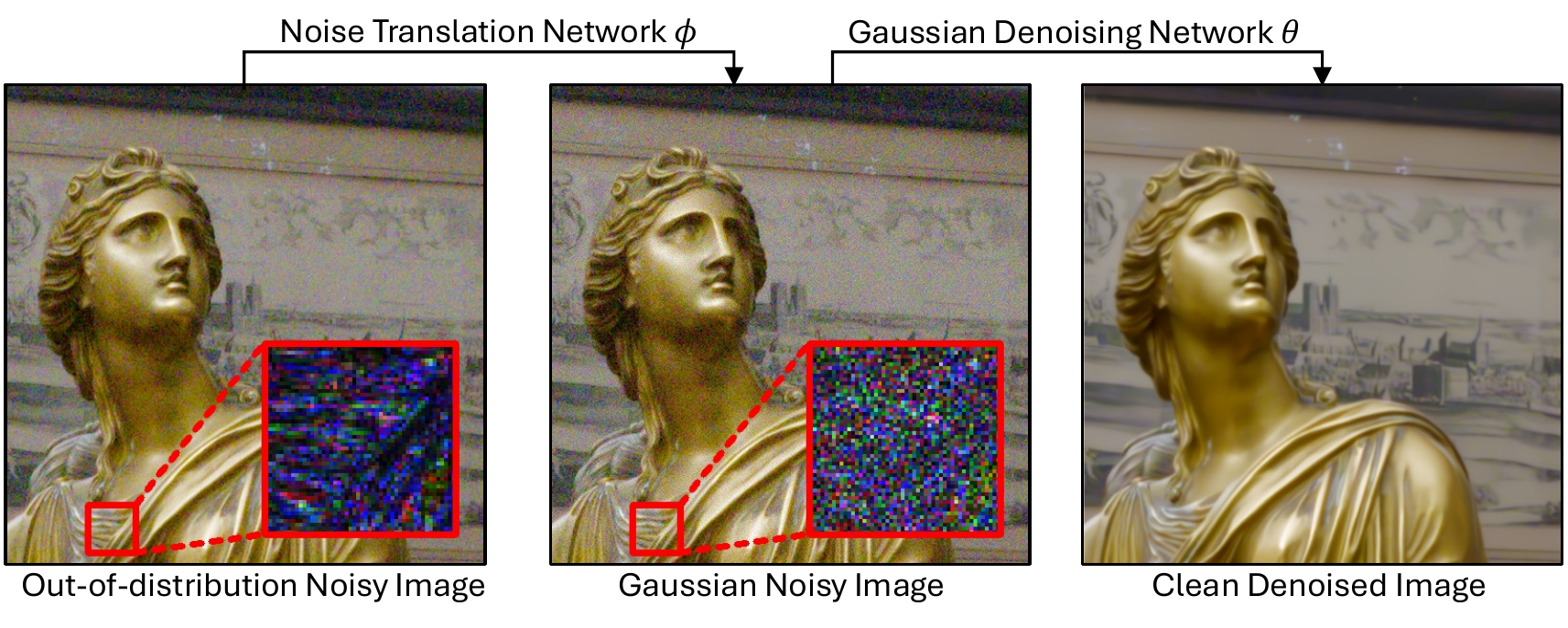}
\vspace{-4mm}
\captionof{figure}{
    Our noise translation network (NTN) first transforms arbitrary real-world noise into Gaussian noise without signal correlation.
    The translated noise can then be effectively removed using any existing Gaussian denoiser.
    The shown noisy image is from the NIND dataset~\citep{Brummer_2019_CVPR_Workshops}, captured at ISO 6400.
    The figure is best viewed by zooming in.
    }
\label{fig:nind_teaser}
\vspace{-2mm}
\end{strip}

\vspace{-1mm}
\begin{abstract}
    Image denoising techniques based on deep learning often struggle with poor generalization performance to out-of-distribution real-world noise.
    To tackle this challenge, we propose a novel noise translation framework that performs denoising on an image with translated noise rather than directly denoising an original noisy image.
    Specifically, our approach translates unknown real-world noise into Gaussian noise, which is spatially uncorrelated and independent of image content, through our noise translation network.
    The translated noisy images are then processed by an image denoising network pretrained to effectively remove Gaussian noise, enabling robust and consistent denoising performance.
    We also design well-motivated loss functions and architectures for the noise translation network by leveraging the mathematical properties of Gaussian noise.
    Experimental results show that the proposed approach substantially improves robustness and generalizability, outperforming state-of-the-art methods across diverse benchmarks.
\end{abstract}

\vspace*{-4mm}
\section{Introduction}
\label{Introduction}

Image denoising aims to restore the pure signal from noisy images and serves as a critical preprocessing step to improve the visual quality of input images, extending the applicability to various downstream tasks.
Recent advances in deep learning facilitate significant performance improvement of image denoising models~\citep{zhang2017beyond, zhang2018ffdnet, guo2019toward, zamir2020learning, zamir2022learning, zamir2022restormer, chen2022simple, zhang2023xformer, Sun_2025_CVPR}.
A common assumption in early approaches was that camera noise could be modeled as a Gaussian distribution~\citep{mao2016image, zhang2017beyond, zhang2018ffdnet}, which simplified the process of generating noisy-clean image pairs by adding synthetic Gaussian noise.
This allowed for the construction of large datasets that could be used to train denoising models in a supervised manner, playing a crucial role in advancing image denoising models.

While denoising models trained on synthetic datasets perform well in controlled environments, they struggle to generalize to real-world scenarios due to the fundamental differences between synthetic and real noise distributions~\citep{guo2019toward}.
In response, researchers collected clean-noisy image pairs from real images~\citep{abdelhamed2018high, xu2018real, yue2019high} to address realistic noises, but denoising models still tend to overfit to the specific noise-signal correlations present in the training data.
Capturing the full spectrum of real-world noise distributions to prevent overfitting is impractical and even unrealistic.

\begin{figure}[t]
    \centering
    \hspace{-4mm}
    \begin{minipage}[c]{0.02\textwidth}
        \centering
        \rotatebox{90}{\fontsize{10pt}{10pt}\selectfont \text{Input \quad \ \quad \ \quad \quad  \quad Output \ \quad}}
    \end{minipage}
    \begin{minipage}[t]{0.148\textwidth}
        \begin{subfigure}[t]{\textwidth}
            \centering
            \includegraphics[width=\linewidth]{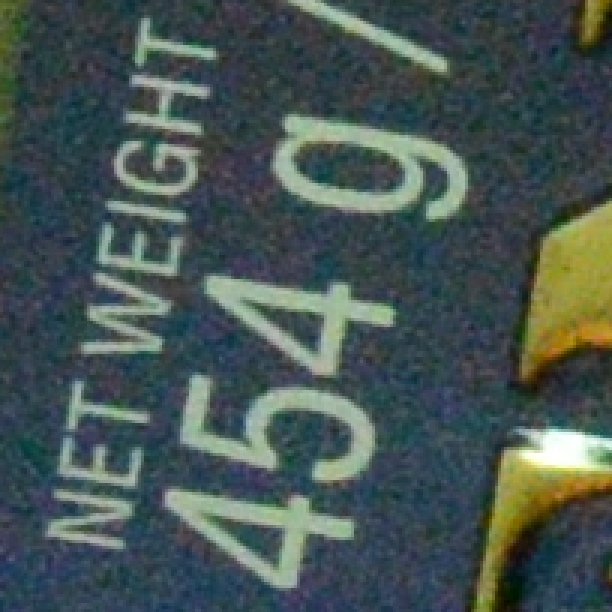}
            \vspace{-4.5mm}
            \caption*{\fontsize{8.5pt}{8.5pt}\selectfont PSNR: 29.63 dB}
        \end{subfigure}
        \vspace{1mm}
        \begin{subfigure}[t]{\textwidth}
            \centering
            \includegraphics[width=\linewidth]{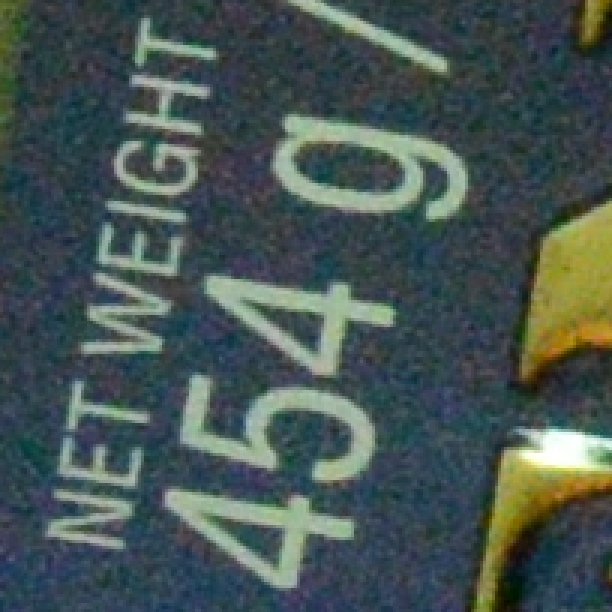}
            \vspace{-4.5mm}
            \caption{\fontsize{8.5pt}{8.5pt}\selectfont Real Noisy Image}
        \end{subfigure}
    \end{minipage}
    \hfill
    \vrule width 0.5pt height 73.5pt depth 86pt
    \vspace{1mm}
    \hfill
    \begin{minipage}[t]{0.148\textwidth}
        \begin{subfigure}[t]{\textwidth}
            \centering
            \includegraphics[width=\linewidth]{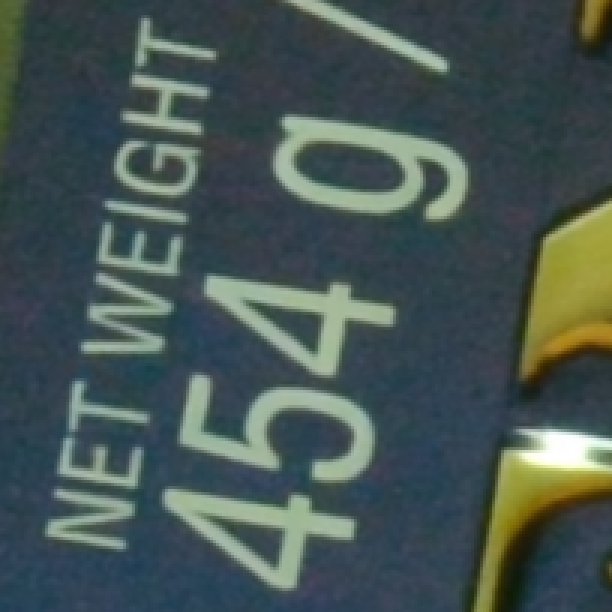}
            \vspace{-4.5mm}
            \caption*{\fontsize{8.5pt}{8.5pt}\selectfont PSNR: 32.73 dB}
        \end{subfigure}
        \vspace{1mm}
        \begin{subfigure}[t]{\textwidth}
            \centering
            \includegraphics[width=\linewidth]{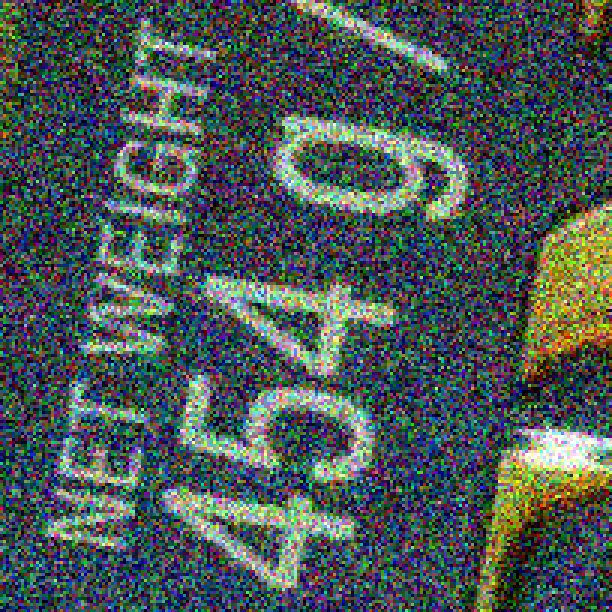}
            \vspace{-4.5mm}
            \caption{\fontsize{8.5pt}{8.5pt}\selectfont Gaussian Added}
        \end{subfigure}
    \end{minipage}
    \hfill
    \vrule width 0.5pt height 73.5pt depth 86pt
    \vspace{1mm}
    \hfill
    \begin{minipage}[t]{0.148\textwidth}
        \begin{subfigure}[t]{\textwidth}
            \centering
            \includegraphics[width=\linewidth]{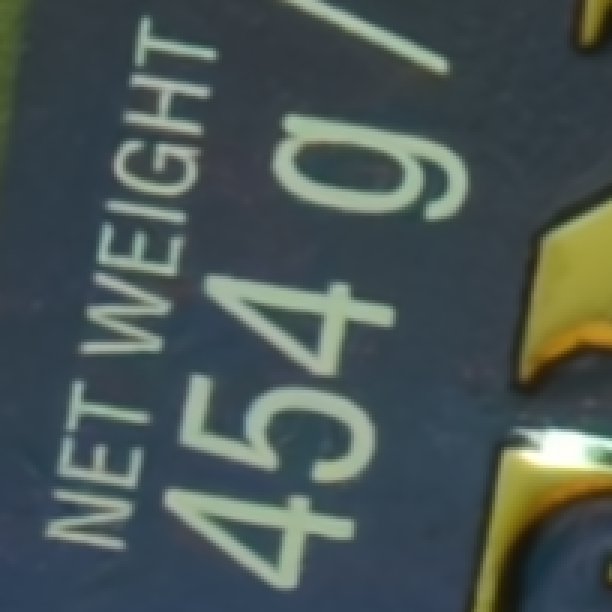}
            \vspace{-4.5mm}
            \caption*{\fontsize{8.5pt}{8.5pt}\selectfont PSNR: \textbf{34.63 dB}}
        \end{subfigure}
        \vspace{1mm}
        \begin{subfigure}[t]{\textwidth}
            \centering
            \includegraphics[width=\linewidth]{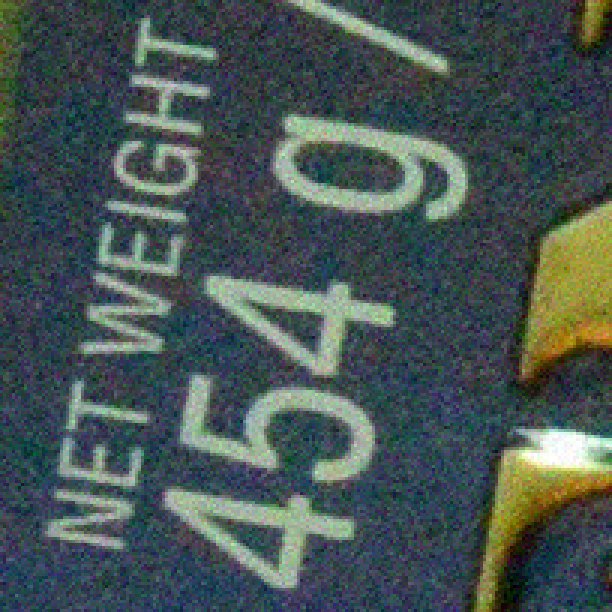}
            \vspace{-4.5mm}
            \caption{\fontsize{8.5pt}{8.5pt}\selectfont Translated \textbf{(Ours)}}
        \end{subfigure}
    \end{minipage}
    \vspace{-4mm}
    \caption{
        Denoising results of Gaussian-trained Restormer~\citep{zamir2022restormer}. 
        From left to right: (1) real noisy image and its denoised output, (2) noisy image with added Gaussian noise and its denoised output, (3) noise-translated image with our framework and its corresponding denoised output, achieving superior quality. 
    }
    \label{fig:motivation1}
    \vspace{-4mm}
\end{figure}

To address this challenge, we propose a novel noise translation framework for image denoising to better generalize to diverse real-world noise using a limited training dataset.
The intuition behind our framework is as follows.
While existing denoising algorithms trained on images with Gaussian noise exhibit limited performance when applied to real-world noisy images, we observed that adding Gaussian noise to these noisy images significantly improves their effectiveness in denoising, as shown in Figure~\ref{fig:motivation1}.
This observation motivated us to explore the idea that, instead of directly denoising unseen real noise, first translating it into known Gaussian noise and then applying denoising could improve the model's ability to generalize across unseen out-of-distribution (OOD) noise.
To this end, we introduce the lightweight \textit{Noise Translation Network}, which, prior to the denoising process, utilizes Gaussian injection blocks to transform arbitrary complex noise into Gaussian noise that is spatially uncorrelated and independent of an input image.
The translated images are then processed by the pretrained denoising networks specialized for Gaussian noise, resulting in the clean denoised images.
Once trained, our noise translation network can be seamlessly paired with any denoising network pretrained on Gaussian noise.
Our experimental results and analysis validate that the proposed framework outperforms existing denoising approaches by huge margins on various benchmarks.

Our key contributions are summarized as follows:
\begin{itemize}
\item We propose a novel noise translation network for robust image denoising that converts unseen complex noise in an input image into Gaussian noise.
Our lightweight noise translation network can seamlessly integrate with any Gaussian-pretrained denoising networks.
\item We employ well-motivated loss functions and architecture for the noise translation network.
Our translation network transforms the noise distribution of the input image into a Gaussian distribution both implicitly and explicitly by leveraging key mathematical properties.
\item We demonstrate the efficacy of our approach through extensive experiments on image denoising benchmarks with diverse noise distributions, achieving significant improvements in terms of robustness and generalization ability.
\end{itemize}

\section{Related Works}
\label{sec:relatedworks}

\paragraph{Image denoising}
In recent years, deep learning has led to significant progress in image restoration~\citep{chen2022simple, zamir2022restormer, kang2025icm, Chen_2025_CVPR, ryou2025beyond}, achieving impressive results by leveraging paired noisy and clean images for training.
DnCNN~\citep{zhang2017beyond} pioneered the use of CNNs for image denoising, which paved the way for further advancements involving residual learning~\citep{gu2019self,liu2019dual,zhang2019residual}, attention mechanisms~\citep{liu2018non,zhang2019residual}, and transformer models~\citep{zamir2022restormer, zhang2023xformer}.
Despite its success, acquiring the noisy-clean pairs required for supervised training remains a significant challenge.
To address this, self-supervised approaches~\citep{lehtinen2018noise2noise, krull2019noise2void, batson2019noise2self, pang2021recorrupted, li2023spatially} have emerged to train networks using only noisy images, but these models typically perform considerably worse than their supervised counterparts.
Additionally, zero-shot approaches~\citep{quan2020self2self, huang2021neighbor2neighbor,mansour2023zero} have been proposed for image denoising even without a training dataset, but they require substantial computational cost at inference, making them impractical for real-time applications.
While previous methods either rely heavily on supervised datasets or bypass them with performance trade-offs, we show that robust generalization is achievable even under limited supervision. 
This makes it suitable for real-world deployment, where collecting large-scale clean references remains a major bottleneck.

\begin{figure*}[t]
    \vspace{-0.1em}
    \centering
    \includegraphics[width=0.9\textwidth]{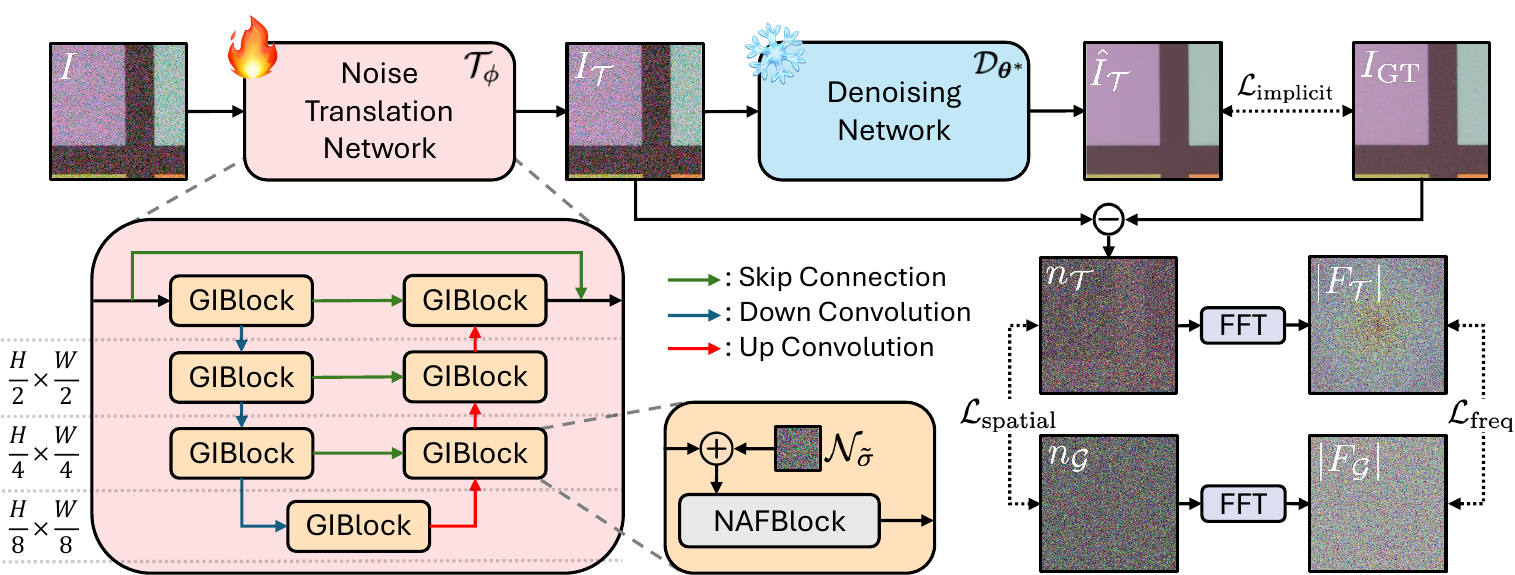}
    \caption{
        Illustration of our training framework for the noise translation network.
    }
    \vspace{-3mm}
    \label{fig:framework}
\end{figure*}

\vspace{-2mm}
\paragraph{Generalization for denoising}

Generalization is a critical challenge in image denoising, as the performance of denoising models often degrades when encountering noise characteristics that were not seen during training.
To handle unseen noise type or levels, DnCNN~\citep{zhang2017beyond} employed blind Gaussian training to adapt to various noise levels, while Mohan~\etal~\cite{Mohan2020Robust} designed a bias-free network to prevent overfitting to noise levels in the training set.
Recent works employed masking-based learning~\citep{chen2023masked} or leveraged the pretrained CLIP encoder~\citep{Jun2024Transfer} to prevent overfitting by encouraging the model to understand global context rather than relying on local patterns.
In addition, IDF~\citep{kim2025idf} introduced an iterative dynamic filtering to better generalize to unseen noise.
While these approaches enhance robustness to unseen noise, they often struggle to produce high-quality images, particularly in complex real-world scenarios.

To handle real-world noise, prior work has focused on building training datasets that reflect realistic noise distributions.
This includes collecting real clean-noisy image pairs~\citep{abdelhamed2018high, xu2018real, yue2019high}, simulating realistic noise via data augmentation~\citep{yue2020dual, jang2021c2n, cai2021learning}, or leveraging adversarial training~\citep{yan2022towards, ryou2024robust}.
However, these approaches are inherently limited by the noise distributions seen during training and often fail to generalize to out-of-distribution (OOD) noise.
Alternatively, noise transformation methods such as the Anscombe transform~\citep{makitalo2012optimal} and pixel-shuffle down-sampling~\citep{zhou2020awgn} aim to simplify denoising by converting complex noise into Gaussian-like noise. 
However, their reliance on fixed transformations limits their adaptability to diverse noise types.
More recently, LAN~\citep{kim2024lan} addressed unseen noise through test-time optimization of pixel-wise offsets. 
While effective, its per-pixel updates hinder scalability to images larger than 256$\times$256.
In contrast, our approach removes the need for test-time training and enables efficient, scalable denoising under a wide range of noise conditions.


\section{Image Denoising through Noise Translation}
\label{sec:method}

This section presents our robust image denoising framework with a novel noise translation process, designed to handle diverse real-world noise.
Our framework first transforms a noisy image $I$ into an image with Gaussian noise \(I_{\mathcal{T}}\), which is then fed into the denoising network to produce the final denoised output $\hat{I}_{\mathcal{T}}$.
The whole process is given by
\begin{equation}
    \hat{I}_{\mathcal{T}} = \mathcal{D}(I_{\mathcal{T}}; \vtheta) =  \mathcal{D}(\mathcal{T}(I;\boldsymbol{\phi}); \vtheta),
\end{equation}
where $\mathcal{T}(\cdot\,; \boldsymbol{\phi})$ denotes the noise translation network parametrized by $\boldsymbol{\phi}$, and $\mathcal{D}(\cdot\,;\vtheta)$ indicates the denoising network with parameters $\vtheta$.
Note that $\mathcal{D}(\cdot\,; \vtheta)$ is specialized in handling Gaussian noise in our framework.
We next discuss how to train the denoising network $\mathcal{D}(\cdot\,; {\vtheta})$ and noise translation network $\mathcal{T}(\cdot\,; \boldsymbol{\phi})$.

\subsection{Image Denoising Network for Gaussian Noise}
We first train a denoising network by the standard supervised learning.  
Our denoising network, $\mathcal{D}(\cdot\,; \vtheta)$, aims to remove the Gaussian noise in an input image $I$ and reconstruct a clean image close to the ground-truth \(I_\text{GT}\).
To achieve this, the model parameters $\vtheta$ are optimized by minimizing the following loss function:
\begin{equation}
    \mathcal{L} = \Vert \mathcal{D}(I; \vtheta) - {I_\text{GT}} \Vert_1.
   \label{eq:sup_denoising}
\end{equation}
For training, the noisy input image $I$ is corrupted by synthetic Gaussian noise.
Detailed procedures for dataset construction are provided in Section~\ref{sub:experimental}.

\subsection{Noise Translation Network} 
\label{sub:Noise_Translation}
We now describe the training procedure for the noise translation network \(\mathcal{T}(\cdot; \boldsymbol{\phi})\).
During this phase, the parameters of the denoising network \(\mathcal{D}(\cdot\,; \vtheta^*)\), trained for Gaussian noise removal, are kept fixed to provide a stable supervisory signal.  
The overall training pipeline is illustrated in Figure~\ref{fig:framework}.

\subsubsection{Implicit Noise Translation Loss}
\label{sub:implicit}

The objective is to transform an arbitrary noisy input image \(I\) into a noise-translated image \(I_{\mathcal{T}}\) that is well-suited for the pretrained denoising network specialized in handling Gaussian noise.  
To this end, we optimize the noise translation network using the \textit{implicit} noise translation loss, which minimizes the discrepancy between the denoised output and the corresponding clean ground-truth image:
\begin{equation}
\mathcal{L}_\text{implicit} = \Vert \hat{I}_{\mathcal{T}} - I_\text{GT} \Vert_1 =  \Vert \mathcal{D}(\mathcal{T}(I; \boldsymbol{\phi}); \vtheta^*) - I_\text{GT} \Vert_1.
\end{equation}
This loss function encourages the translation network to map arbitrary unseen noise into a form that can be effectively handled by the pretrained Gaussian Denoiser  \(\mathcal{D}(\cdot\,;\vtheta^*)\).
A natural choice for constructing training pairs \((I, I_{\text{GT}})\) for training the noise translation network is to use real noisy-clean image pairs.  
Accordingly, we exclusively use noisy-clean pairs from the real noise dataset SIDD~\cite{abdelhamed2018high}.

\subsubsection{Explicit Noise Translation Loss}
\label{sub:explicit}
Although the implicit noise translation loss encourages the translated image to be well-aligned with the pretrained denoising network, it does not explicitly enforce that the translated noise follows an ideal Gaussian distribution, as it lacks direct control over the noise characteristics.  
To address this, we introduce an explicit loss that promotes the translated noise to follow a Gaussian distribution.

Let \(n_{\mathcal{T}} = I_{\mathcal{T}} - I_{GT} \in \mathbb{R}^{H \times W \times C}\) represent the translated noise, and let \(n_{\mathcal{G}}\in \mathbb{R}^{H \times W \times C}\) be a random variable following a Gaussian distribution \(\mathcal{N}(\hat{\mu}, \hat{\sigma}^2)\), where $\hat{\mu}$ and $\hat{\sigma}$ denote the empirical mean and standard deviation calculated from all elements of \(n_{\mathcal{T}}\), respectively.
We adjust the distribution of \( n_{\mathcal{T}} \) to align with the distribution of \( n_{\mathcal{G}} \).
To this end, we employ the Wasserstein distance between the two distributions as a loss function, which is given by
\begin{equation}
    \mathcal{L}_\text{spatial} \equiv d_{W_1}(n_{\mathcal{T}}, n_\mathcal{G}),
    \label{eq:explicit_loss1}
\end{equation}
where $d_{W_1}(\cdot, \cdot)$ is 1-Wasserstein distance, also known as the Earth Mover's Distance.
To compute this, we first flatten each channel in $n_{\mathcal{T}}$ and $n_\mathcal{G}$ over the spatial dimensions into one-dimensional vectors, and then sort them in an ascending order.
Let \(\mathbf{X}^c \equiv (X_1^c, X_2^c, \dots, X_{H\times W}^c)\) and \(\mathbf{Y}^c \equiv (Y_1^c, Y_2^c, \dots, Y_{H\times W}^c)\) denote the ordered values of $n_{\mathcal{T}}$ and \( n_\mathcal{G} \) for the $c^\text{th}$ channel, respectively.
The 1-Wasserstein distance is then given by the following simple function\footnote{Please refer to Appendix for the detailed proof.}:
%
\begin{equation}
    d_{W_1}(n_{\mathcal{T}}, n_\mathcal{G}) = \frac{1}{H \cdot W \cdot C} \sum_{c=1}^C \sum_{i=1}^{H \cdot W} \vert X_{(i)}^c - Y_{(i)}^c \vert.
    \label{eq:explicit_wasserstein1}
\end{equation}

This loss encourages the translated noise \(n_{\mathcal{T}}\) to follow a Gaussian distribution element-wise, but it is still insufficient to ensure that \(n_{\mathcal{T}}\) is spatially uncorrelated.
%
To handle the spatial correlation, we convert the signals of $n_{\mathcal{T}}$ and $n_\mathcal{G}$ into the frequency domain using their respective channel-wise Fourier transforms, which are given by
\begin{align}
    F_{\mathcal{T}}^c(u, v) &= \sum_{x=0}^{H-1} \sum_{y=0}^{W-1} n_{\mathcal{T}}(x, y, c) e^{-2\pi i \left(\frac{ux}{H} + \frac{vy}{W}\right)}, \\
    F_\mathcal{G}^c(u, v) &= \sum_{x=0}^{H-1} \sum_{y=0}^{W-1} n_\mathcal{G}(x, y, c) e^{-2\pi i \left(\frac{ux}{H} + \frac{vy}{W}\right)},
\end{align}
where $(u, v)$ are the frequency domain coordinates and $c$ is a channel index.
Since $n_\mathcal{G}$ is spatially uncorrelated Gaussian noise, the real and imaginary parts of the Fourier coefficients, $F_\mathcal{G}^c(u, v)$, also follow \textit{i.i.d.} Gaussian distributions with zero mean and the same variance.
Consequently, the magnitude of the Fourier coefficients, $|F_\mathcal{G}^c(u, v)|$, follows a Rayleigh distribution as
\begin{equation}
\hspace{-1.9mm}  p_R(|F_\mathcal{G}^c(u, v)|; \sigma) = \frac{|F_\mathcal{G}^c(u, v)|}{\sigma^2} \exp\left(-\frac{|F_\mathcal{G}^c(u, v)|^2}{2\sigma^2}\right),
\end{equation}
which implies that $|F_\mathcal{T}^c(u, v)|$ should also follow a Rayleigh distribution, guiding $n_\mathcal{T}$ to be spatially uncorrelated.
To this end, similar to Eqs.~(\ref{eq:explicit_loss1}) and~(\ref{eq:explicit_wasserstein1}), we minimize the difference between the distributions of $|F_\mathcal{T}^c(u, v)|$ and $|F_\mathcal{G}^c(u, v)|$ by utilizing 1-Wasserstein distance, which is defined as
%
%
\begin{equation}
    \mathcal{L}_\text{freq} \equiv d_{W_1}(\vert F_{\mathcal{T}} \vert, \vert F_\mathcal{G} \vert) = \frac{1}{H \cdot W \cdot C} \sum_{c=1}^C \sum_{i=1}^{H \cdot W} \vert \tilde{X}^c_{(i)} - \tilde{Y}^c_{(i)} \vert,
    \label{eq:explicit_wasserstein2}
\end{equation}
where $\tilde{\mathbf{X}}^c\equiv(\tilde{X}^c_1, \tilde{X}^c_2, \dots, \tilde{X}^c_{H\times W})$ and $\tilde{\mathbf{Y}}^c\equiv(\tilde{Y}^c_1, \tilde{Y}^c_2, \dots, \tilde{Y}^c_{H\times W})$ are the sorted values of flattened magnitude of Fourier coefficients $|F_{\mathcal{T}}^c(u,v)|$ and $|F_\mathcal{G}^c(u,v)|$, respectively.
The full explicit noise translation loss is defined by $\mathcal{L}_\text{spatial}$ and $\mathcal{L}_\text{freq}$ as
\begin{equation}
    \label{eq:spatial_frequency}
    \mathcal{L}_\text{explicit} = \mathcal{L}_\text{spatial} + \beta \cdot \mathcal{L}_\text{freq},
\end{equation}
where $\beta$ controls the contribution of the two Wasserstein distances.
This loss function explicitly guides the translated noise to follow Gaussian distribution.
The total loss function for training the noise translation network is given by
\begin{equation}
    \label{eq:total_loss}
    \mathcal{L}_\text{total} = \mathcal{L}_\text{implicit} + \alpha \cdot \mathcal{L}_\text{explicit},
\end{equation}
where $\alpha$ balances the influence of two loss terms.


\subsubsection{Gaussian Injection Block}
\label{sub:Gaussian_injection}

As shown in Figure~\ref{fig:framework}, our noise translation network is based on a lightweight U-Net architecture, where each layer contains a Gaussian Injection Block (GIBlock).  
Each GIBlock combines a Nonlinear Activation-Free (NAF) block~\citep{chen2022simple} with our key design, the injection of Gaussian noise, to consistently impose a Gaussian prior throughout the network.

While injecting Gaussian noise into the input can provide a Gaussian prior to some extent, it inevitably distorts the input signal.  
In contrast, our method injects Gaussian noise internally into sub-blocks of the network, allowing the network to receive the Gaussian prior gradually without perturbing the original image for translating unseen real noise effectively.  
Additionally, the residual connections between the input \(I\) and the output \(I_{\mathcal{T}}\) further mitigate any signal distortion introduced by the injected noise.  
Ablation studies confirm that Gaussian noise injection is essential for enabling the noise translation network to reliably map unseen noise to a Gaussian distribution during inference.



\begin{table*}[t]
    \centering
    \caption{
        Quantitative comparison with state-of-the-art real-world denoising networks on the SIDD validation set and various out-of-distribution (OOD) real-world benchmarks.
        The results are presented in terms of PSNR$\uparrow$ (dB) and SSIM$\uparrow$.
        Models marked with an asterisk (*) are evaluated using official out-of-the-box models, while a dagger ($\dagger$) indicates models fine-tuned on the SIDD training dataset.
    }
    \vspace{-2mm}
    \setlength{\tabcolsep}{6pt}
    \renewcommand{\arraystretch}{0.85}
    \scalebox{0.8}{
        \begin{tabular}{l l c c c c c c c c c c c}
            \toprule
            \multicolumn{3}{c}{}                                              & \multicolumn{1}{c}{In-distribution} & \multicolumn{8}{c}{Out-of-distribution} &                                                                                                                             \\
            \cmidrule(lr){4-4} \cmidrule(lr){5-12}
            \multicolumn{1}{c}{} & \multicolumn{1}{c}{Denoising Algo.}                               & Metric                              & SIDD                                    & Poly        & CC          & HighISO     & iPhone      & Huawei      & OPPO        & Sony        & Xiaomi      & OOD Avg.    \\
            \midrule\midrule

            \multirow{9}{*}{\rotatebox{90}{\textit{\large Self-supervised}}} & \multirow{2}{*}{R2R*~\citep{pang2021recorrupted} }                & PSNR                                & 35.09                                   & 36.84       & 35.28       & 37.37       & 39.25       & 38.35       & 39.38       & 41.55       & 35.39       & 37.93       \\
            & & SSIM                                & 0.9154                                  & 0.9726      & 0.9758      & 0.9716      & 0.9614      & 0.9667      & 0.9744      & 0.9735      & 0.9668      & 0.9703      \\
            \cmidrule{2-13}
            & \multirow{2}{*}{AP-BSN*~\citep{lee2022ap} }                       & PSNR                                & 36.35                                   & 35.91       & 33.15       & 36.70       & 39.90       & 37.06       & 39.11       & 40.13       & 33.39       & 36.92       \\
            & & SSIM                                & 0.9285                                  & 0.9755      & 0.9734      & 0.9781      & 0.9773      & 0.9633      & 0.9750      & 0.9805      & 0.9552      & 0.9723      \\
            \cmidrule{2-13}
            & \multirow{2}{*}{SSID*~\citep{li2023spatially} }                   & PSNR                                & 37.43                                   & 37.17       & 34.95       & 38.28       & 40.95       & 37.27       & 39.17       & 43.03       & 34.27       & 38.14       \\
            & & SSIM                                & 0.9343                                  & 0.9803      & 0.9806      & 0.9811      & 0.9823      & 0.9656      & 0.9743      & 0.9883      & 0.9566      & 0.9762      \\
            \cmidrule{2-13}
            & \multirow{2}{*}{APR-RD*~\citep{kim2025apr} }                      & PSNR                                & 38.06                                   & 37.01       & 35.83       & 38.75       & 40.40       & 36.75       & 38.93       & 43.35       & 33.74       & 38.10       \\
            & & SSIM                                & 0.9470                                  & 0.9781      & 0.9824      & 0.9823      & 0.9763      & 0.9590      & 0.9713      & 0.9861      & 0.9483      & 0.9730      \\
            \midrule\midrule
            \multirow{9}{*}{\rotatebox{90}{\textit{\large Generalization}}} & \multirow{2}{*}{Mask-Denoising$\dagger$~\citep{chen2023masked} }  & PSNR                                & 38.91                                   & 37.45       & 35.64       & 38.12       & 40.54       & 37.99       & 39.55       & 44.04       & 35.12       & 38.56       \\
            & & SSIM                                & 0.9529                                  & 0.9808      & 0.9810      & 0.9793      & 0.9801      & 0.9694      & 0.9790      & 0.9899      & 0.9697      & 0.9786      \\
            \cmidrule{2-13}
            & \multirow{2}{*}{Clip-Denoising$\dagger$~\citep{Jun2024Transfer} } & PSNR                                & 38.03                                   & 37.49       & 35.92       & 38.26       & 40.27       & 38.07       & 39.73       & 43.42       & 35.11       & 38.53       \\
            & & SSIM                                & 0.9446                                  & 0.9795      & 0.9819      & 0.9782      & 0.9730      & 0.9702      & 0.9789      & 0.9857      & 0.9690      & 0.9771      \\
            \cmidrule{2-13}
            & \multirow{2}{*}{AFM*~\citep{ryou2024robust}}                      & PSNR                                & 38.29                                   & 37.71       & 36.81       & 39.12       & 40.56       & 38.33       & 40.13       & {44.66}     & 35.25       & 39.07       \\
            & & SSIM                                & 0.9474                                  & 0.9800      & 0.9828      & 0.9797      & 0.9769      & 0.9679      & 0.9795      & {0.9901}    & 0.9665      & 0.9745      \\
            \cmidrule{2-13}
            & \multirow{2}{*}{IDF$\dagger$~\citep{kim2025idf} }                 & PSNR                                & 37.01                                   & 36.42       & 33.62       & 37.17       & 39.79       & 36.23       & 38.74       & 43.12       & 33.69       & 37.35       \\
            & & SSIM                                & 0.9354                                  & 0.9764      & 0.9754      & 0.9764      & 0.9771      & 0.9593      & 0.9754      & 0.9885      & 0.9563      & 0.9731      \\

            \midrule\midrule
            \multirow{9}{*}{\rotatebox{90}{\textit{\large Ours}}} & \multirow{2}{*}{NAFNet*~\citep{chen2022simple}}                   & PSNR                                & {39.97}                                 & 37.17       & 35.69       & 38.32       & 40.25       & 37.73       & 39.64       & 43.65       & 34.99       & 38.43       \\
            & & SSIM                                & {0.9600}                                & 0.9717      & 0.9811      & 0.9788      & 0.9707      & 0.9680      & 0.9786      & 0.9829      & 0.9685      & 0.9750      \\
            \cmidrule{2-13}
            & \multirow{2}{*}{\bf{NAFNet + NTN}}                                     & PSNR                                & 39.24                                   & \bf{38.72}  & \bf{37.84}  & \bf{40.00}  & \bf{42.08}  & \bf{39.83}  & \bf{40.55}  & \bf{44.34}  & \bf{36.17}  & \bf{39.94}  \\
            & & SSIM                                & 0.9570                                  & \bf{0.9855} & \bf{0.9877} & \bf{0.9856} & \bf{0.9812} & \bf{0.9782} & \bf{0.9801} & \bf{0.9875} & \bf{0.9749} & \bf{0.9826} \\

            \cmidrule{2-13}
            \noalign{\vskip-2.5pt}
            \cmidrule{2-13}
            & \multirow{2}{*}{XFormer*~\citep{zhang2023xformer}}                & PSNR                                & {39.98}                                 & 37.45       & 35.95       & 37.86       & 39.99       & 38.35       & 39.61       & 43.89       & 35.54       & 38.58       \\
            & & SSIM                                & {0.9603}                                & 0.9778      & 0.9782      & 0.9741      & 0.9732      & 0.9678      & 0.9779      & \bf{0.9892} & 0.9706      & 0.9761      \\

            \cmidrule{2-13}
            & \multirow{2}{*}{\bf{XFormer + NTN}}                                    & PSNR                                & 39.10                                   & \bf{38.75}  & \bf{37.82}  & \bf{40.29}  & \bf{42.20}  & \bf{39.89}  & \bf{40.70}  & \bf{44.44}  & \bf{36.25}  & \bf{40.04}  \\
            & & SSIM                                & 0.9557                                  & \bf{0.9856} & \bf{0.9862} & \bf{0.9860} & \bf{0.9826} & \bf{0.9779} & \bf{0.9807} & {0.9886}    & \bf{0.9751} & \bf{0.9828} \\
            \bottomrule
        \end{tabular}
    }
    \vspace{-2mm}
    \label{tab:real_main}
\end{table*}

\section{Experiments}
\label{sec:exp}
We demonstrate the effectiveness of the proposed approach on various benchmarks, evaluating performance and conducting analysis on both ID and OOD datasets.
This section also provides an in-depth analysis of our algorithm, including detailed ablation studies and qualitative assessments.

\subsection{Experimental Settings}
\label{sub:experimental}
\paragraph{Training details}
\begin{figure*}[tb]
\centering

\begin{minipage}{\textwidth}
\centering
\begin{minipage}{0.25\textwidth}
\centering
\includegraphics[width=\textwidth]{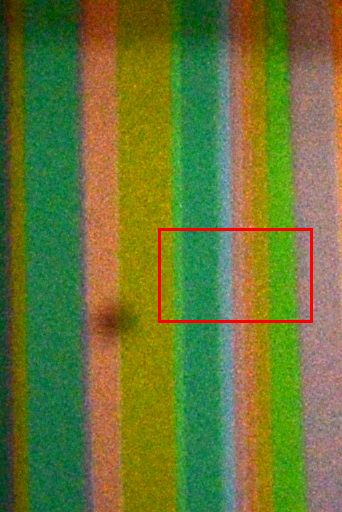}\\[-1mm]
\small CC Noisy
\end{minipage}%
\hspace{-0.5mm}%
\begin{minipage}{0.752\textwidth}
\centering
\vspace{-0.6mm}
\begin{minipage}{0.24\textwidth}
\centering
\includegraphics[width=\textwidth]{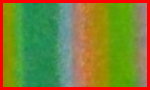}\\[-1mm]
\small R2R 37.03dB
\end{minipage}
\begin{minipage}{0.24\textwidth}
\centering
\includegraphics[width=\textwidth]{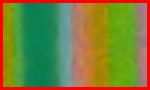}\\[-1mm]
\small AP-BSN 35.43dB
\end{minipage}
\begin{minipage}{0.24\textwidth}
\centering
\includegraphics[width=\textwidth]{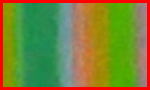}\\[-1mm]
\small SSID 38.48dB
\end{minipage}
\begin{minipage}{0.24\textwidth}
\centering
\includegraphics[width=\textwidth]{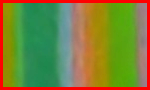}\\[-1mm]
\small APR-RD 39.77dB
\end{minipage}

\vspace{1mm}

\begin{minipage}{0.24\textwidth}
\centering
\includegraphics[width=\textwidth]{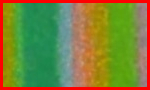}\\[-1mm]
\small Mask-DN 38.05dB
\end{minipage}
\begin{minipage}{0.24\textwidth}
\centering
\includegraphics[width=\textwidth]{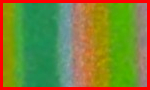}\\[-1mm]
\small CLIP-DN 37.87dB
\end{minipage}
\begin{minipage}{0.24\textwidth}
\centering
\includegraphics[width=\textwidth]{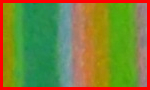}\\[-1mm]
\small AFM 39.28dB
\end{minipage}
\begin{minipage}{0.24\textwidth}
\centering
\includegraphics[width=\textwidth]{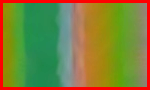}\\[-1mm]
\small IDF 35.80dB
\end{minipage}

\vspace{1mm}

\begin{minipage}{0.24\textwidth}
\centering
\includegraphics[width=\textwidth]{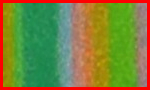}\\[-1mm]
\small NAFNet 37.98dB
\end{minipage}
\begin{minipage}{0.24\textwidth}
\centering
\includegraphics[width=\textwidth]{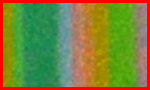}\\[-1mm]
\small XFormer 36.85dB
\end{minipage}
\begin{minipage}{0.24\textwidth}
\centering
\includegraphics[width=\textwidth]{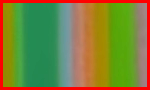}\\[-1mm]
\small \textbf{Ours 42.19dB}
\end{minipage}
\begin{minipage}{0.24\textwidth}
\centering
\includegraphics[width=\textwidth]{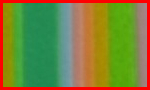}\\[-1mm]
\small Ground Truth
\end{minipage}
\end{minipage}

\vspace{1mm}
\end{minipage}

\begin{minipage}{\textwidth}
\centering
\begin{minipage}{0.25\textwidth}
\centering
\includegraphics[width=\textwidth]{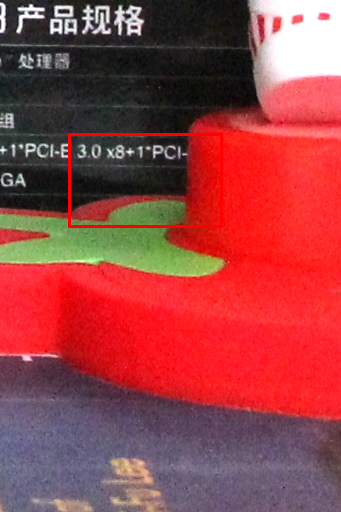}\\[-1mm]
\small HighISO Noisy
\end{minipage}%
\hspace{-0.5mm}%
\begin{minipage}{0.752\textwidth}
\centering
\vspace{-0.6mm}
\begin{minipage}{0.24\textwidth}
\centering
\includegraphics[width=\textwidth]{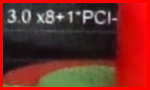}\\[-1mm]
\small R2R 34.52dB
\end{minipage}
\begin{minipage}{0.24\textwidth}
\centering
\includegraphics[width=\textwidth]{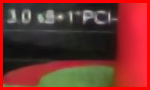}\\[-1mm]
\small AP-BSN 28.26dB
\end{minipage}
\begin{minipage}{0.24\textwidth}
\centering
\includegraphics[width=\textwidth]{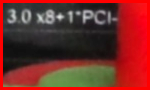}\\[-1mm]
\small SSID 35.59dB
\end{minipage}
\begin{minipage}{0.24\textwidth}
\centering
\includegraphics[width=\textwidth]{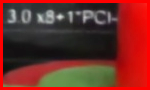}\\[-1mm]
\small APR-RD 35.70dB
\end{minipage}

\vspace{1mm}

\begin{minipage}{0.24\textwidth}
\centering
\includegraphics[width=\textwidth]{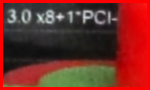}\\[-1mm]
\small Mask-DN 36.83dB
\end{minipage}
\begin{minipage}{0.24\textwidth}
\centering
\includegraphics[width=\textwidth]{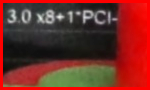}\\[-1mm]
\small CLIP-DN 36.97dB
\end{minipage}
\begin{minipage}{0.24\textwidth}
\centering
\includegraphics[width=\textwidth]{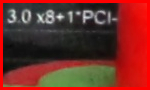}\\[-1mm]
\small AFM 38.30dB
\end{minipage}
\begin{minipage}{0.24\textwidth}
\centering
\includegraphics[width=\textwidth]{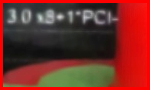}\\[-1mm]
\small IDF 33.80dB
\end{minipage}

\vspace{1mm}

\begin{minipage}{0.24\textwidth}
\centering
\includegraphics[width=\textwidth]{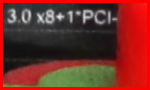}\\[-1mm]
\small NAFNet 37.39dB
\end{minipage}
\begin{minipage}{0.24\textwidth}
\centering
\includegraphics[width=\textwidth]{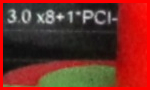}\\[-1mm]
\small XFormer 37.14dB
\end{minipage}
\begin{minipage}{0.24\textwidth}
\centering
\includegraphics[width=\textwidth]{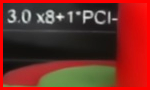}\\[-1mm]
\small \textbf{Ours 40.43dB}
\end{minipage}
\begin{minipage}{0.24\textwidth}
\centering
\includegraphics[width=\textwidth]{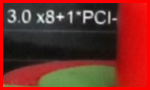}\\[-1mm]
\small Ground Truth
\end{minipage}
\end{minipage}

\end{minipage}
\vspace{-2mm}
\caption{Comparison between the qualitative results of various denoising networks including ours (noise translation network with pretrained NAFNet), on the out-of-distribution (OOD) datasets.
Our result displays cleaner outputs compared to other state-of-the-art networks that are directly trained from a real-noise dataset.
Zoom in for better comparison.}
\vspace{-2mm}
\label{fig:qual_comparison}
\end{figure*}

Our framework is model-agnostic and can be applied to various denoising networks.
To demonstrate the effectiveness of the proposed approach, we adopt NAFNet~\citep{chen2022simple} and Xformer~\citep{zhang2023xformer}, pretrained on BSD400~\citep{martin2001database}, WED~\citep{ma2016waterloo}, and SIDD-Medium~\citep{abdelhamed2018high} datasets.
BSD400 and WED consist of clean images while SIDD provides real noisy-clean image pairs.
During the training of a denoising network, noisy images are created by adding Gaussian noise with a standard deviation of 15 to clean images from BSD400 and WED and noisy images from SIDD.
The noise translation network is trained subsequently, using SIDD images with data augmentation of random Gaussian noise sampled from $[0, 15]$.
The hyperparameters are set as follows: a Gaussian noise level $\tilde{\sigma}= 100$, $\alpha=5\times10^{-2}$, and $\beta=2\times10^{-3}$.
Additional training details such as batch size or training resources are discussed in Appendix.

\vspace{-2mm}
\paragraph{Evaluation}

To evaluate the generalization performance of our framework, we employ various real-world image denoising benchmarks.
We conduct experiments using SIDD validation dataset~\citep{abdelhamed2018high}, Poly~\citep{xu2018real}, CC~\citep{nam2016holistic}, HighISO~\citep{yue2019high}, iPhone, Huawei, Oppo, Sony, and Xiaomi~\citep{kong2023comparison}.
The SIDD validation dataset consists of 256$\times$256 images, while the Poly and CC datasets contain images with a resolution of $512 \times 512$ pixels.
Images from HighISO, iPhone, Huawei, Oppo, Sony, and Xiaomi are $1024 \times 1024$ pixels in size.

\subsection{Results}

\paragraph{Performance on real noise}
Table~\ref{tab:real_main} presents the performance of the proposed approach applied to existing denoising networks, NAFNet~\citep{chen2022simple} and Xformer~\citep{zhang2023xformer}.
We also compare their results with those from self-supervised image denoising methods, including R2R~\citep{pang2021recorrupted}, AP-BSN~\citep{lee2022ap}, SSID~\citep{li2023spatially} and APR-RD~\citep{kim2025apr} and generalizable image denoising methods, including Mask-Denoising~\citep{chen2023masked}, Clip-Denoising~\citep{Jun2024Transfer}, AFM~\citep{ryou2024robust}, and IDF~\citep{kim2025idf}.
We used the officially published, SIDD-pretrained models for comparisons if available.
For the methods without such pretrained models such as Mask-Denoising, Clip-Denoising, and IDF, we trained these models on the SIDD training dataset until convergence to ensure a fair comparison.

As shown in Table~\ref{tab:real_main}, incorporating our noise translation framework, denoted by NAFNet + NTN and Xformer + NTN, achieves substantial average PSNR gains of 1.51dB and 1.46dB, respectively, in out-of-distribution (OOD) scenarios.
Furthermore, our method demonstrates superior real-world image denoising generalization performance compared to other self-supervised and generalizable denoising methods.
Additional comparisons—including LAN~\cite{kim2024lan} applied to the same NAFNet architecture, as well as results on Restormer~\citep{zamir2022restormer} and KBNet~\citep{Zhang2023kbnet} with NTN—are provided in the Appendix.

\paragraph{Qualitative results on OOD benchmarks}
Figure~\ref{fig:qual_comparison} illustrates qualitative results for denoising models on several real-world OOD datasets, including CC~\citep{nam2016holistic}, and HighISO~\citep{yue2019high}.
Our model demonstrates significant superiority over other denoising methods in these challenging scenarios.
Qualitative results on every benchmark listed in Table~\ref{tab:real_main} are available in the Appendix.

\subsection{Analysis}

In this section, we present a comprehensive analysis using our noise translation network combined with a pretrained NAFNet as a denoising network (NAFNet + NTN).


\begin{table}[t]
    \centering
    \caption{
        Ablation study of the Gaussian injection block (GIBlock) and the explicit noise translation loss.
        Full results on individual OOD datasets are provided in the Appendix.
    }
    \vspace{-2.5mm}
    \setlength{\tabcolsep}{12pt}
    \renewcommand{\arraystretch}{0.85}
    \scalebox{0.8}{
        \begin{tabular}{l c c c}
            \toprule
                                                                               & Metric & SIDD     & OOD Avg.        \\
            \midrule
            \midrule
            \multirow{2}{*}{Baseline translation}                              & PSNR   & 39.35    & 39.27           \\
                                                                               & SSIM   & 0.9573   & 0.9774          \\
            \midrule
            \multirow{2}{*}{+ GIBlock}                                         & PSNR   & 39.05    & 39.61           \\
                                                                               & SSIM   & 0.9556   & 0.9801          \\
            \midrule
            \multirow{2}{*}{{+ Explicit loss ($\mathcal{L}_\text{explicit}$)}} & PSNR   & {39.33}  & {39.61}         \\
                                                                               & SSIM   & {0.9572} & {0.9809}        \\
            \midrule
            \multirow{2}{*}{+ Both \textbf{(Ours)}}                            & PSNR   & 39.24    & \textbf{39.94}  \\
                                                                               & SSIM   & 0.9570   & \textbf{0.9826} \\
            \bottomrule
        \end{tabular}
    }
    \label{tab:real_ablation}
    \vspace{-4mm}
\end{table}

\begin{table*}[tb]
	\centering
	\caption{
		Quantitative results based on variations in the noisy input to the pretrained denoising network on the in-distribution SIDD validation set and other out-of-distribution real-world benchmarks.
		$I$, $I + \mathcal{N}_5$, $I + \mathcal{N}_{10}$, and $I + \mathcal{N}_{15}$ represent the noisy input images with additional Gaussian noise levels of 0, 5, 10, and 15, respectively, which are fed into the pretrained Gaussian denoising network.
	}
	\vspace{-2mm}
	\setlength{\tabcolsep}{9pt}
	\renewcommand{\arraystretch}{0.9}
	\scalebox{0.8}{
		\begin{tabular}{c c c c c c c c c c c c c}
			\toprule
			                                        &        & \multicolumn{1}{c}{In-distribution} & \multicolumn{8}{c}{Out-of-distribution} &                                                                                                               \\
			\cmidrule(lr){3-3}\cmidrule(lr){4-11}
			\multicolumn{1}{c}{Input}               & Metric & SIDD                                & ~~Poly~~                                & ~~~CC~~~    & HighISO~    & iPhone      & Huawei      & OPPO        & Sony        & Xiaomi      & OOD Avg.    \\
			\midrule\midrule
			\multirow{2}{*}{$I$}                    & PSNR   & 37.77                               & 15.24                                   & 33.76       & 21.18       & 40.13       & 8.68        & 8.45        & 6.35        & 9.33        & 17.89       \\
			                                        & SSIM   & 0.9360                              & 0.3466                                  & 0.9139      & 0.5232      & 0.9734      & 0.1218      & 0.1138      & 0.0245      & 0.1845      & 0.4002      \\

			\midrule
			\multirow{2}{*}{$I + \mathcal{N}_5$}    & PSNR   & 38.15                               & 27.07                                   & 34.97       & 32.30       & 15.28       & 16.99       & 13.79       & 12.82       & 14.96       & 22.93       \\
			                                        & SSIM   & 0.9436                              & 0.7010                                  & 0.9211      & 0.8227      & 0.2392      & 0.3943      & 0.2564      & 0.1912      & 0.3540      & 0.5359      \\

			\midrule
			\multirow{2}{*}{$I + \mathcal{N}_{10}$} & PSNR   & 38.76                               & \underline{38.27}                       & \underline{37.33}     & \underline{39.40}     & 40.95       & \underline{39.44}     & \underline{39.98}     & 42.96       & \underline{35.91}     & \underline{39.22}     \\
			                                        & SSIM   & 0.9536                              & 0.9795                                  & \underline{0.9850}    & \underline{0.9825}    & 0.9638      & \underline{0.9762}    & 0.9768      & 0.9758      & \underline{0.9728}      & 0.9740      \\

			\midrule
			\multirow{2}{*}{$I + \mathcal{N}_{15}$} & PSNR   & \underline{39.16}                   & 38.08                                   & 36.26       & 38.85       & \underline{41.12}     & 38.71       & 39.69       & \underline{43.42}     & 35.25       & 38.95       \\
			                                        & SSIM   & \underline{0.9565}                  & \underline{0.9834}                      & 0.9829      & 0.9808      & \underline{0.9811} & 0.9719      & \underline{0.9770}    & \bf{0.9886} & 0.9680      & \underline{0.9767}    \\

			\midrule
			\multirow{2}{*}{\textbf{$I_{\mathcal{T}}$}}      & PSNR   & \bf{39.24}                          & \bf{38.72}                              & \bf{37.84}  & \bf{40.00}  & \bf{42.08}  & \bf{39.83}  & \bf{40.55}  & \bf{44.34}  & \bf{36.17}  & \bf{39.94}  \\
			                                        & SSIM   & \bf{0.9570}                         & \bf{0.9855}                             & \bf{0.9877} & \bf{0.9856} & \bf{0.9812} & \bf{0.9782} & \bf{0.9801} & \underline{0.9875} & \bf{0.9749} & \bf{0.9826} \\

			\bottomrule
		\end{tabular}
	}
	\label{tab:real_add_gaussian}
\end{table*}

\paragraph{Ablation study}
Table~\ref{tab:real_ablation} presents the ablation study results of the proposed algorithm.
The baseline model is trained using only the implicit noise translation loss, excluding both the Gaussian noise injection block (GIBlock) and the explicit noise translation loss.
The results demonstrate that both components (GIBlock and explicit loss) are essential for enhancing performance: GIBlock promotes the translation of complex noise into Gaussian noise, while the explicit loss provides a clear constraint that guides the generation toward \textit{i.i.d.} Gaussian noise.

\paragraph{Comparisons with simple Gaussian noise additions}
We evaluate the effectiveness of our noise translation network by comparing it to directly adding Gaussian noise to the input.
As shown in Table~\ref{tab:real_add_gaussian}, simply adding Gaussian noise yields reasonable generalization.
However, some datasets perform better with \(I + \mathcal{N}_{10}\), while others with \(I + \mathcal{N}_{15}\), indicating that noise characteristics vary across datasets or even individual images.
This variation underscores the limitation of fixed noise and the need for a more adaptive approach.
Our noise translation network can adaptively transform input noise into ideal Gaussian noise, achieving consistent performance gains across all datasets.

%

\begin{table}[t]
    \centering
    \caption{
        Transferability of our noise translation network.
        Once trained with NAFNet~\citep{chen2022simple}, it integrates seamlessly with other denoising networks.
        We present performance only in PSNR$\uparrow$.
        Full results on each OOD dataset are provided in the Appendix.
    }
    \vspace{-2mm}
    \setlength{\tabcolsep}{10pt}
    \scalebox{0.8}{
        \begin{tabular}{c c c c}
            \toprule
            $\mathcal{D}_{\theta^*}$ for testing                  & $\mathcal{D}_{\theta^*}$ for training $\mathcal{T}_{\phi}$ & SIDD  & OOD Avg. \\

            \midrule
            \multirow{2}{*}{Xformer~\citep{zhang2023xformer}}     & \multirow{1}{*}{NAFNet~\citep{chen2022simple}}                                    & 39.17 & 39.94    \\
                                                                  & \multirow{1}{*}{Xformer~\citep{zhang2023xformer}}                                   & 39.10 & 40.04    \\
            \bottomrule
        \end{tabular}
    }
    \label{tab:ablation_cross}
    \vspace{-2mm}
\end{table}

\vspace{-2mm}
\paragraph{Transferability of noise translation network}
The noise translation network is functionally decoupled from the denoising network, allowing a single trained noise translation model to be paired with various pretrained denoising networks.
 Table~\ref{tab:ablation_cross} demonstrates this property by applying the noise translation network, trained with pretrained NAFNet~\citep{chen2022simple}, to other architecture, Xformer~\citep{zhang2023xformer}.
The performance remains comparable to that of individually trained noise translation networks for each denoiser.
This suggests that a one-time trained noise translation network can generalize effectively across diverse pretrained denoising networks without additional training.


\begin{figure}[t]
    \centering
    \begin{tabular}{@{}c@{\hspace{0.1em}}c@{\hspace{0.1em}}c@{\hspace{0.1em}}c@{}}
        \parbox[t]{0.12\textwidth}{
            \centering
        \includegraphics[scale=0.147]{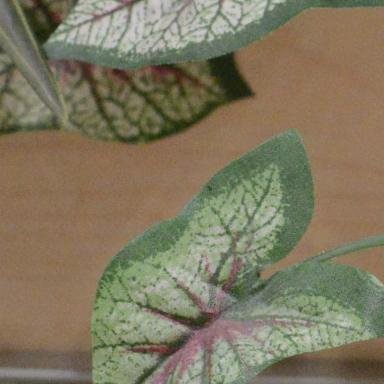}          \\
        } &
        \hspace{-1.2mm}
        \parbox[t]{0.12\textwidth}{
            \centering
        \includegraphics[scale=0.147]{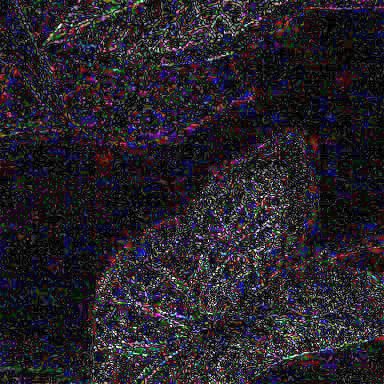}       \\
        } &
        \hspace{-1.2mm}
        \parbox[t]{0.12\textwidth}{
            \centering
        \includegraphics[scale=0.147]{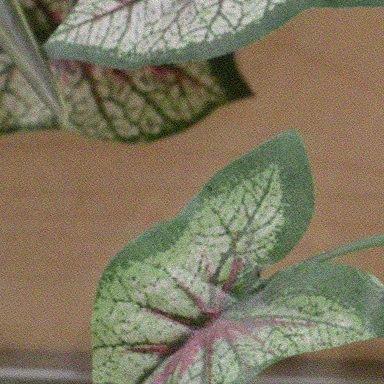}        \\
        } &
        \hspace{-1.2mm}
        \parbox[t]{0.12\textwidth}{
            \centering
        \includegraphics[scale=0.147]{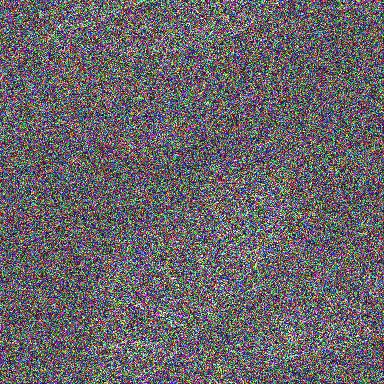} \\

        }                                                                                                \\
        \vspace{-5mm}
    \end{tabular}
    \begin{tabular}{@{}c@{\hspace{0.1em}}c@{\hspace{0.1em}}c@{\hspace{0.1em}}c@{}}
        \parbox[t]{0.12\textwidth}{
            \centering
        \includegraphics[scale=0.147]{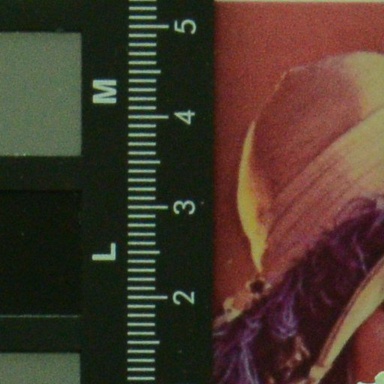}          \\
            \fontsize{8pt}{8pt}\selectfont Real Noisy Image
        } &
        \hspace{-1.2mm}
        \parbox[t]{0.12\textwidth}{
            \centering
        \includegraphics[scale=0.147]{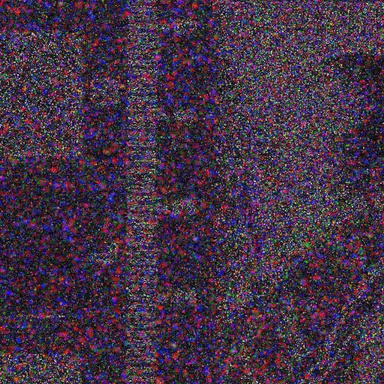}       \\
            \fontsize{8pt}{8pt}\selectfont Real Noise
        } &
        \hspace{-1.2mm}
        \parbox[t]{0.12\textwidth}{
            \centering
        \includegraphics[scale=0.147]{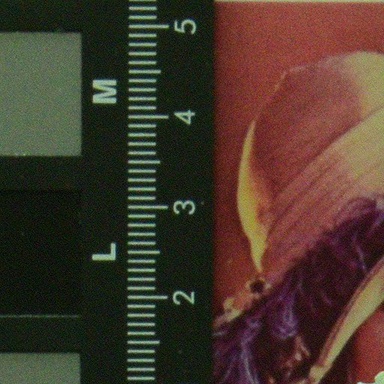}        \\
            \fontsize{8pt}{8pt}\selectfont Translated Image
        } &
        \hspace{-1.2mm}
        \parbox[t]{0.12\textwidth}{
            \centering
        \includegraphics[scale=0.147]{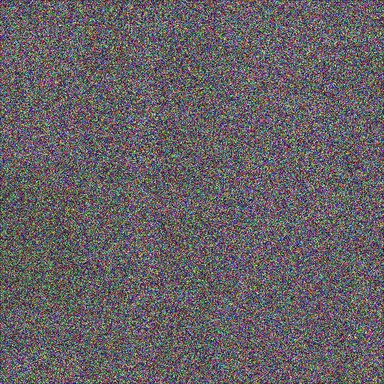} \\
            \fontsize{8pt}{8pt}\selectfont Translated Noise

        }                                                                                  \\
    \end{tabular}
    \vspace{-2mm}
    \caption{
        Visual results of noise translation. 
        The noisy image in the top row is from the Poly~\citep{xu2018real} dataset, while the one in the bottom row is from the CC~\citep{nam2016holistic} dataset. 
        For enhanced visualization, noise is displayed as the absolute value, scaled by a factor of 10.
    }
    \label{fig:translation}
\end{figure}

\begin{figure}[t]

    \centering
    \begin{subfigure}[t]{0.2323\textwidth}
        \centering
        \includegraphics[width=0.975\linewidth]{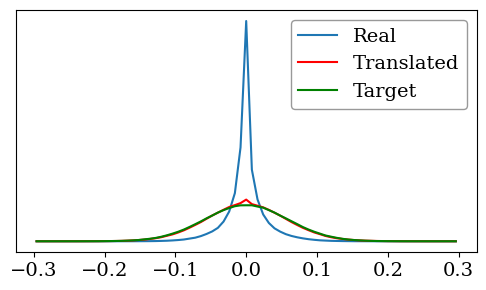}
        \caption{Poly - Spatial}
        \label{fig:subfig-poly-spatial}
    \end{subfigure}
    \hspace{0.1em}
    \begin{subfigure}[t]{0.2277\textwidth}
        
        \centering
        \includegraphics[width=\linewidth]{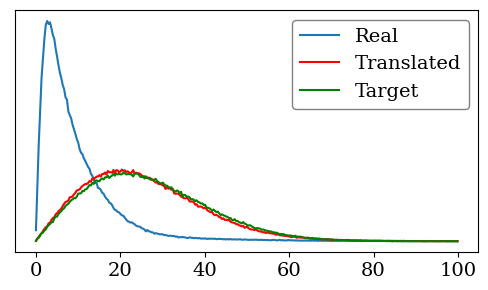}
        \caption{Poly - Frequency}
        
        \label{fig:subfig-poly-frequency}
    \end{subfigure}

    \begin{subfigure}[t]{0.2323\textwidth}
        \centering
        \includegraphics[width=0.975\linewidth]{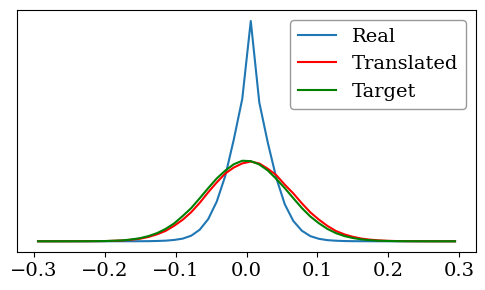}
        \caption{CC - Spatial}
        \label{fig:subfig-cc-spatial}
    \end{subfigure}
    \hspace{0.1em}
    \begin{subfigure}[t]{0.2277\textwidth}
        \centering
        \includegraphics[width=\linewidth]{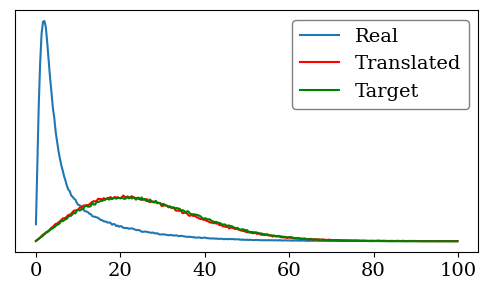}
        \caption{CC - Frequency}
        \label{fig:subfig-cc-frequency}
    \end{subfigure}

    \vspace{-2mm}
    \caption{
        Histogram of noise distribution before (Real) and after noise translation network in both spatial and frequency domains.
        Target noise corresponds to the Gaussian noise with a level of 15, which the denoising network has been pretrained to remove.
    }
    \vspace{-2mm}
    \label{fig:histogram}
\end{figure}

\begin{figure}[t]

    \centering
    \begin{subfigure}[t]{0.23\textwidth}
    \includegraphics[width=0.975\textwidth]{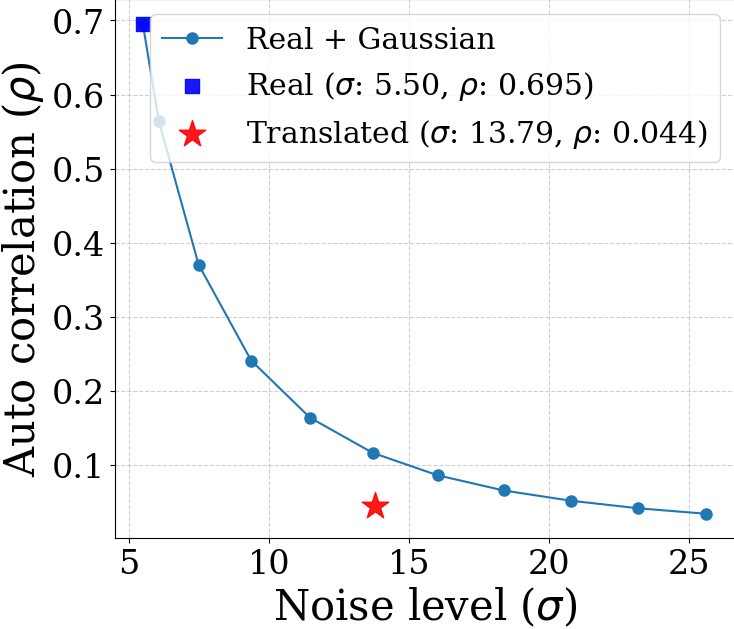}
        \caption{CC}
        \label{fig:subfig-poly-spatial}
        
    \end{subfigure}
    \hspace{0.1em}
    \begin{subfigure}[t]{0.23\textwidth}
        \includegraphics[width=0.975\textwidth]{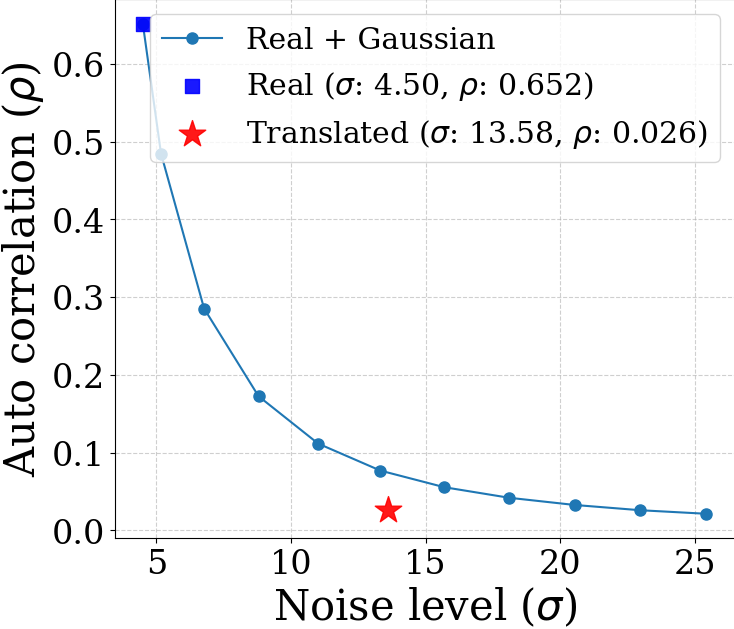}
            \caption{HighISO}
            \label{fig:subfig-poly-spatial}
            
    \end{subfigure}

    \vspace{-2mm}
    \caption{
        Analysis of noise characteristics. 
        The blue curve illustrates the changes in noise level and auto-correlation resulting from adding various levels of Gaussian noise to the real noise. 
        The results are averaged over the CC~\citep{nam2016holistic} and HighISO~\citep{yue2019high} datasets.
        }
    \vspace{-2mm}
    \label{fig:var_ac}
\end{figure}

\vspace{-2mm}
\paragraph{Analysis of translated noise}
Figure~\ref{fig:translation} visualizes the effect of our noise translation.
Real noise shows strong spatial correlation, which our network effectively removes, producing noise that closely resembles ideal Gaussian noise.

Figure~\ref{fig:histogram} shows the noise distributions of the two images in Figure~\ref{fig:translation}.
In the spatial domain, ideal Gaussian noise follows a normal distribution; in the frequency domain, it follows a Rayleigh distribution (see Section~\ref{sub:explicit}).
Real noise significantly deviates from these targets, whereas translated noise closely matches them, demonstrating effective alignment with spatially uncorrelated, \textit{i.i.d.} Gaussian noise.

Additionally, Figure~\ref{fig:var_ac} plots the noise level versus auto-correlation for each dataset, comparing three cases: (1) real noise, (2) real noise with added Gaussian noise, and (3) translated noise produced by our method. 
While adding Gaussian noise to real noise increases the overall noise level and reduces auto-correlation, our translated noise achieves substantially lower auto-correlation at comparable noise levels.
This indicates that our translation process more effectively decorrelates the noise.

\vspace{-2mm}
\paragraph{Evaluating Trade-offs in ID performance}

\begin{figure}[t!]
    \centering
    \scalebox{0.58}{
        \begin{tabular}{@{}ccc@{}}
            \begin{subfigure}[t]{0.255\textwidth}
                \centering
                \includegraphics[width=\linewidth]{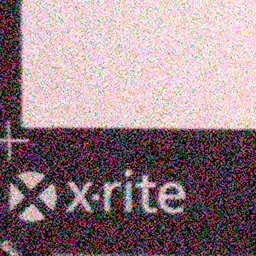}
                \caption*{\fontsize{13pt}{13pt}\selectfont SIDD Noisy}
            \end{subfigure}                         &
            \hspace{-0.7em}
            \begin{subfigure}[t]{0.255\textwidth}
                \centering
                \includegraphics[width=\linewidth]{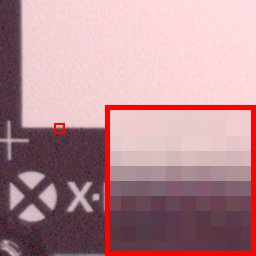}
                \caption*{\fontsize{13pt}{13pt}\selectfont Ground Truth}
            \end{subfigure}                    &
            \hspace{-0.7em}
            \begin{subfigure}[t]{0.255\textwidth}
                \centering
                \includegraphics[width=\linewidth]{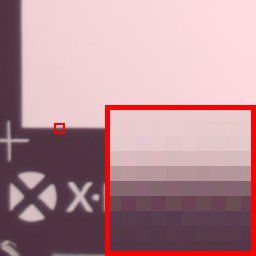}
                \caption*{\fontsize{13pt}{13pt}\selectfont Restormer 33.17 dB}
            \end{subfigure} \\
            \begin{subfigure}[t]{0.255\textwidth}
                \centering
                \includegraphics[width=\linewidth]{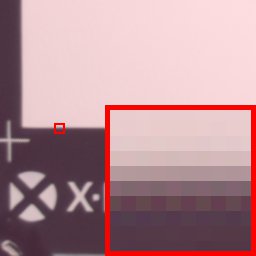}
                \caption*{\fontsize{13pt}{13pt}\selectfont NAFNet 33.09 dB}
            \end{subfigure}  &
            \hspace{-0.7em}
            \begin{subfigure}[t]{0.255\textwidth}
                \centering
                \includegraphics[width=\linewidth]{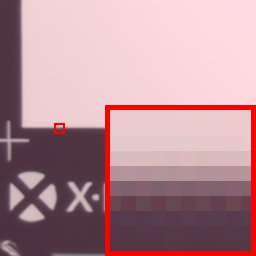}
                \caption*{\fontsize{13pt}{13pt}\selectfont Xformer 33.41 dB}
            \end{subfigure} &
            \hspace{-0.7em}
            \begin{subfigure}[t]{0.255\textwidth}
                \centering
                \includegraphics[width=\linewidth]{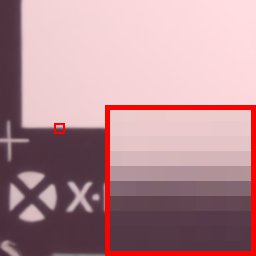}
                \caption*{\fontsize{13pt}{13pt}\selectfont Ours 33.01 dB}
            \end{subfigure}
        \end{tabular}
    }
    \vspace{-2mm}
    \caption{Denoised results of an in-distribution SIDD~\cite{abdelhamed2018high} noisy image.
        Only our approach effectively avoids overfitting, preventing unnecessary zipper artifacts prevalent in the training set, while preserving the visual quality.
        Zoom in for detailed comparison.}
    \vspace{-1mm}
    \label{fig:zipper}
\end{figure}

Figure~\ref{fig:zipper} displays enlarged qualitative results on the SIDD validation dataset.
Although our method may appear to exhibit a trade-off in in-distribution performance, this primarily stems from the pronounced overfitting of other models, which often reconstruct irrelevant artifacts inherent in the training data.
In contrast, our approach avoids such overfitting while still producing high-quality reconstructions.

If improving in-distribution performance on a specific dataset is desired, the denoising network can be fine-tuned. 
We demonstrate this by freezing the trained translation network and fine-tuning only the denoising network on the SIDD dataset.
As shown in Table~\ref{tab:real_finetuning}, this procedure yields an improvement of approximately 0.3dB on SIDD while having negligible impact on out-of-distribution performance.
\vspace{-2mm}
\paragraph{Computational efficiency}
Table~\ref{tab:params_flops} presents a comparison of our noise translation network with other image denoising networks including NAFNet~\citep{chen2022simple} and Xformer~\citep{zhang2023xformer} in terms of the number of parameters and multiply-accumulate operations (MACs).
Our noise translation network is substantially smaller in both parameter count and MACs than the image denoising networks, adding only negligible computational overhead during inference.

\begin{table}[t!]
	\centering
	\caption{
		Impact of SIDD~\citep{abdelhamed2018high} fine-tuning.
		We present the performance of NAFNet and Xformer when the noise translation network is frozen and only the denoising network is fine-tuned.}
	\vspace{-2mm}
	\setlength{\tabcolsep}{7pt}
	\renewcommand{\arraystretch}{0.9}
	\scalebox{0.8}{

		\begin{tabular}{l c c c}
			\toprule
			                                   & Metric & SIDD            & OOD Avg.         \\
			\midrule
			\midrule
			\multirow{2}{*}{{NAFNet + NTNet}}  & PSNR   & 39.56 (+0.32)   & 39.95 (+0.01)    \\
			                                   & SSIM   & 0.9580 (+0.001) & 0.9830 (+0.0004) \\
			\midrule
			\multirow{2}{*}{{Xformer + NTNet}} & PSNR   & 39.43 (+0.33)   & 39.97 (-0.07)    \\
			                                   & SSIM   & 0.9589 (+0.0032) & 0.9903 (+0.0075) \\
			\bottomrule
		\end{tabular}
	}
	\label{tab:real_finetuning}
\end{table}


\begin{table}[t]
	\centering
	\caption{
		Parameter counts and MACs of denoising networks and our noise translation network. We report the number of parameters and MACs at inference, estimated with an input size of 256$\times$256.
	}
	\vspace{-2mm}
	\setlength{\tabcolsep}{12pt}
	\renewcommand{\arraystretch}{0.9}
	\scalebox{0.8}{
		\begin{tabular}{l c c}
			\toprule
			\multicolumn{1}{l}{Architecture}     & Parameters (M) & MACs (G) \\
			\midrule\midrule
			NAFNet~\citep{chen2022simple}        & \ \ 29.10          & \ \ 16.23    \\
			Xformer~\citep{zhang2023xformer}     & \ \ 25.12          & 142.68   \\

			\midrule
			Noise Translation Network            & \ \ \  0.29           & \ \ \  1.07     \\

			\bottomrule
		\end{tabular}
	}
	\vspace{-2mm}
	\label{tab:params_flops}
\end{table}

\section{Conclusion}
\label{sec:conclusion}
We presented a novel noise translation framework for robust image denoising.
Our framework allows us to effectively remove various unseen real noise, even with a limited amount of training data.
By employing the noise translation network, arbitrary out-of-distribution (OOD) noise is translated into Gaussian noise for which an image denoising network has learned during training.
The noise translation network is designed with well-motivated loss functions and architecture, enabling effective noise translation while preserving image contents. 
Our experiments demonstrate that the proposed approach significantly outperforms state-of-the-art denoising models in diverse OOD real-noise benchmarks.
Finally, we highlight that the generalization issue in real-world scenarios remains a critical challenge in image denoising, and our approach offers a promising solution.


\paragraph{Acknowledgements}
This work was supported by the National Research Foundation of Korea (NRF) [RS-2022-NR070855, Trustworthy Artificial Intelligence] and the Institute of Information \& Communications Technology Planning \& Evaluation (IITP) [RS2022-II220959 (No.2022-0-00959), 
(Part 2) Few-Shot Learning of Causal Inference in Vision and Language for Decision Making; 
RS-2025-25442338, AI star Fellowship Support Program (Seoul National Univ.);
No.RS-2021-II211343, Artificial Intelligence Graduate School Program (Seoul National University)], funded by the Korea government (MSIT).

{
    \small
    \bibliographystyle{ieeenat_fullname}
    \bibliography{main}
}


\clearpage


\makeatletter
\twocolumn[
    \centering
    \Large
    \textbf{\thetitle}\\
    \vspace{1em}
    Supplementary Material \\
    \vspace{2.0em}
    \normalsize
    \@author
    \vspace{1.0em}
]
\makeatother
\begingroup
\renewcommand\thefootnote{} 
\footnotetext{\protect\hspace{-0.5em}${}^*$indicates equal contribution.}
\endgroup

\section{Proof on Wasserstein Distance}
\label{sec:proof_wasserstein}

Let $P$ and $Q$ represent two probability distributions over $\mathbb{R}^d$. We use $X \sim P$ and $Y \sim Q$ to denote random variables with the distributions $P$ and $Q$, respectively.
The $p$-Wasserstein distance between two probability measures $P$ and $Q$ is defined as follows:
\begin{equation}
    W_p(P, Q) = \left( \inf_{J \in \mathcal{J}(P, Q)} \int \|x - y\|^p dJ(x, y) \right)^{1/p},
\end{equation}
where $\mathcal{J}(P, Q)$ is the set of all joint distributions (or couplings) $J$ on $(X, Y)$ that have marginals $P$ and $Q$.
This formulation describes the minimum cost of transporting mass from distribution $P$ to distribution $Q$ using the coupling $J$, with the cost measured as the $p$-th power of the distance between points $x$ and $y$.

In the Monge formulation, the goal is to find a transport map $T: \mathbb{R}^d \to \mathbb{R}^d$ such that the push-forward of $P$ under $T$, denoted as $T_\# P$, equals $Q$. This problem can be mathematically formulated as:
\begin{equation}
    \inf_T \int |x - T(x)|^p dP(x),
\end{equation}
where the map $T$ moves the distribution $P$ to $Q$. However, an optimal map $T$ may not always exist.
In such cases, the Kantorovich formulation is used, allowing mass at each point to be split and transported to multiple locations, leading to a coupling-based approach.

For the specific case of $p = 1$, known as the Earth Mover's Distance, the dual formulation of the Wasserstein distance can be expressed as:
\begin{equation}
    W_1(P, Q) = \sup_{f \in F} \left( \int f(x) dP(x) - \int f(x) dQ(x) \right),
\end{equation}
where $F$ represents the set of all Lipschitz continuous functions $f: \mathbb{R}^d \to \mathbb{R}$ such that $|f(y) - f(x)| \leq \|x - y\|$ for all $x, y \in \mathbb{R}^d$.
Then, the $1$-Wasserstein distance is given by:
\begin{equation}
    W_1(P, Q) = \int_0^1 |F^{-1}(z) - G^{-1}(z)| dz,
\end{equation} where $F^{-1}$ and $G^{-1}$ denote the quantile functions (inverse CDFs) of $P$ and $Q$, respectively.

When $P$ and $Q$ are empirical distributions based on the datasets, ${X_1, X_2, \dots, X_n}$ and ${Y_1, Y_2, \dots, Y_n}$, each of size $n$, the Wasserstein distance can be computed as a function of the order statistics:
\begin{equation}
    \label{eq:final_wasserstein}
    W_1(P, Q) = \sum_{i=1}^n |X_{(i)} - Y_{(i)}|,
\end{equation}
where $X_{(i)}$ and $Y_{(i)}$ denote the $i$-th order statistics of the datasets ${X_1, X_2, \dots, X_n}$ and ${Y_1, Y_2, \dots, Y_n}$.

In our approach, we utilize (\ref{eq:final_wasserstein}) to formulate $\mathcal{L}_\text{spatial}$ in Eq. (5) and $\mathcal{L}_\text{freq}$ in Eq. (10), which are employed during training to explicitly transport the translated noise distribution towards the target Gaussian distribution.
We refer to~\citep{villani2003topics} for a detailed discussion on Wasserstein distances and optimal transport.

\section{Additional training details}

The denoising models are trained for 200K iterations with a batch size of 32, except for Restormer and Xformer, where the batch size is reduced to 4 due to the limitation of computational resources.
The noise translation network is trained for 5K iterations with a batch size of 4.
The explicit noise translation loss was adopted only after half (2.5k) iterations, as early translated noise exhibits large deviations that destabilize optimization.
Denoising network and noise translation network are trained with the AdamW~\citep{loshchilov2018decoupled} optimizer with an initial learning rate of $10^{-3}$, which is reduced to  $10^{-7}$ and  $10^{-5}$ by a cosine annealing schedule, respectively.
Each image is randomly cropped to $256 \times 256$ for training.
All trainings were conducted using two NVIDIA RTX A6000 GPUs.


\begin{table*}[t]
    \centering
    \caption{
        Demonstration of model-agnostic applicability. 
        We compare representative state-of-the-art denoising networks with their counterparts integrated with our proposed framework (+ NTN). 
        The results are presented in terms of PSNR$\uparrow$ (dB) and SSIM$\uparrow$.
        Models marked with an asterisk (*) are evaluated using official out-of-the-box models.
    }
    \vspace{-2mm}
    \setlength{\tabcolsep}{9pt}
    \renewcommand{\arraystretch}{0.85}
    \scalebox{0.75}{
        \begin{tabular}{l c c c c c c c c c c c}
            \toprule
            \multicolumn{2}{c}{}                                   & \multicolumn{1}{c}{In-distribution} & \multicolumn{8}{c}{Out-of-distribution} &                                                                                                                             \\
            \cmidrule(lr){3-3} \cmidrule(lr){4-11}
            \multicolumn{1}{c}{Denoising Algo.}                    & Metric                              & SIDD                                    & Poly        & CC          & HighISO     & iPhone      & Huawei      & OPPO        & Sony        & Xiaomi      & OOD Avg.    \\
            \midrule\midrule

            \multirow{2}{*}{Restormer*~\citep{zamir2022restormer}} & PSNR                                & {40.02}                                 & 37.66       & 36.33       & 38.29       & 40.13       & 38.42       & 39.56       & 44.19       & 35.65       & 38.78       \\
                                                                   & SSIM                                & {0.9603}                                & 0.9793      & 0.9807      & 0.9756      & 0.9734      & 0.9675      & 0.9773      & \bf{0.9894} & 0.9710      & 0.9768      \\
            \midrule
            \multirow{2}{*}{\bf{Restormer + NTN}}                  & PSNR                                & {39.22}                                 & \bf{38.76}  & \bf{37.68}  & \bf{40.14}  & \bf{41.85}  & \bf{39.72}  & \bf{40.64}  & \bf{44.29}  & \bf{36.19}  & \bf{39.91}  \\
                                                                   & SSIM                                & {0.9569}                                & \bf{0.9854} & \bf{0.9866} & \bf{0.9857} & \bf{0.9786} & \bf{0.9766} & \bf{0.9800} & {0.9871}    & \bf{0.9750} & \bf{0.9819} \\
            \midrule\midrule
            \multirow{2}{*}{NAFNet*~\citep{chen2022simple}}        & PSNR                                & {39.97}                                 & 37.17       & 35.69       & 38.32       & 40.25       & 37.73       & 39.64       & 43.65       & 34.99       & 38.43       \\
                                                                   & SSIM                                & {0.9600}                                & 0.9717      & 0.9811      & 0.9788      & 0.9707      & 0.9680      & 0.9786      & 0.9829      & 0.9685      & 0.9750      \\

            \midrule
            \multirow{2}{*}{\bf{NAFNet + NTN}}                     & PSNR                                & {39.24}                                 & \bf{38.72}  & \bf{37.84}  & \bf{40.00}  & \bf{42.08}  & \bf{39.83}  & \bf{40.55}  & \bf{44.34}  & \bf{36.17}  & \bf{39.94}  \\
                                                                   & SSIM                                & {0.9570}                                & \bf{0.9855} & \bf{0.9877} & \bf{0.9856} & \bf{0.9812} & \bf{0.9782} & \bf{0.9801} & \bf{0.9875} & \bf{0.9749} & \bf{0.9826} \\
            \midrule\midrule

            \multirow{2}{*}{KBNet*~\citep{Zhang2023kbnet}}         & PSNR                                & {40.35}                                 & 36.82       & 35.23       & 38.09       & 38.00       & 35.18       & 37.80       & 41.79       & 34.25       & 37.15       \\
                                                                   & SSIM                                & {0.9623}                                & 0.9789      & 0.9810      & 0.9788      & 0.9533      & 0.9462      & 0.9663      & 0.9790      & 0.9644      & 0.9685      \\

            \midrule
            \multirow{2}{*}{\bf{KBNet + NTN}}                      & PSNR                                & {39.13}                                 & \bf{38.62}  & \bf{37.61}  & \bf{39.89}  & \bf{41.76}  & \bf{39.80}  & \bf{40.57}  & \bf{44.22}  & \bf{36.07}  & \bf{39.82}  \\
                                                                   & SSIM                                & {0.9564}                                & \bf{0.9843} & \bf{0.9861} & \bf{0.9849} & \bf{0.9761} & \bf{0.9777} & \bf{0.9798} & \bf{0.9845} & \bf{0.9743} & \bf{0.9810} \\
            \midrule\midrule
            \multirow{2}{*}{XFormer*~\citep{zhang2023xformer}}     & PSNR                                & {39.98}                                 & 37.45       & 35.95       & 37.86       & 39.99       & 38.35       & 39.61       & 43.89       & 35.54       & 38.58       \\
                                                                   & SSIM                                & {0.9603}                                & 0.9778      & 0.9782      & 0.9741      & 0.9732      & 0.9678      & 0.9779      & \bf{0.9892} & 0.9706      & 0.9761      \\

            \midrule
            \multirow{2}{*}{\bf{XFormer + NTN}}                    & PSNR                                & {39.10}                                 & \bf{38.75}  & \bf{37.82}  & \bf{40.29}  & \bf{42.20}  & \bf{39.89}  & \bf{40.70}  & \bf{44.44}  & \bf{36.25}  & \bf{40.04}  \\
                                                                   & SSIM                                & {0.9557}                                & \bf{0.9856} & \bf{0.9862} & \bf{0.9860} & \bf{0.9826} & \bf{0.9779} & \bf{0.9807} & {0.9886}    & \bf{0.9751} & \bf{0.9828} \\

            \bottomrule
        \end{tabular}
    }
    \label{tab:real_main_supp}
    \vspace{-2mm}
\end{table*}

\begin{table*}[t]
	\centering
	\caption{
		Quantitative comparison of additional state-of-the-art image denoising methods on the SIDD validation set and multiple real-noise benchmarks.
		We present the results in terms of PSNR$\uparrow$ (dB) and SSIM$\uparrow$.
		Methods marked with an asterisk (*) are evaluated using official out-of-the-box models.
	}
	\vspace{-2mm}
	\setlength{\tabcolsep}{6.5pt}
	\renewcommand{\arraystretch}{0.85}
	\scalebox{0.75}{
		\begin{tabular}{c l c c c c c c c c c c c c  c c c}
			\toprule
			                             & \multicolumn{1}{c}{Denoising method}                     & Metric & SIDD       & ~~Poly~~    & ~~~CC~~~    & HighISO~    & iPhone      & Huawei      & OPPO        & Sony        & Xiaomi      & Total Avg.  \\
			\midrule
			\midrule

			\multirow{6.8}{*}{GAN-based} & \multirow{2}{*}{DANet*~\citep{yue2020dual}}              & PSNR   & 39.47      & 37.53       & 36.16       & 38.35       & 40.46       & 38.40       & 39.85       & 44.36       & 35.57       & 38.90       \\
			                             &                                                          & SSIM   & 0.9570     & 0.9800      & 0.9815      & 0.9770      & 0.9751      & 0.9691      & 0.9789      & 0.9895      & 0.9706      & 0.9754      \\
			\cmidrule{2-13}
			                             & \multirow{2}{*}{C2N + DIDN*~\citep{jang2021c2n}}         & PSNR   & 35.36      & 37.60       & 36.92       & 38.82       & 40.48       & 38.56       & 40.08       & \bf{44.63}  & 35.44       & 38.65       \\
			                             &                                                          & SSIM   & 0.9321     & 0.9781      & 0.9813      & 0.9765      & 0.9745      & 0.9679      & 0.9793      & \bf{0.9899} & 0.9684      & 0.9720      \\
			\cmidrule{2-13}
			                             & \multirow{2}{*}{PNGAN + MIRNet*~\citep{cai2021learning}} & PSNR   & \bf{39.98} & 37.41       & 36.10       & 38.24       & 39.93       & 38.02       & 39.56       & 43.15       & 35.24       & 38.62       \\
			                             &                                                          & SSIM   & \bf{0.959} & 0.9783      & 0.9810      & 0.9763      & 0.9711      & 0.9678      & 0.9782      & 0.9860      & 0.9690      & 0.9741      \\

			\midrule
			\multirow{9}{*}{Others}      & \multirow{2}{*}{PDDenoising*~\citep{zhou2020awgn}}       & PSNR   & 33.95      & 37.10       & 35.82       & 37.09       & 39.13       & 37.69       & 38.75       & 43.60       & 35.03       & 37.57       \\
			                             &                                                          & SSIM   & 0.8986     & 0.9716      & 0.9742      & 0.9628      & 0.9611      & 0.9592      & 0.9707      & 0.9866      & 0.9626      & 0.9608      \\

			\cmidrule{2-13}
			                             & \multirow{2}{*}{CLIPDenoising*~\citep{Jun2024Transfer}}  & PSNR   & 34.79      & 37.54       & 36.30       & 38.01       & 40.09       & 38.74       & 39.56       & 42.94       & 35.50       & 38.16       \\
			                             &                                                          & SSIM   & 0.8982     & 0.9794      & 0.9809      & 0.9771      & 0.9685      & 0.9715      & 0.9769      & 0.9824      & 0.9707      & 0.9672      \\
			\cmidrule{2-13}
			                             & \multirow{2}{*}{MaskDenoising*~\citep{chen2023masked}}   & PSNR   & 28.66      & 34.56       & 33.87       & 34.61       & 36.54       & 34.89       & 35.30       & 37.89       & 33.46       & 34.42       \\
			                             &                                                          & SSIM   & 0.7127     & 0.9553      & 0.9703      & 0.9649      & 0.9273      & 0.9586      & 0.9593      & 0.9354      & 0.9531      & 0.9263      \\
			\cmidrule{2-13}
			                             & \multirow{2}{*}{IDF*~\citep{kim2025idf}}                 & PSNR   & 31.73      & 37.22       & 36.14       & 37.50       & 40.36       & 37.77       & 38.64       & 42.58       & 34.95       & 37.43       \\
			                             &                                                          & SSIM   & 0.8011     & 0.9790      & 0.9809      & 0.9759      & 0.9776      & 0.9649      & 0.9728      & 0.9848      & 0.9617      & 0.9554      \\
			\cmidrule{1-13}
			\multirow{2}{*}{Ours}        & \multirow{2}{*}{\bf{NAFNet + NTN}}                       & PSNR   & {39.24}    & \bf{38.72}  & \bf{37.84}  & \bf{40.00}  & \bf{42.08}  & \bf{39.83}  & \bf{40.55}  & {44.34}     & \bf{36.17}  & \bf{39.86}  \\
			                             &                                                          & SSIM   & {0.9570}   & \bf{0.9855} & \bf{0.9877} & \bf{0.9856} & \bf{0.9812} & \bf{0.9782} & \bf{0.9801} & {0.9875}    & \bf{0.9749} & \bf{0.9797} \\

			\bottomrule
		\end{tabular}

	}
	\label{tab:supp_real}
	\vspace{-4mm}
\end{table*}

\begin{table*}[t]
	\centering
	\caption{
		Quantitative comparisons with state-of-the-art generalization methods on the SIDD validation set and multiple real-noise benchmarks.
		We present the results in terms of PSNR$\uparrow$ (dB) and SSIM$\uparrow$.
		Methods marked with a dagger (\dag) indicate evaluations performed on reproduced models under standardized conditions.
	}
	\vspace{-2mm}
	\setlength{\tabcolsep}{6.2pt}
    \renewcommand{\arraystretch}{0.85}
	\scalebox{0.75}{
		\begin{tabular}{c l c c c c c c c c c c c c  c c c}
			\toprule
			\multicolumn{3}{c}{}                                        & \multicolumn{1}{c}{In-distribution}                                 & \multicolumn{8}{c}{Out-of-distribution} &                                                                                                                                       \\
			\cmidrule(lr){4-4} \cmidrule(lr){5-12}
			\multicolumn{1}{c}{Baseline}                                & \multicolumn{1}{l}{Generalization \ }                               & Metric                                  & SIDD     & ~~Poly~~    & ~~~CC~~~    & HighISO~    & iPhone      & Huawei      & OPPO        & Sony       & Xiaomi      & OOD Avg.    \\
			\midrule
			\midrule
			\multirow{9}{*}{\ \ \ NAFNet~\citep{chen2022simple} \ \ \ } & \multirow{2}{*}{None}                                               & PSNR                                    & {39.97}  & 37.17       & 35.69       & 38.32       & 40.25       & 37.73       & 39.64       & 43.65      & 34.99       & 38.43       \\
			                                                            &                                                                     & SSIM                                    & {0.9600} & 0.9717      & 0.9811      & 0.9788      & 0.9707      & 0.9680      & 0.9786      & 0.9829     & 0.9685      & 0.9750      \\
			\cmidrule{2-13}
			                                                            & \multirow{2}{*}{+ AFM\textsuperscript{\dag}~\citep{ryou2024robust}} & PSNR                                    & 39.83    & {37.28}     & {36.31}     & {38.40}     & 39.68       & {38.21}     & {39.84}     & 43.31      & {35.39}     & {38.55}     \\
			                                                            &                                                                     & SSIM                                    & 0.9592   & 0.9696      & {0.9815}    & 0.9783      & 0.9545      & {0.9683}    & {0.9788}    & 0.9637     & {0.9709}    & 0.9707      \\
			\cmidrule{2-13}
			                                                            & \multirow{2}{*}{+ LAN\textsuperscript{\dag}~\citep{kim2024lan}}     & PSNR                                    & 39.88    & 37.17       & 35.59       & 38.21       & {40.48}     & 37.53       & 39.42       & {43.66}    & 34.83       & 38.36       \\
			                                                            &                                                                     & SSIM                                    & 0.9594   & {0.9792}    & 0.9808      & {0.9787}    & {0.9766}    & 0.9680      & 0.9773      & \bf{0.9883}   & 0.9667      & {0.9770}    \\
			\cmidrule{2-13}
			                                                            & \multirow{2}{*}{\textbf{+ NTN}}                                     & PSNR                                    & 39.24    & \bf{38.72}  & \bf{37.84}  & \bf{40.00}  & \bf{42.08}  & \bf{39.83}  & \bf{40.55}  & \bf{44.34} & \bf{36.17}  & \bf{39.94}  \\
			                                                            &                                                                     & SSIM                                    & 0.9570   & \bf{0.9855} & \bf{0.9877} & \bf{0.9856} & \bf{0.9812} & \bf{0.9782} & \bf{0.9801} & {0.9875}   & \bf{0.9749} & \bf{0.9826} \\
			\bottomrule
		\end{tabular}
	}
	\vspace{-2mm}
	\label{tab:supp_real_2}
\end{table*}

\section{Additional experimental results}

\subsection{Model-agnostic applicability}
To demonstrate the model-agnostic nature of our proposed noise translation framework, we applied our method to distinct architectural paradigms, specifically the transformer-based Restormer~\citep{zamir2022restormer} and the kernel-based KBNet~\citep{Zhang2023kbnet}, as noted in the main paper. 
Integrating our training framework into these baselines yielded consistent performance improvements in real-world denoising tasks, as detailed in Table~\ref{tab:real_main_supp}.
These results verify that our approach is not tailored to a specific backbone network but functions as a highly compatible framework capable of boosting the performance of diverse denoising architectures.

\subsection{Comparison with additional denoising methods}
Table~\ref{tab:supp_real} presents quantitative comparisons between additional image denoising methods and our approach.
Generation-based methods such as DANet~\citep{yue2020dual}, C2N~\citep{jang2021c2n}, and PNGAN~\citep{cai2021learning} utilize the SIDD dataset for training, focusing on synthetic-to-real noise generation.
PDDenoising~\citep{zhou2020awgn} applies pixel-shuffle down-sampling to denoise complex real noise with a simple Gaussian denoiser. 
However, pixel-shuffling alters image semantics, and optimal shuffling parameters vary per image, making optimization challenging and yielding poor performance.
MaskDenoising~\citep{chen2023masked} and IDF~\citep{kim2025idf} are trained solely on Gaussian noise ($\sigma = 15$), leading to notably poor performance on real-world noise datasets.
CLIPDenoising~\citep{Jun2024Transfer} leverages the CLIP encoder and incorporates additional training on synthetic noise generated using Poisson-Gaussian models for sRGB denoising.
However, none of these methods outperform our approach across most benchmarks, underscoring the efficacy of our noise translation framework in handling real-world noise scenarios.


\begin{table*}[t!]
	\centering
	\caption{
		Expanded table of Table 2: Gaussian injection and explicit noise translation loss.
	}
	\vspace{-2mm}
	\setlength{\tabcolsep}{7.5pt}
    \renewcommand{\arraystretch}{0.85}
	\scalebox{0.75}{
		\begin{tabular}{l c c c c c c c c c c c}
			\toprule
			 &        & \multicolumn{1}{c}{In-distribution} & \multicolumn{8}{c}{Out-of-distribution} &                                                                                                               \\
			\cmidrule(lr){3-3}\cmidrule(lr){4-11}
			 & Metric & SIDD                                & Poly                                    & CC          & HighISO     & iPhone      & Huawei      & OPPO        & Sony        & Xiaomi      & OOD Avg.    \\
			\midrule
			\midrule

			\multirow{2}{*}{\begin{tabular}{c}w/o GIBlock\end{tabular}}
			 & PSNR   & 39.35                               & 38.32                                   & 37.25       & 39.22       & 40.80       & 39.24       & 39.75       & 43.86       & 35.74       & 39.27       \\
			 & SSIM   & 0.9573                              & 0.9820                                  & 0.9864      & 0.9794      & 0.9700      & 0.9727      & 0.9745      & {0.9857}    & 0.9683      & 0.9774      \\
			\midrule

			\multirow{2}{*}{\begin{tabular}{c}w/o Explicit Loss\end{tabular}}
			 & PSNR   & 39.05                               & {38.54}                                 & {37.58}     & {39.79}     & {41.53}     & {39.68}     & {40.40}     & {43.89}     & {36.00}     & {39.61}     \\
			 & SSIM   & 0.9556                              & {0.9835}                                & {0.9866}    & {0.9844}    & {0.9737}    & {0.9773}    & {0.9790}    & 0.9827      & {0.9737}    & {0.9801}    \\
			\midrule

			\multirow{2}{*}{\begin{tabular}{c}w/o adaptive gating ($\mathcal{L}_\text{explicit}$)\end{tabular}}
			 & PSNR   & 39.33                               & 38.59                                   & 37.10       & 39.35       & 41.69       & 39.56       & 40.27       & \bf{44.34}  & 35.95       & 39.61       \\
			 & SSIM   & 0.9572                              & 0.9846                                  & 0.9864      & 0.9821      & 0.9793      & 0.9753      & 0.9782      & \bf{0.9888} & 0.9720      & 0.9808      \\
			\midrule

			\multirow{2}{*}{\begin{tabular}{c}Full (\bf{Ours})\end{tabular}}
			 & PSNR   & 39.24                               & \bf{38.72}                              & \bf{37.84}  & \bf{40.00}  & \bf{42.08}  & \bf{39.83}  & \bf{40.55}  & \bf{44.34}  & \bf{36.17}  & \bf{39.94}  \\
			 & SSIM   & 0.9570                              & \bf{0.9855}                             & \bf{0.9877} & \bf{0.9856} & \bf{0.9812} & \bf{0.9782} & \bf{0.9801} & {0.9875}    & \bf{0.9749} & \bf{0.9826} \\
			\bottomrule
		\end{tabular}
	}
	\label{tab:real_ablation_full}
	\vspace{-4mm}
\end{table*}

\begin{table*}[t]
	\centering
	\caption{
		Expanded table of Table 4: Adaptability of the noise translation network.
	}
	\vspace{-2mm}
	\setlength{\tabcolsep}{6pt}
	\renewcommand{\arraystretch}{0.85}
	\scalebox{0.75}{
		\begin{tabular}{c c  c c c c c c c c c c c c}
			\toprule
			                                     &                                                            &        & \multicolumn{1}{c}{In-distribution} & \multicolumn{8}{c}{Out-of-distribution} &                                                                                       \\
			\cmidrule(lr){4-4}\cmidrule(lr){5-12}
			$\mathcal{D}_{\theta^*}$ for testing & $\mathcal{D}_{\theta^*}$ for training $\mathcal{T}_{\phi}$ & Metric & SIDD                                & ~~Poly~~                                & ~~~CC~~~ & HighISO~ & iPhone   & Huawei   & OPPO     & Sony     & Xiaomi   & OOD Avg. \\
			\midrule\midrule

			\multirow{9}{*}{Restormer}           & \multirow{2}{*}{Restormer}                                 & PSNR   & {39.22}                             & {38.76}                                 & {37.68}  & {40.14}  & {41.85}  & {39.72}  & {40.64}  & {44.29}  & {36.19}  & {39.91}  \\
			                                     &                                                            & SSIM   & {0.9569}                            & {0.9854}                                & {0.9866} & {0.9857} & {0.9786} & {0.9766} & {0.9800} & {0.9871} & {0.9750} & {0.9819} \\
			\cmidrule(lr){2-13}
			                                     & \multirow{2}{*}{NAFNet}                                    & PSNR   & 39.16                               & 38.65                                   & 37.34    & 39.98    & 41.92    & 39.73    & 40.56    & 43.95    & 36.19    & 39.79    \\
			                                     &                                                            & SSIM   & 0.9564                              & 0.9851                                  & 0.9861   & 0.9854   & 0.9811   & 0.9770   & 0.9800   & 0.9871   & 0.9750   & 0.9821   \\
			\cmidrule(lr){2-13}
			                                     & \multirow{2}{*}{KBNet}                                     & PSNR   & 39.21                               & 38.53                                   & 37.05    & 39.88    & 41.93    & 39.65    & 40.53    & 43.81    & 36.14    & 39.69    \\
			                                     &                                                            & SSIM   & 0.9567                              & 0.9848                                  & 0.9851   & 0.9850   & 0.9817   & 0.9766   & 0.9799   & 0.9873   & 0.9743   & 0.9819   \\
			\cmidrule(lr){2-13}
			                                     & \multirow{2}{*}{Xformer}                                   & PSNR   & 39.12                               & 38.73                                   & 37.80    & 40.11    & 41.65    & 39.64    & 40.42    & 44.11    & 36.18    & 39.83    \\
			                                     &                                                            & SSIM   & 0.9559                              & 0.9850                                  & 0.9867   & 0.9852   & 0.9774   & 0.9763   & 0.9795   & 0.9866   & 0.9746   & 0.9814   \\

			\midrule

			\multirow{9}{*}{NAFNet}              & \multirow{2}{*}{Restormer}                                 & PSNR   & 39.16                               & 38.61                                   & 38.00    & 40.07    & 41.95    & 39.71    & 40.41    & 44.05    & 36.22    & 39.88    \\
			                                     &                                                            & SSIM   & 0.9565                              & 0.9846                                  & 0.9877   & 0.9856   & 0.9800   & 0.9775   & 0.9793   & 0.9858   & 0.9748   & 0.9819   \\
			\cmidrule(lr){2-13}
			                                     & \multirow{2}{*}{NAFNet}                                    & PSNR   & {39.24}                             & {38.72}                                 & {37.84}  & {40.00}  & {42.08}  & {39.83}  & {40.55}  & {44.34}  & {36.17}  & {39.94}  \\
			                                     &                                                            & SSIM   & {0.9570}                            & {0.9855}                                & {0.9877} & {0.9856} & {0.9812} & {0.9782} & {0.9801} & {0.9875} & {0.9749} & {0.9826} \\

			\cmidrule(lr){2-13}
			                                     & \multirow{2}{*}{KBNet}                                     & PSNR   & 39.17                               & 38.67                                   & 37.54    & 39.83    & 41.95    & 39.73    & 40.45    & 44.18    & 36.11    & 39.81    \\
			                                     &                                                            & SSIM   & 0.9565                              & 0.9853                                  & 0.9870   & 0.9849   & 0.9814   & 0.9777   & 0.9798   & 0.9875   & 0.9741   & 0.9822   \\
			\cmidrule(lr){2-13}
			                                     & \multirow{2}{*}{Xformer}                                   & PSNR   & 39.06                               & 38.66                                   & 38.04    & 40.07    & 41.93    & 39.72    & 40.46    & 44.11    & 36.17    & 39.89    \\
			                                     &                                                            & SSIM   & 0.9560                              & 0.9848                                  & 0.9877   & 0.9852   & 0.9792   & 0.9773   & 0.9794   & 0.9857   & 0.9743   & 0.9817   \\

			\midrule

			\multirow{9}{*}{KBNet}               & \multirow{2}{*}{Restormer}                                 & PSNR   & {39.14}                             & {38.59}                                 & {37.79}  & {40.14}  & {41.77}  & {39.67}  & {40.45}  & {44.02}  & {36.22}  & {39.83}  \\
			                                     &                                                            & SSIM   & {0.9564}                            & {0.9843}                                & {0.9863} & {0.9855} & {0.9776} & {0.9773} & {0.9798} & {0.9850} & {0.9751} & {0.9814} \\
			\cmidrule(lr){2-13}
			                                     & \multirow{2}{*}{NAFNet}                                    & PSNR   & {39.12}                             & {38.62}                                 & {37.67}  & {39.95}  & {41.88}  & {39.73}  & {40.51}  & {44.16}  & {36.16}  & {39.84}  \\
			                                     &                                                            & SSIM   & {0.9563}                            & {0.9849}                                & {0.9863} & {0.9853} & {0.9797} & {0.9776} & {0.9800} & {0.9861} & {0.9747} & {0.9818} \\
			\cmidrule(lr){2-13}
			                                     & \multirow{2}{*}{KBNet}                                     & PSNR   & {39.13}                             & {38.62}                                 & {37.61}  & {39.89}  & {41.76}  & {39.80}  & {40.57}  & {44.22}  & {36.07}  & {39.82}  \\
			                                     &                                                            & SSIM   & {0.9564}                            & {0.9843}                                & {0.9861} & {0.9849} & {0.9761} & {0.9777} & {0.9798} & {0.9845} & {0.9743} & {0.9810} \\
			\cmidrule(lr){2-13}
			                                     & \multirow{2}{*}{Xformer}                                   & PSNR   & {39.08}                             & {38.64}                                 & {37.74}  & {40.08}  & {41.76}  & {39.69}  & {40.48}  & {44.09}  & {36.18}  & {39.83}  \\
			                                     &                                                            & SSIM   & {0.9560}                            & {0.9845}                                & {0.9862} & {0.9851} & {0.9776} & {0.9772} & {0.9797} & {0.9849} & {0.9747} & {0.9812} \\

			\midrule

			\multirow{9}{*}{Xformer}             & \multirow{2}{*}{Restormer}                                 & PSNR   & 39.09                               & 38.60                                   & 37.81    & 40.25    & 42.12    & 39.72    & 40.51    & 44.10    & 36.28    & 39.92    \\
			                                     &                                                            & SSIM   & 0.9558                              & 0.9851                                  & 0.9859   & 0.9861   & 0.9831   & 0.9776   & 0.9801   & 0.9885   & 0.9753   & 0.9827   \\
			\cmidrule(lr){2-13}
			                                     & \multirow{2}{*}{NAFNet}                                    & PSNR   & 39.17                               & 38.62                                   & 37.73    & 40.18    & 42.29    & 39.75    & 40.56    & 44.27    & 36.19    & 39.95    \\
			                                     &                                                            & SSIM   & 0.9563                              & 0.9856                                  & 0.9860   & 0.9860   & 0.9851   & 0.9775   & 0.9806   & 0.9896   & 0.9750   & 0.9832   \\
			\cmidrule(lr){2-13}
			                                     & \multirow{2}{*}{KBNet}                                     & PSNR   & 39.10                               & 38.51                                   & 37.43    & 39.98    & 42.12    & 39.69    & 40.47    & 43.98    & 36.15    & 39.79    \\
			                                     &                                                            & SSIM   & 0.9557                              & 0.9851                                  & 0.9854   & 0.9852   & 0.9844   & 0.9774   & 0.9801   & 0.9889   & 0.9743   & 0.9826   \\
			\cmidrule(lr){2-13}
			                                     & \multirow{2}{*}{Xformer}                                   & PSNR   & {39.10}                             & {38.75}                                 & {37.82}  & {40.29}  & {42.20}  & {39.89}  & {40.70}  & {44.44}  & {36.25}  & {40.04}  \\
			                                     &                                                            & SSIM   & {0.9557}                            & {0.9856}                                & {0.9862} & {0.9860} & {0.9826} & {0.9779} & {0.9807} & {0.9886} & {0.9751} & {0.9828} \\

			\bottomrule
		\end{tabular}
	}
	\vspace{-2.5mm}
	\label{tab:cross_full}
\end{table*}
\subsection{Comparison with generalization methods}
Table~\ref{tab:supp_real_2} showcases results from our reproduction of the AFM~\citep{ryou2024robust} and LAN~\citep{kim2024lan} methods using their officially released code, in combination with NAFNet, to enable a direct comparison with our proposed method.
Applying LAN to other denoising architectures from our experiments, such as Restormer~\citep{zamir2022restormer}, KBNet~\citep{Zhang2023kbnet}, and Xformer~\citep{zhang2023xformer}, was infeasible due to excessive memory demands of LAN when processing images with resolutions exceeding 512 $\times$ 512.
Additionally, our reproduction experiments revealed that LAN's adaptation methodology not only requires substantial computational resources but also fails to improve generalization performance at resolutions exceeding 256 $\times$ 256.
In contrast, our noise translation framework consistently delivers robust and superior generalization performance across diverse out-of-distribution benchmarks, outperforming both AFM and LAN.

\subsection{Expanded Table: Gaussian Injection Block and Explicit Noise Translation Loss}
Table~\ref{tab:real_ablation_full} provides the complete results for the OOD benchmarks in Table 2 of the main paper. 
These results further demonstrate the effectiveness of the Gaussian injection block and explicit noise translation loss within the proposed framework.

\subsection{Expanded Table: Adaptability of the Noise Translation Network}
Table~\ref{tab:cross_full} extends the results of Table 4 in the main paper, analyzing various combinations of denoising networks used during the training and inference phases within our framework.
In this context, \(\mathcal{D}_{\theta^*}\) for testing refers to the pretrained denoising network utilized during inference, while \(\mathcal{D}_{\theta^*}\) for training \(\mathcal{T}_{\phi}\) denotes the pretrained denoising network employed for training the noise translation network.
For instance, when \(\mathcal{D}_{\theta^*}\) for testing is NAFNet and \(\mathcal{D}_{\theta^*}\) for training \(\mathcal{T}_{\phi}\) is KBNet, the noise translation network is trained to map real-world noise to align with KBNet's prior but is evaluated with NAFNet during inference.
Notably, when \( D_{\theta^*}\) for testing and \(\mathcal{D}_{\theta^*}\) for training \(\mathcal{T}_{\phi}\) correspond to different pretrained denoising models, our framework maintains high performance by effectively translating noise distributions to align with the priors of the inference model.
This adaptability implies that as new denoising architectures are introduced, they can be seamlessly incorporated into our framework, potentially achieving even greater performance improvements without the need to retrain the noise translation network.
Such flexibility ensures the practical applicability of our noise translation framework in real-world scenarios.


\begin{table*}[t]
	\centering
	\caption{
		Quantitative comparisons of real-world denoising performance across various training set configurations.
		Methods marked with an asterisk (*) are evaluated using official out-of-the-box models.
	}
	\vspace{-2mm}
	\setlength{\tabcolsep}{6pt}
	\renewcommand{\arraystretch}{0.85}
	\scalebox{0.75}{
		\begin{tabular}{c l c c c c c c c c c c c c  c c c}
			\toprule
			\multicolumn{1}{c}{Denoising network}                                & \multicolumn{1}{c}{Trained noise} & Metric & SIDD     & ~~Poly~~    & ~~~CC~~~    & HighISO~    & iPhone      & Huawei      & OPPO        & Sony        & Xiaomi      & Total Avg.  \\
			\midrule
			\midrule
			\multirow{6.8}{*}{\ \ \ Restormer~\citep{zamir2022restormer} \ \ \ } & \multirow{2}{*}{SIDD*}            & PSNR   & {40.02}  & 37.66       & 36.33       & 38.29       & 40.13       & 38.42       & 39.56       & 44.19       & 35.65       & 38.92       \\
			                                                                     &                                   & SSIM   & {0.9603} & 0.9793      & 0.9807      & 0.9756      & 0.9734      & 0.9675      & 0.9773      & \bf{0.9894} & 0.9710      & 0.9749      \\
			\cmidrule{2-13}
			                                                                     & \multirow{2}{*}{SIDD + Gaussian}  & PSNR   & 39.67    & 37.61       & 35.52       & 38.08       & 40.38       & 38.51       & 39.83       & 43.72       & 35.54       & 38.76       \\
			                                                                     &                                   & SSIM   & 0.9586   & 0.9800      & 0.9771      & 0.9765      & 0.9748      & 0.9690      & 0.9795      & 0.9865      & 0.9714      & 0.9748      \\
			\cmidrule{2-13}
			                                                                     & \multirow{2}{*}{\bf{Ours}}        & PSNR   & {39.22}  & \bf{38.76}  & \bf{37.68}  & \bf{40.14}  & \bf{41.85}  & \bf{39.72}  & \bf{40.64}  & \bf{44.29}  & \bf{36.19}  & \bf{39.83}  \\
			                                                                     &                                   & SSIM   & {0.9569} & \bf{0.9854} & \bf{0.9866} & \bf{0.9857} & \bf{0.9786} & \bf{0.9766} & \bf{0.9800} & {0.9871}    & \bf{0.9750} & \bf{0.9791} \\

			\midrule
			\midrule
			\multirow{6.8}{*}{\ \ \ NAFNet~\citep{chen2022simple} \ \ \ }        & \multirow{2}{*}{SIDD*}            & PSNR   & {39.97}  & 37.17       & 35.69       & 38.32       & 40.25       & 37.73       & 39.64       & 43.65       & 34.99       & 38.60       \\
			                                                                     &                                   & SSIM   & {0.9600} & 0.9717      & 0.9811      & 0.9788      & 0.9707      & 0.9680      & 0.9786      & 0.9829      & 0.9685      & 0.9734      \\
			\cmidrule{2-13}
			                                                                     & \multirow{2}{*}{SIDD + Gaussian}  & PSNR   & 39.61    & 37.93       & 36.10       & 38.20       & 40.66       & 37.92       & 39.66       & 43.49       & 35.65       & 38.80       \\
			                                                                     &                                   & SSIM   & 0.9582   & 0.9809      & 0.9797      & 0.9788      & 0.9779      & 0.9704      & 0.9815      & \bf{0.9886} & 0.9716      & 0.9764      \\
			\cmidrule{2-13}
			                                                                     & \multirow{2}{*}{\bf{Ours}}        & PSNR   & {39.24}  & \bf{38.72}  & \bf{37.84}  & \bf{40.00}  & \bf{42.08}  & \bf{39.83}  & \bf{40.55}  & \bf{44.34}  & \bf{36.17}  & \bf{39.86}  \\
			                                                                     &                                   & SSIM   & {0.9570} & \bf{0.9855} & \bf{0.9877} & \bf{0.9856} & \bf{0.9812} & \bf{0.9782} & \bf{0.9801} & {0.9875} & \bf{0.9749} & \bf{0.9797} \\
			\bottomrule
		\end{tabular}
	}
	\vspace{-3.5mm}

	\label{tab:supp_dataset}
\end{table*}

\begin{table*}[t]
	\centering
	\caption{
		Quantitative results of applying noise translation framework to denoising model  pre-trained on pure synthetic Gaussian noise.
	}
	\vspace{-2mm}
	\setlength{\tabcolsep}{9.5pt}
    \renewcommand{\arraystretch}{0.85}
	\scalebox{0.75}{
		\begin{tabular}{l l c c c c c c c c c c c c  c c c}
			\toprule
			\multicolumn{1}{c}{Denoising methods}                          & Metric & SIDD        & ~~Poly~~    & ~~~CC~~~    & HighISO~    & iPhone      & Huawei      & OPPO        & Sony        & Xiaomi      & Total Avg.  \\
			\midrule
			\midrule

			\multirow{2}{*}{Restormer-Gaussian~\citep{zamir2022restormer}} & PSNR   & 23.99       & 35.97       & 33.41       & {35.31}     & 38.02       & 36.94       & 37.84       & 42.90       & 34.49       & 35.43       \\
			                                                               & SSIM   & 0.4990      & 0.9585      & 0.9563      & 0.9464      & 0.9448      & 0.9502      & 0.9629      & 0.9831      & 0.9554      & 0.9063      \\
			\cmidrule{1-12}
			\multirow{2}{*}{\textbf{+ Ours}}                               & PSNR   & \bf{39.08}  & \bf{37.88}  & \bf{35.75}  & \bf{39.33}  & \bf{41.21}  & \bf{38.44}  & \bf{39.96}  & \bf{43.18}  & \bf{35.63}  & \bf{38.94}  \\
			                                                               & SSIM   & \bf{0.9542} & \bf{0.9832} & \bf{0.9800} & \bf{0.9827} & \bf{0.9823} & \bf{0.9719} & \bf{0.9790} & \bf{0.9835} & \bf{0.9720} & \bf{0.9765} \\


			\bottomrule
		\end{tabular}

	}
	\vspace{-2mm}
	\label{tab:supp_pure_gaussian}
\end{table*}


\subsection{Comparison with na\"ive combination of real and synthetic dataset for training}
We compare our method with the naive approach of combining Gaussian noise dataset with the SIDD dataset for training.
The results are presented in Table~\ref{tab:supp_dataset}.
The Gaussian noise dataset used in this comparison is identical to the one which is used for our method.
The results demonstrate that naively combining real-noise and Gaussian noise datasets during training does not yield significant improvements in generalization performance.
This highlights the superiority and efficacy of our method in enhancing denoising performance across diverse noise distributions.

\subsection{Incorporating a pure Gaussian pretrained denoising network}
We conduct an experiment on the officially published denoising model, Restormer~\citep{zamir2022restormer}, which was trained for blind gaussian noise removal.
By leveraging it in our framework, we trained the corresponding noise translation network with the proposed methodology to evaluate the overall performance.
The results are provided in Table~\ref{tab:supp_pure_gaussian}, which demonstrate that even when using a pure Gaussian pretrained out-of-the-box model within our framework, the real-world denoising performance undergoes significant improvement.
This enhancement can be attributed to our noise translation network, which effectively adapts the real-world noise into a form closer to the Gaussian noise that the pretrained model is adept at handling.
Note that, in our pretraining approach, the denoising network is trained on Gaussian noise augmented with the same real-world noise used to train the noise translation network.
This distinction in pretraining strategies explains the observed performance gap between the two results.

\subsection{Effects of hyperparameters}

\begin{table*}[t]
    \centering
    \caption{
        Sensitivity results on various hyperparameters in our framework.
    }
    \vspace{-2mm}
    \label{tab:Hyperparameters}
    \setlength{\tabcolsep}{11pt}
    \renewcommand{\arraystretch}{0.85}
    \scalebox{0.8}{
        \begin{tabular}{c c c | c c c | c c c | c c c}
            \toprule
            \multicolumn{3}{c|}{Pretraining Level} & \multicolumn{3}{c|}{Noise Injection Level} & \multicolumn{3}{c|}{Explicit Loss Weight} & \multicolumn{3}{c}{Spatial-Frequency Ratio}                                                                                                                            \\
            \midrule
            $\sigma$                               & SIDD                                       & OOD Avg.                                  & $\tilde{\sigma}$                            & SIDD           & OOD Avg.       & $\alpha$ & SIDD           & OOD Avg.       & $\beta$ & SIDD           & OOD Avg.       \\
            \midrule
            5                                      & 39.37                                      & 38.96                                     & 0                                           & 39.43          & 39.37          & 0        & 39.07          & 39.74          & 0       & 39.13          & 39.84          \\
            10                                     & 39.29                                      & 39.64                                     & 1                                           & 39.38          & 39.54          & 0.001    & 39.08          & 39.76          & 0.0005  & 39.16          & 39.88          \\
            15                                     & 39.24                                      & \textbf{39.94}                            & 5                                           & 39.09          & 39.78          & 0.005    & 39.11          & 39.80          & 0.001   & 39.18          & 39.91          \\
            20                                     & 38.88                                      & 39.53                                     & 15                                          & 39.15          & 39.86          & 0.01     & 39.14          & 39.85          & 0.002   & \textbf{39.24} & \textbf{39.94} \\
            25                                     & 38.73                                      & 39.21                                     & 50                                          & 39.15          & 39.90          & 0.05     & \textbf{39.24} & \textbf{39.94} & 0.005   & 39.17          & 39.94          \\
                                                   &                                            &                                           & 100                                         & \textbf{39.24} & \textbf{39.94} & 0.1      & 39.19          & 39.92          & 0.01    & 39.09          & 39.77          \\
                                                   &                                            &                                           & 200                                         & 39.19          & 39.93          & 0.5      & 39.13          & 39.07          & 0.02    & 39.00          & 38.97          \\
            \bottomrule
        \end{tabular}
    }
    \vspace{-3mm}
\end{table*}

Table~\ref{tab:Hyperparameters} presents the ablative results of our hyperparameters, including pretraining noise level ($\sigma$), noise injection level ($\tilde{\sigma}$), explicit loss weight ($\alpha$), and spatial frequency ratio ($\beta$).
The pretraining level part of Table~\ref{tab:Hyperparameters} shows the ablation results for the Gaussian noise levels added to create noisy input during denoising network pretraining.
As the noise level increased, the performance on in-distribution (ID) consistently decreased.
For out-of-distribution (OOD), the performance improved until noise level of 15, beyond which it began to degrade, due to oversmoothing effects caused by learning to handle strong noise.
The noise injection level part presents the ablation results for Gaussian noise injection levels.
Increasing the noise level led to a decline in ID performance, while OOD performance improved up to a certain point.
The explicit loss weight and spatial-frequency ratio parts show the ablation results for the $\alpha$ and $\beta$ values in Eq. (11) and (10), respectively.
Overall, the ablation experiments determined the optimal hyperparameters as follows: a pretraining noise level $\sigma=15$, a Gaussian noise injection level $\tilde{\sigma}= 100$, $\alpha=5\times10^{-2}$, and $\beta=2\times10^{-3}$.

\subsection{Training Stability}
To evaluate the stability of training our noise translation network, we conducted five independent training runs using different random seeds and measured the PSNR metrics across multiple datasets.
The standard deviations for each dataset were as follows: SIDD (0.023), Poly (0.007), CC (0.021), HighISO (0.009), iPhone (0.008), Huawei (0.010), OPPO (0.016), Sony (0.022), and Xiaomi (0.020).
The average standard deviation across all datasets was 0.013, confirming the robustness and stability of training with our framework.



\section{Qualitative Analysis}
\subsection{Additional OOD datasets and SIDD dataset}
\label{sec:supple_qual}
\begin{figure*}[tb]
\centering

\begin{minipage}{\textwidth}
\centering
\begin{minipage}{0.25\textwidth}
\centering
\includegraphics[width=\textwidth]{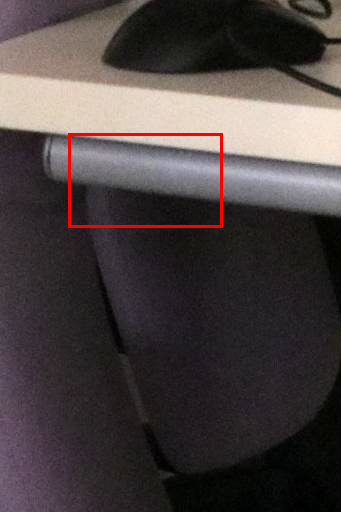}\\[-1mm]
\small Poly Noisy
\end{minipage}%
\hspace{-0.5mm}%
\begin{minipage}{0.752\textwidth}
\centering
\vspace{-0.6mm}
\begin{minipage}{0.24\textwidth}
\centering
\includegraphics[width=\textwidth]{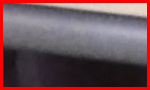}\\[-1mm]
\small R2R 39.84dB
\end{minipage}
\begin{minipage}{0.24\textwidth}
\centering
\includegraphics[width=\textwidth]{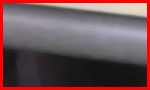}\\[-1mm]
\small AP-BSN 42.13dB
\end{minipage}
\begin{minipage}{0.24\textwidth}
\centering
\includegraphics[width=\textwidth]{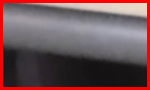}\\[-1mm]
\small SSID 42.14dB
\end{minipage}
\begin{minipage}{0.24\textwidth}
\centering
\includegraphics[width=\textwidth]{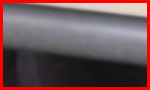}\\[-1mm]
\small APR-RD 42.16dB
\end{minipage}

\vspace{1mm}

\begin{minipage}{0.24\textwidth}
\centering
\includegraphics[width=\textwidth]{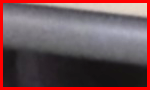}\\[-1mm]
\small Mask-DN 41.03dB
\end{minipage}
\begin{minipage}{0.24\textwidth}
\centering
\includegraphics[width=\textwidth]{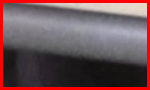}\\[-1mm]
\small CLIP-DN 41.06dB
\end{minipage}
\begin{minipage}{0.24\textwidth}
\centering
\includegraphics[width=\textwidth]{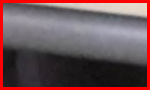}\\[-1mm]
\small AFM 41.51dB
\end{minipage}
\begin{minipage}{0.24\textwidth}
\centering
\includegraphics[width=\textwidth]{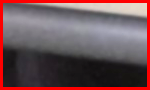}\\[-1mm]
\small IDF 40.72dB
\end{minipage}

\vspace{1mm}

\begin{minipage}{0.24\textwidth}
\centering
\includegraphics[width=\textwidth]{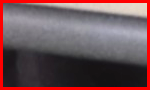}\\[-1mm]
\small NAFNet 40.71dB
\end{minipage}
\begin{minipage}{0.24\textwidth}
\centering
\includegraphics[width=\textwidth]{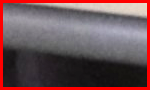}\\[-1mm]
\small XFormer 40.35dB
\end{minipage}
\begin{minipage}{0.24\textwidth}
\centering
\includegraphics[width=\textwidth]{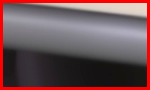}\\[-1mm]
\small \textbf{Ours 43.81dB}
\end{minipage}
\begin{minipage}{0.24\textwidth}
\centering
\includegraphics[width=\textwidth]{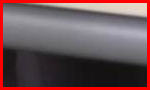}\\[-1mm]
\small Ground Truth
\end{minipage}
\end{minipage}

\vspace{1mm}
\end{minipage}

\begin{minipage}{\textwidth}
\centering
\begin{minipage}{0.25\textwidth}
\centering
\includegraphics[width=\textwidth]{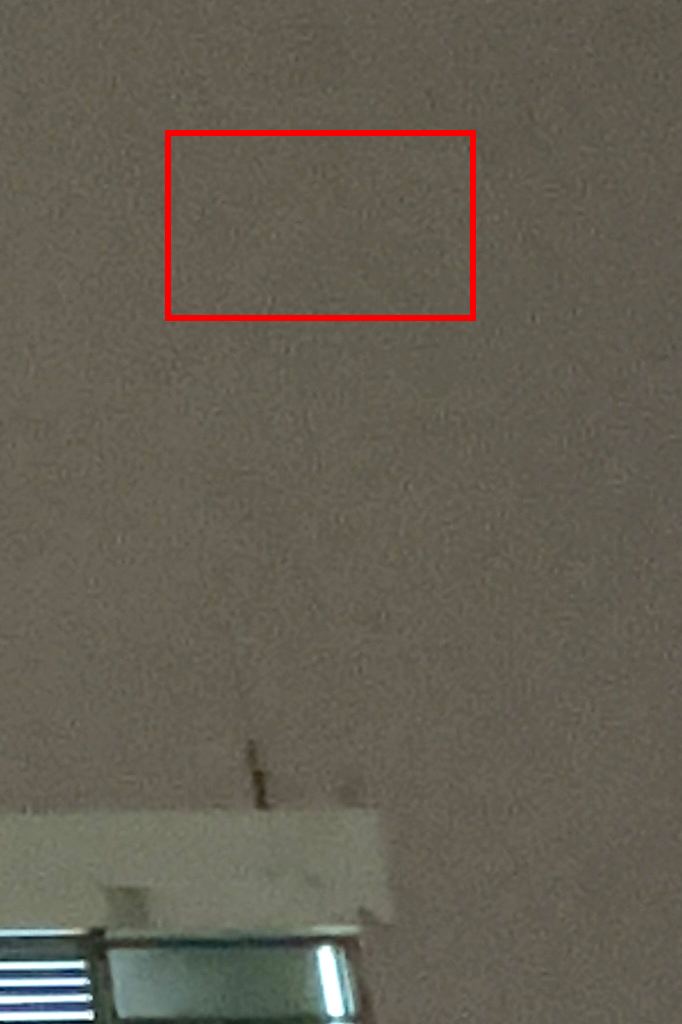}\\[-1mm]
\small iPhone Noisy
\end{minipage}%
\hspace{-0.5mm}%
\begin{minipage}{0.752\textwidth}
\centering
\vspace{-0.6mm}
\begin{minipage}{0.24\textwidth}
\centering
\includegraphics[width=\textwidth]{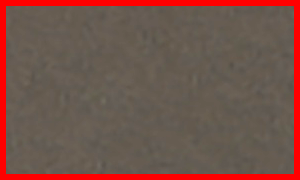}\\[-1mm]
\small R2R 40.94dB
\end{minipage}
\begin{minipage}{0.24\textwidth}
\centering
\includegraphics[width=\textwidth]{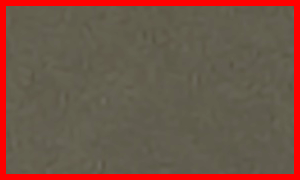}\\[-1mm]
\small AP-BSN 42.61dB
\end{minipage}
\begin{minipage}{0.24\textwidth}
\centering
\includegraphics[width=\textwidth]{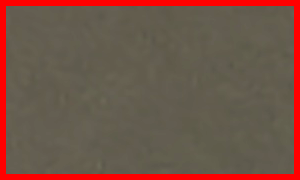}\\[-1mm]
\small SSID 43.93dB
\end{minipage}
\begin{minipage}{0.24\textwidth}
\centering
\includegraphics[width=\textwidth]{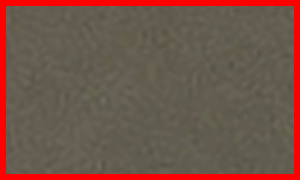}\\[-1mm]
\small APR-RD 41.70dB
\end{minipage}

\vspace{1mm}

\begin{minipage}{0.24\textwidth}
\centering
\includegraphics[width=\textwidth]{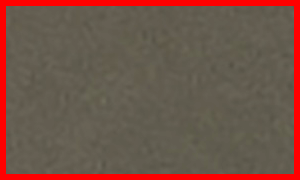}\\[-1mm]
\small Mask-DN 42.60dB
\end{minipage}
\begin{minipage}{0.24\textwidth}
\centering
\includegraphics[width=\textwidth]{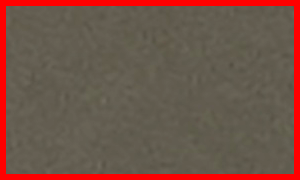}\\[-1mm]
\small CLIP-DN 42.13dB
\end{minipage}
\begin{minipage}{0.24\textwidth}
\centering
\includegraphics[width=\textwidth]{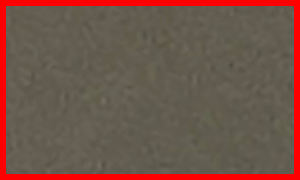}\\[-1mm]
\small AFM 41.85dB
\end{minipage}
\begin{minipage}{0.24\textwidth}
\centering
\includegraphics[width=\textwidth]{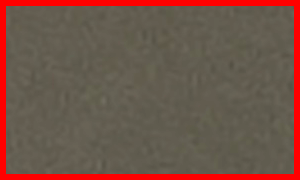}\\[-1mm]
\small IDF 41.82dB
\end{minipage}

\vspace{1mm}

\begin{minipage}{0.24\textwidth}
\centering
\includegraphics[width=\textwidth]{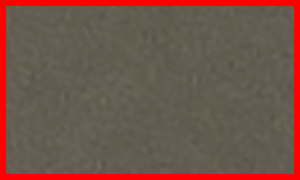}\\[-1mm]
\small NAFNet 42.45dB
\end{minipage}
\begin{minipage}{0.24\textwidth}
\centering
\includegraphics[width=\textwidth]{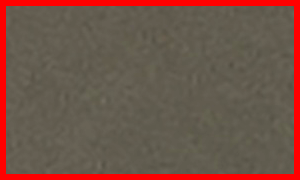}\\[-1mm]
\small XFormer 41.80dB
\end{minipage}
\begin{minipage}{0.24\textwidth}
\centering
\includegraphics[width=\textwidth]{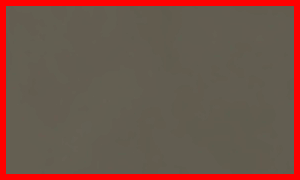}\\[-1mm]
\small \textbf{Ours 47.07dB}
\end{minipage}
\begin{minipage}{0.24\textwidth}
\centering
\includegraphics[width=\textwidth]{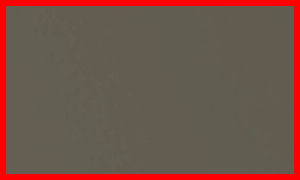}\\[-1mm]
\small Ground Truth
\end{minipage}
\end{minipage}

\end{minipage}
\vspace{-2mm}
\caption{Comparison between the qualitative results of various denoising networks including ours (noise translation network with pretrained NAFNet), on the out-of-distribution (OOD) datasets (Poly and iPhone).
Our result displays cleaner outputs compared to other state-of-the-art networks that are directly trained from a real-noise dataset.
Zoom in for better comparison.}
\vspace{-2mm}
\label{fig:qual_comparison2}
\end{figure*}

\begin{figure*}[tb]
\centering

\begin{minipage}{\textwidth}
\centering
\begin{minipage}{0.25\textwidth}
\centering
\includegraphics[width=\textwidth]{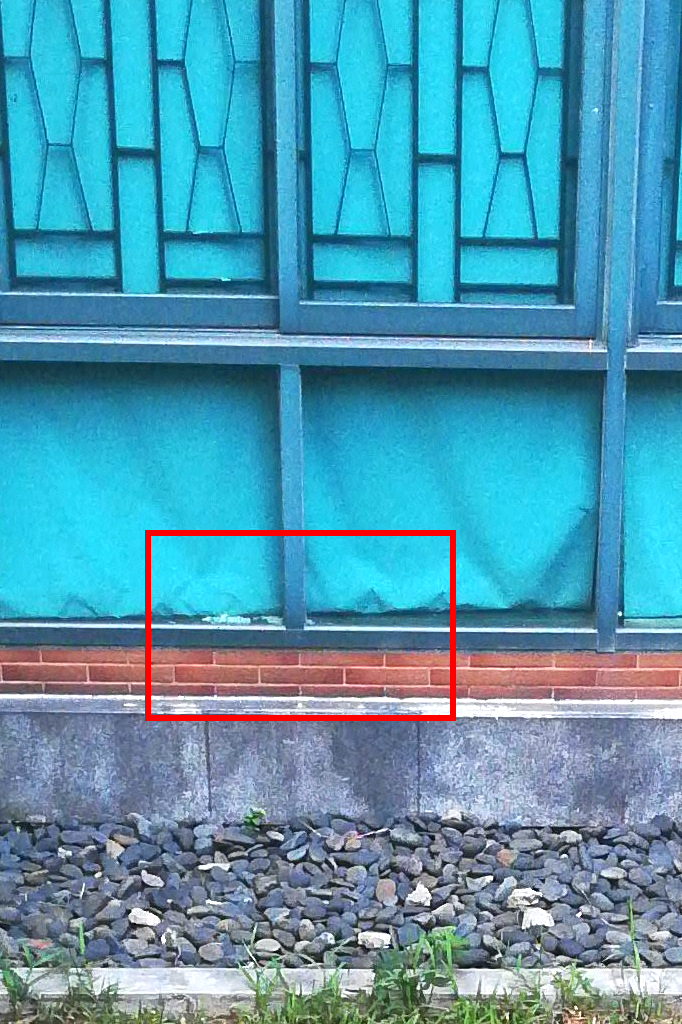}\\[-1mm]
\small Huawei Noisy
\end{minipage}%
\hspace{-0.5mm}%
\begin{minipage}{0.752\textwidth}
\centering
\vspace{-0.6mm}
\begin{minipage}{0.24\textwidth}
\centering
\includegraphics[width=\textwidth]{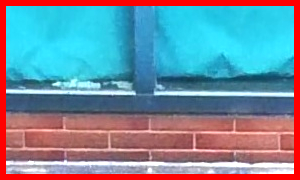}\\[-1mm]
\small R2R 33.51dB
\end{minipage}
\begin{minipage}{0.24\textwidth}
\centering
\includegraphics[width=\textwidth]{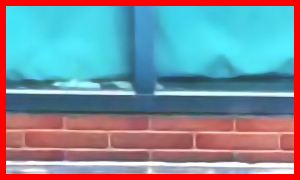}\\[-1mm]
\small AP-BSN 27.67dB
\end{minipage}
\begin{minipage}{0.24\textwidth}
\centering
\includegraphics[width=\textwidth]{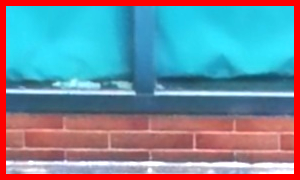}\\[-1mm]
\small SSID 29.47dB
\end{minipage}
\begin{minipage}{0.24\textwidth}
\centering
\includegraphics[width=\textwidth]{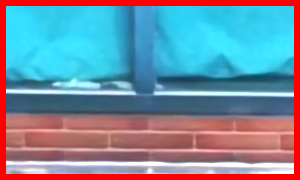}\\[-1mm]
\small APR-RD 29.21dB
\end{minipage}

\vspace{1mm}

\begin{minipage}{0.24\textwidth}
\centering
\includegraphics[width=\textwidth]{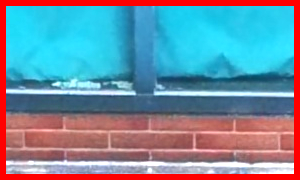}\\[-1mm]
\small Mask-DN 31.93dB
\end{minipage}
\begin{minipage}{0.24\textwidth}
\centering
\includegraphics[width=\textwidth]{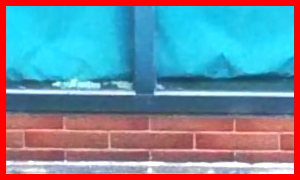}\\[-1mm]
\small CLIP-DN 31.45dB
\end{minipage}
\begin{minipage}{0.24\textwidth}
\centering
\includegraphics[width=\textwidth]{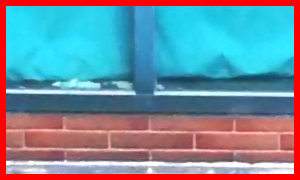}\\[-1mm]
\small AFM 32.33dB
\end{minipage}
\begin{minipage}{0.24\textwidth}
\centering
\includegraphics[width=\textwidth]{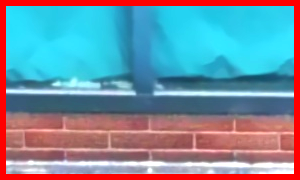}\\[-1mm]
\small IDF 28.33dB
\end{minipage}

\vspace{1mm}

\begin{minipage}{0.24\textwidth}
\centering
\includegraphics[width=\textwidth]{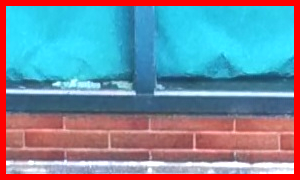}\\[-1mm]
\small NAFNet 31.92dB
\end{minipage}
\begin{minipage}{0.24\textwidth}
\centering
\includegraphics[width=\textwidth]{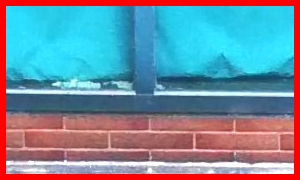}\\[-1mm]
\small XFormer 33.59dB
\end{minipage}
\begin{minipage}{0.24\textwidth}
\centering
\includegraphics[width=\textwidth]{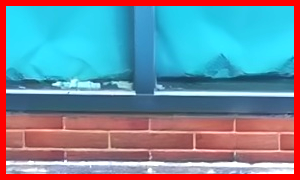}\\[-1mm]
\small \textbf{Ours 33.95dB}
\end{minipage}
\begin{minipage}{0.24\textwidth}
\centering
\includegraphics[width=\textwidth]{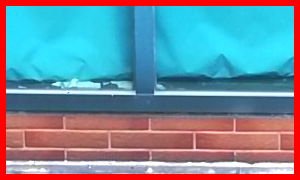}\\[-1mm]
\small Ground Truth
\end{minipage}
\end{minipage}

\vspace{1mm}
\end{minipage}

\begin{minipage}{\textwidth}
\centering
\begin{minipage}{0.25\textwidth}
\centering
\includegraphics[width=\textwidth]{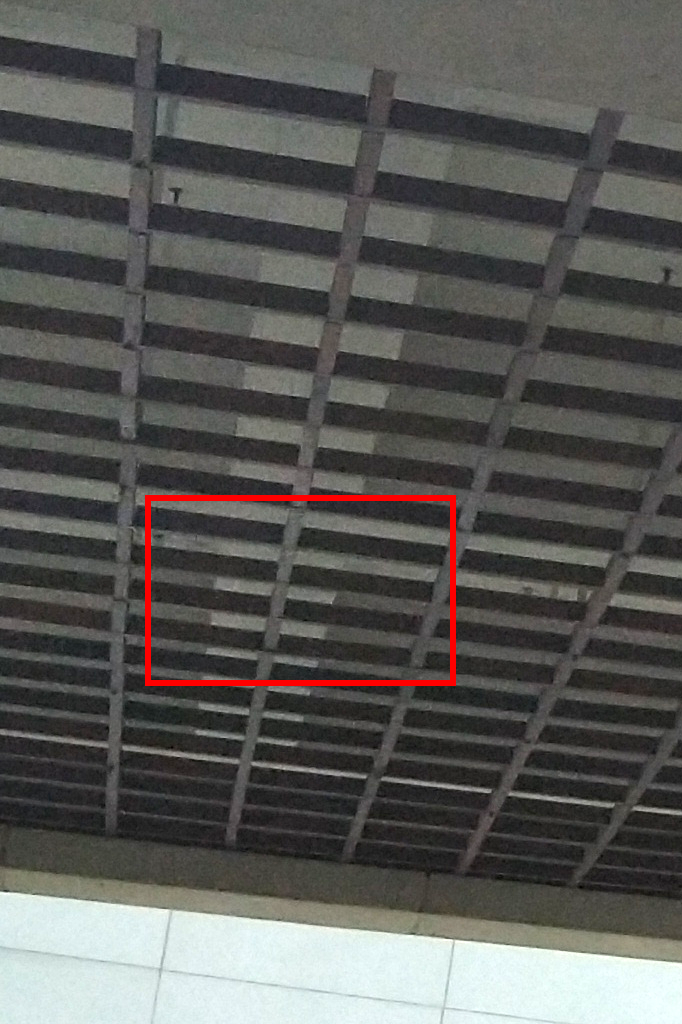}\\[-1mm]
\small OPPO Noisy
\end{minipage}%
\hspace{-0.5mm}%
\begin{minipage}{0.752\textwidth}
\centering
\vspace{-0.6mm}
\begin{minipage}{0.24\textwidth}
\centering
\includegraphics[width=\textwidth]{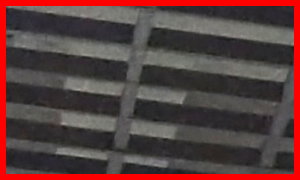}\\[-1mm]
\small R2R 39.19dB
\end{minipage}
\begin{minipage}{0.24\textwidth}
\centering
\includegraphics[width=\textwidth]{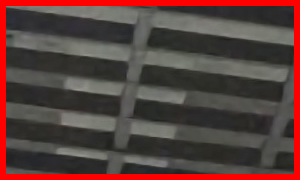}\\[-1mm]
\small AP-BSN 40.36dB
\end{minipage}
\begin{minipage}{0.24\textwidth}
\centering
\includegraphics[width=\textwidth]{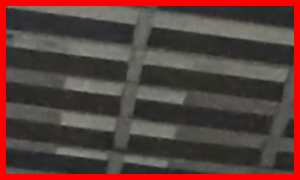}\\[-1mm]
\small SSID 40.07dB
\end{minipage}
\begin{minipage}{0.24\textwidth}
\centering
\includegraphics[width=\textwidth]{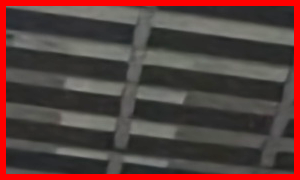}\\[-1mm]
\small APR-RD 40.34dB
\end{minipage}

\vspace{1mm}

\begin{minipage}{0.24\textwidth}
\centering
\includegraphics[width=\textwidth]{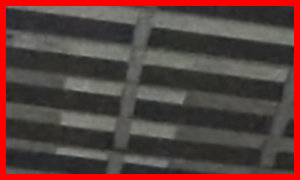}\\[-1mm]
\small Mask-DN 39.16dB
\end{minipage}
\begin{minipage}{0.24\textwidth}
\centering
\includegraphics[width=\textwidth]{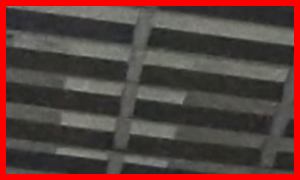}\\[-1mm]
\small CLIP-DN 39.58dB
\end{minipage}
\begin{minipage}{0.24\textwidth}
\centering
\includegraphics[width=\textwidth]{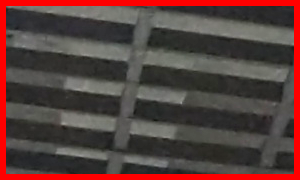}\\[-1mm]
\small AFM 39.69dB
\end{minipage}
\begin{minipage}{0.24\textwidth}
\centering
\includegraphics[width=\textwidth]{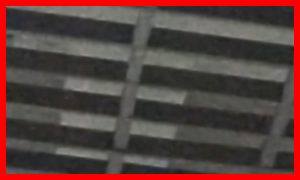}\\[-1mm]
\small IDF 38.70dB
\end{minipage}

\vspace{1mm}

\begin{minipage}{0.24\textwidth}
\centering
\includegraphics[width=\textwidth]{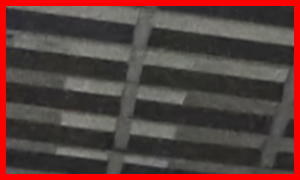}\\[-1mm]
\small NAFNet 39.02dB
\end{minipage}
\begin{minipage}{0.24\textwidth}
\centering
\includegraphics[width=\textwidth]{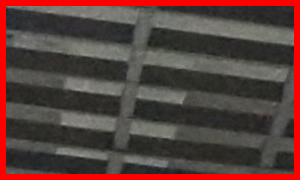}\\[-1mm]
\small XFormer 39.14dB
\end{minipage}
\begin{minipage}{0.24\textwidth}
\centering
\includegraphics[width=\textwidth]{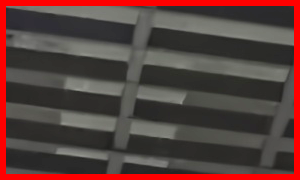}\\[-1mm]
\small \textbf{Ours 42.19dB}
\end{minipage}
\begin{minipage}{0.24\textwidth}
\centering
\includegraphics[width=\textwidth]{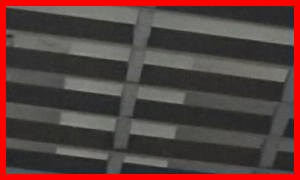}\\[-1mm]
\small Ground Truth
\end{minipage}
\end{minipage}

\end{minipage}
\vspace{-2mm}
\caption{Comparison between the qualitative results of various denoising networks including ours (noise translation network with pretrained NAFNet), on the out-of-distribution (OOD) datasets (Huawei and OPPO).
Our result displays cleaner outputs compared to other state-of-the-art networks that are directly trained from a real-noise dataset.
Zoom in for better comparison.}
\vspace{-2mm}
\label{fig:qual_comparison3}
\end{figure*}

\begin{figure*}[tb]
\centering

\begin{minipage}{\textwidth}
\centering
\begin{minipage}{0.25\textwidth}
\centering
\includegraphics[width=\textwidth]{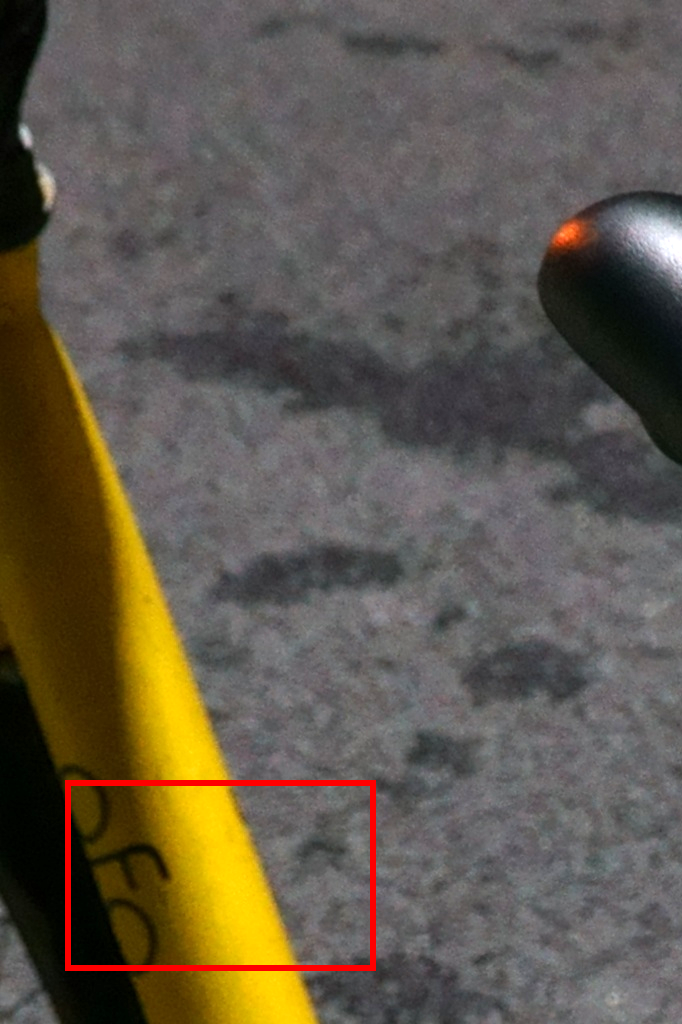}\\[-1mm]
\small Sony Noisy
\end{minipage}%
\hspace{-0.5mm}%
\begin{minipage}{0.752\textwidth}
\centering
\vspace{-0.6mm}
\begin{minipage}{0.24\textwidth}
\centering
\includegraphics[width=\textwidth]{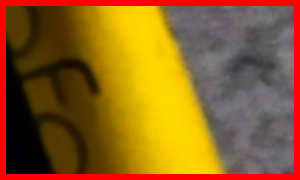}\\[-1mm]
\small R2R 40.01dB
\end{minipage}
\begin{minipage}{0.24\textwidth}
\centering
\includegraphics[width=\textwidth]{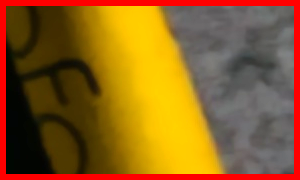}\\[-1mm]
\small AP-BSN 39.48dB
\end{minipage}
\begin{minipage}{0.24\textwidth}
\centering
\includegraphics[width=\textwidth]{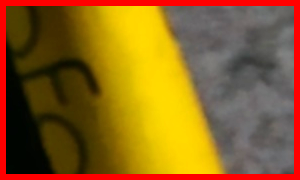}\\[-1mm]
\small SSID 40.19dB
\end{minipage}
\begin{minipage}{0.24\textwidth}
\centering
\includegraphics[width=\textwidth]{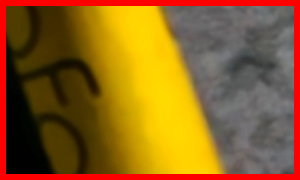}\\[-1mm]
\small APR-RD 39.58dB
\end{minipage}

\vspace{1mm}

\begin{minipage}{0.24\textwidth}
\centering
\includegraphics[width=\textwidth]{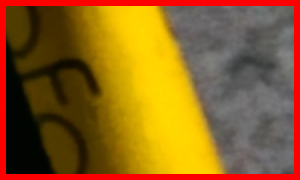}\\[-1mm]
\small Mask-DN 41.78dB
\end{minipage}
\begin{minipage}{0.24\textwidth}
\centering
\includegraphics[width=\textwidth]{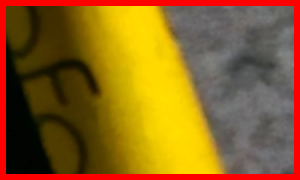}\\[-1mm]
\small CLIP-DN 41.52dB
\end{minipage}
\begin{minipage}{0.24\textwidth}
\centering
\includegraphics[width=\textwidth]{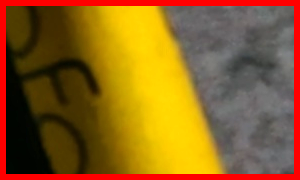}\\[-1mm]
\small AFM 41.20dB
\end{minipage}
\begin{minipage}{0.24\textwidth}
\centering
\includegraphics[width=\textwidth]{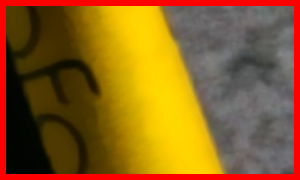}\\[-1mm]
\small IDF 40.11dB
\end{minipage}

\vspace{1mm}

\begin{minipage}{0.24\textwidth}
\centering
\includegraphics[width=\textwidth]{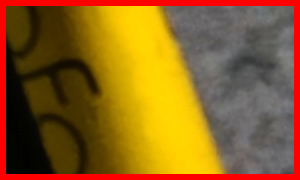}\\[-1mm]
\small NAFNet 41.62dB
\end{minipage}
\begin{minipage}{0.24\textwidth}
\centering
\includegraphics[width=\textwidth]{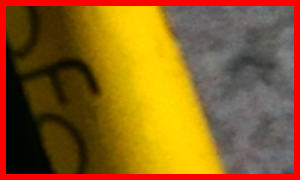}\\[-1mm]
\small XFormer 41.48dB
\end{minipage}
\begin{minipage}{0.24\textwidth}
\centering
\includegraphics[width=\textwidth]{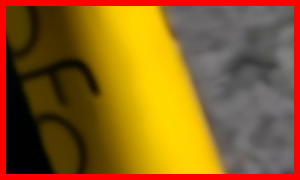}\\[-1mm]
\small \textbf{Ours 41.72dB}
\end{minipage}
\begin{minipage}{0.24\textwidth}
\centering
\includegraphics[width=\textwidth]{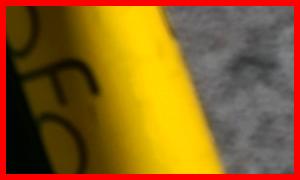}\\[-1mm]
\small Ground Truth
\end{minipage}
\end{minipage}

\vspace{1mm}
\end{minipage}

\begin{minipage}{\textwidth}
\centering
\begin{minipage}{0.25\textwidth}
\centering
\includegraphics[width=\textwidth]{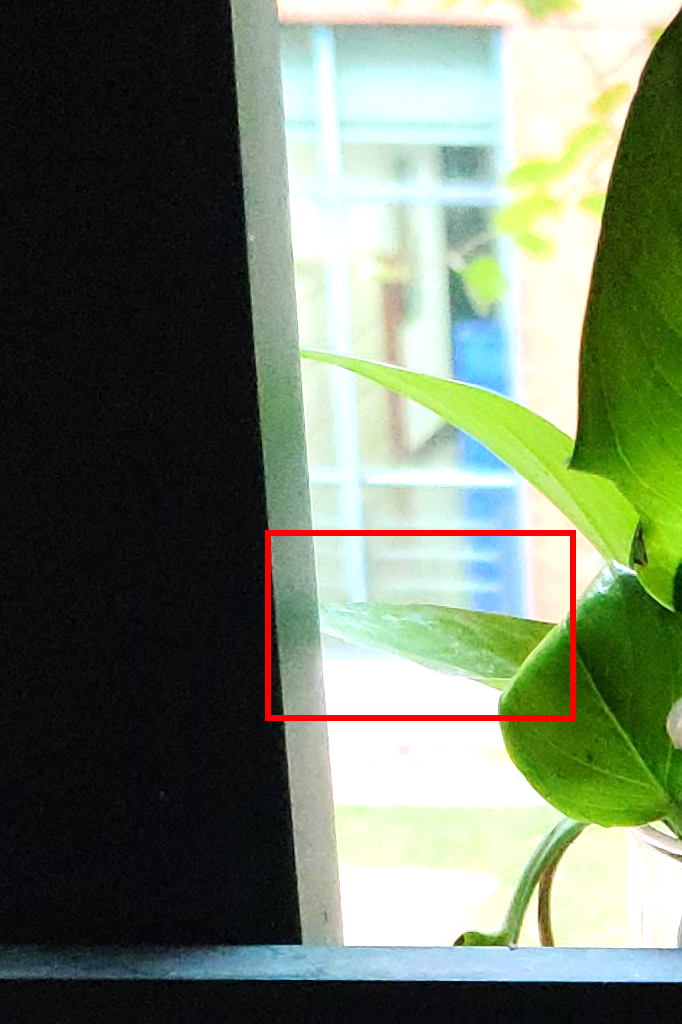}\\[-1mm]
\small Xiaomi Noisy
\end{minipage}%
\hspace{-0.5mm}%
\begin{minipage}{0.752\textwidth}
\centering
\vspace{-0.6mm}
\begin{minipage}{0.24\textwidth}
\centering
\includegraphics[width=\textwidth]{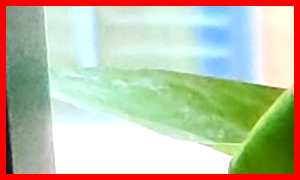}\\[-1mm]
\small R2R 37.99dB
\end{minipage}
\begin{minipage}{0.24\textwidth}
\centering
\includegraphics[width=\textwidth]{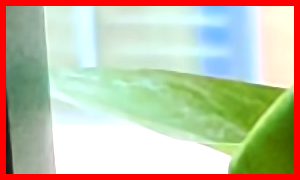}\\[-1mm]
\small AP-BSN 34.70dB
\end{minipage}
\begin{minipage}{0.24\textwidth}
\centering
\includegraphics[width=\textwidth]{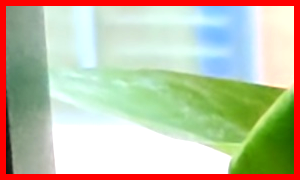}\\[-1mm]
\small SSID 36.94dB
\end{minipage}
\begin{minipage}{0.24\textwidth}
\centering
\includegraphics[width=\textwidth]{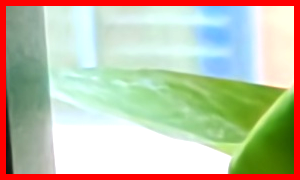}\\[-1mm]
\small APR-RD 37.97dB
\end{minipage}

\vspace{1mm}

\begin{minipage}{0.24\textwidth}
\centering
\includegraphics[width=\textwidth]{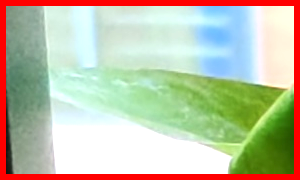}\\[-1mm]
\small Mask-DN 37.91dB
\end{minipage}
\begin{minipage}{0.24\textwidth}
\centering
\includegraphics[width=\textwidth]{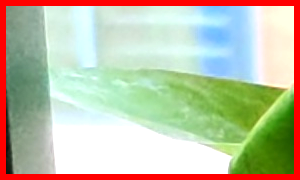}\\[-1mm]
\small CLIP-DN 38.07dB
\end{minipage}
\begin{minipage}{0.24\textwidth}
\centering
\includegraphics[width=\textwidth]{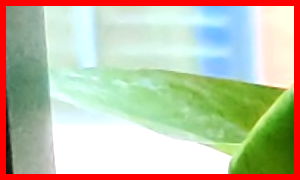}\\[-1mm]
\small AFM 38.76dB
\end{minipage}
\begin{minipage}{0.24\textwidth}
\centering
\includegraphics[width=\textwidth]{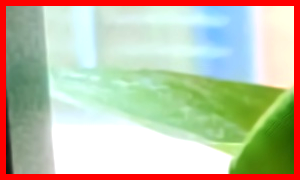}\\[-1mm]
\small IDF 34.96dB
\end{minipage}

\vspace{1mm}

\begin{minipage}{0.24\textwidth}
\centering
\includegraphics[width=\textwidth]{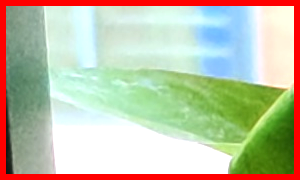}\\[-1mm]
\small NAFNet 37.96dB
\end{minipage}
\begin{minipage}{0.24\textwidth}
\centering
\includegraphics[width=\textwidth]{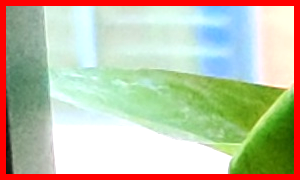}\\[-1mm]
\small XFormer 38.32dB
\end{minipage}
\begin{minipage}{0.24\textwidth}
\centering
\includegraphics[width=\textwidth]{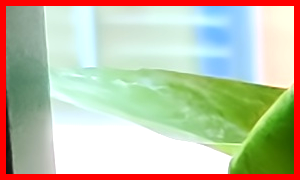}\\[-1mm]
\small \textbf{Ours 39.65dB}
\end{minipage}
\begin{minipage}{0.24\textwidth}
\centering
\includegraphics[width=\textwidth]{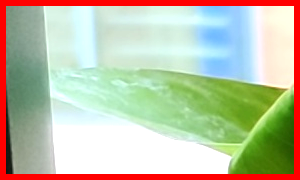}\\[-1mm]
\small Ground Truth
\end{minipage}
\end{minipage}

\end{minipage}
\vspace{-2mm}
\caption{Comparison between the qualitative results of various denoising networks including ours (noise translation network with pretrained NAFNet), on the out-of-distribution (OOD) datasets (Sony and Xiaomi).
Our result displays cleaner outputs compared to other state-of-the-art networks that are directly trained from a real-noise dataset.
Zoom in for better comparison.}
\vspace{-2mm}
\label{fig:qual_comparison4}
\end{figure*}

\begin{figure*}[t]
\centering

\begin{minipage}{\textwidth}
\centering
\begin{minipage}{0.25\textwidth}
\centering
\includegraphics[width=\textwidth]{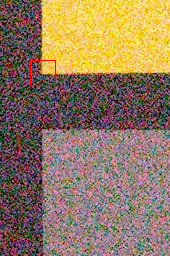}\\[-1mm]
\small SIDD Noisy
\end{minipage}%
\hspace{-0.5mm}%
\begin{minipage}{0.752\textwidth}
\centering
\vspace{-0.6mm}
\begin{minipage}{0.24\textwidth}
\centering
\includegraphics[width=\textwidth]{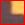}\\[-1mm]
\small R2R 28.98dB
\end{minipage}
\begin{minipage}{0.24\textwidth}
\centering
\includegraphics[width=\textwidth]{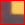}\\[-1mm]
\small AP-BSN 30.65dB
\end{minipage}
\begin{minipage}{0.24\textwidth}
\centering
\includegraphics[width=\textwidth]{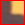}\\[-1mm]
\small SSID 31.84dB
\end{minipage}
\begin{minipage}{0.24\textwidth}
\centering
\includegraphics[width=\textwidth]{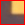}\\[-1mm]
\small APR-RD 32.29dB
\end{minipage}

\vspace{1mm}

\begin{minipage}{0.24\textwidth}
\centering
\includegraphics[width=\textwidth]{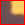}\\[-1mm]
\small Mask-DN 32.92dB
\end{minipage}
\begin{minipage}{0.24\textwidth}
\centering
\includegraphics[width=\textwidth]{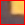}\\[-1mm]
\small CLIP-DN 30.41dB
\end{minipage}
\begin{minipage}{0.24\textwidth}
\centering
\includegraphics[width=\textwidth]{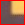}\\[-1mm]
\small AFM 32.19dB
\end{minipage}
\begin{minipage}{0.24\textwidth}
\centering
\includegraphics[width=\textwidth]{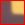}\\[-1mm]
\small IDF 31.18dB
\end{minipage}

\vspace{1mm}

\begin{minipage}{0.24\textwidth}
\centering
\includegraphics[width=\textwidth]{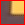}\\[-1mm]
\small NAFNet 34.54dB
\end{minipage}
\begin{minipage}{0.24\textwidth}
\centering
\includegraphics[width=\textwidth]{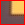}\\[-1mm]
\small XFormer 34.90dB
\end{minipage}
\begin{minipage}{0.24\textwidth}
\centering
\includegraphics[width=\textwidth]{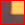}\\[-1mm]
\small \textbf{Ours 34.12dB}
\end{minipage}
\begin{minipage}{0.24\textwidth}
\centering
\includegraphics[width=\textwidth]{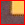}\\[-1mm]
\small Ground Truth
\end{minipage}
\end{minipage}

\vspace{1mm}
\end{minipage}
\vspace{-2mm}
\caption{
  Comparison between the qualitative results of various denoising networks including ours (noise translation network with pretrained NAFNet), on the SIDD dataset.
  Our result displays competitive visual quality with real noise.
}
\vspace{-2mm}
\label{fig:qualitative_sidd}
\end{figure*}

Qualitative results on all datasets in the main paper are shown in Figures~\ref{fig:qual_comparison2}, ~\ref{fig:qual_comparison3}, ~\ref{fig:qual_comparison4} and ~\ref{fig:qualitative_sidd}.
Our method significantly surpasses the PSNR scores of other denoising models on all OOD datasets (iPhone, Huawei, OPPO, Sony, Xiaomi).
For the in-distribution SIDD, our method produces clean results without generating unnecessary zipper artifacts.

\subsection{NIND (Natural Image Noise Dataset)}
\begin{figure*}[t]

  \centering
  \scalebox{0.95}{
    \begin{tabular}{@{}c@{\hspace{0.3em}}c@{\hspace{0.3em}}c@{}}
      \parbox[t]{0.333\textwidth}{
        \centering
      \includegraphics[width=\linewidth]{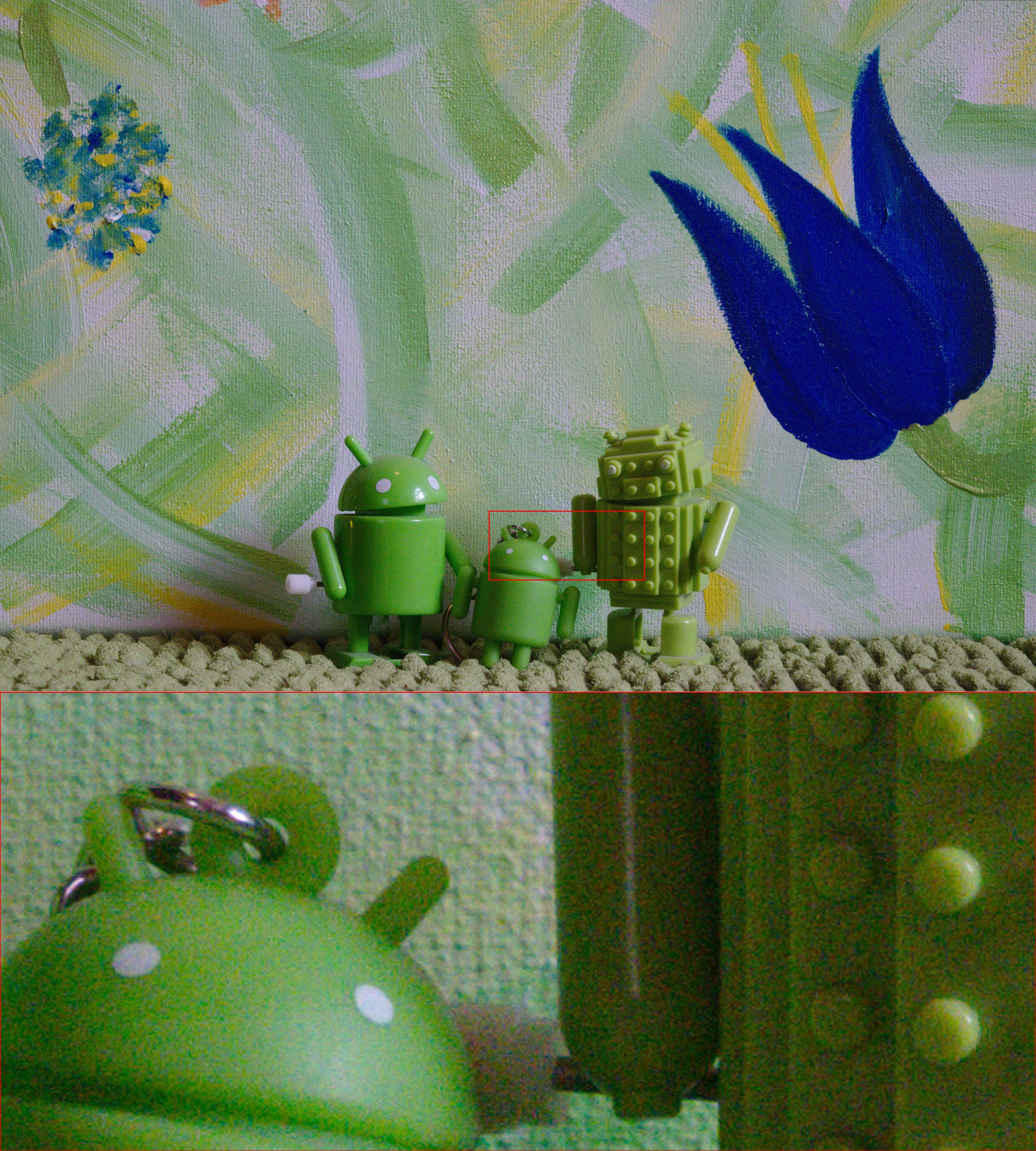}                 \\
        \fontsize{8.5pt}{8.5pt}\selectfont High ISO
      } &
      \parbox[t]{0.333\textwidth}{
        \centering
      \includegraphics[width=\linewidth]{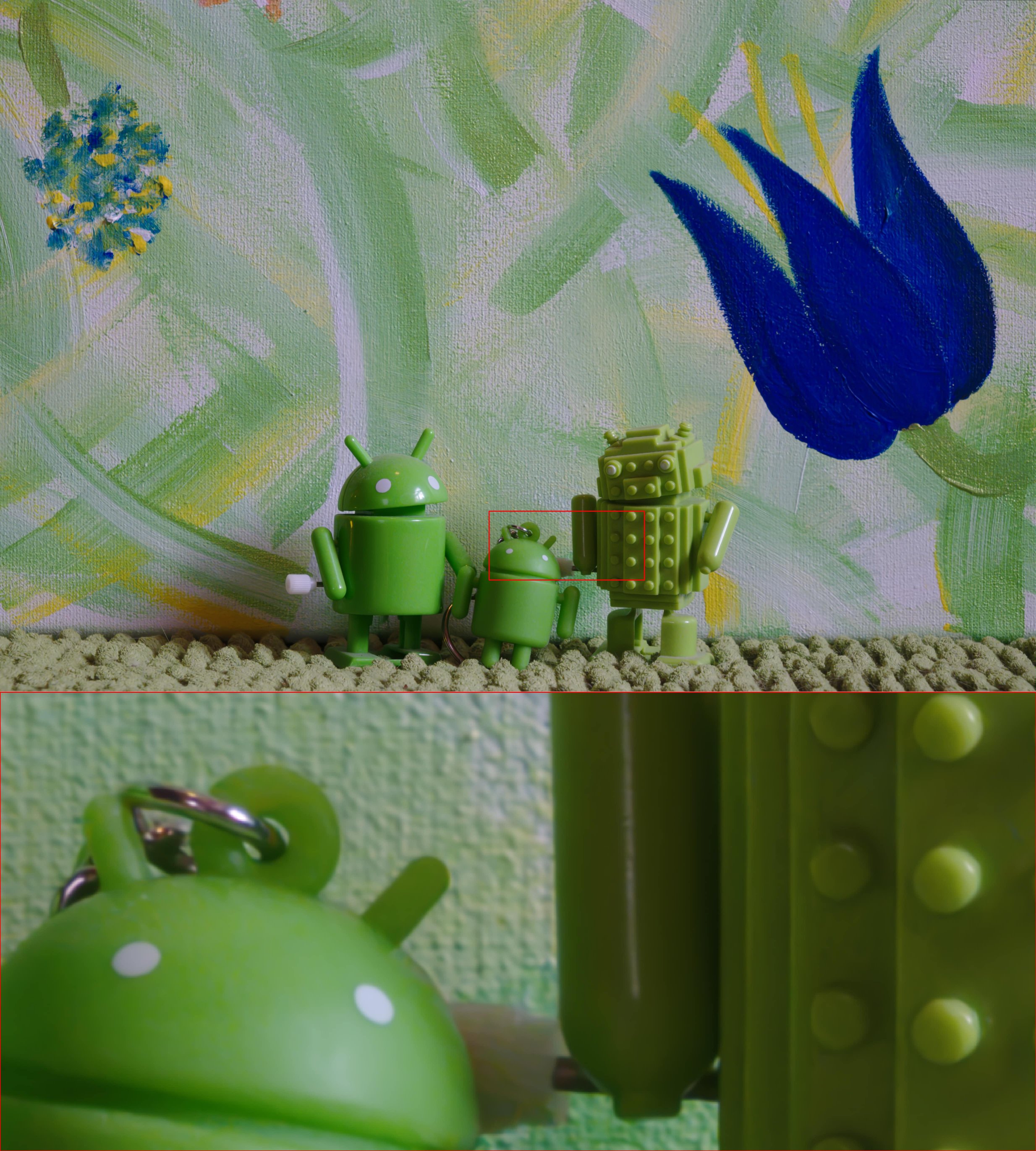} \\
        \fontsize{8.5pt}{8.5pt}\selectfont Denoised High ISO (\bf{Ours})
      } &
      \parbox[t]{0.333\textwidth}{
        \centering
      \includegraphics[width=\linewidth]{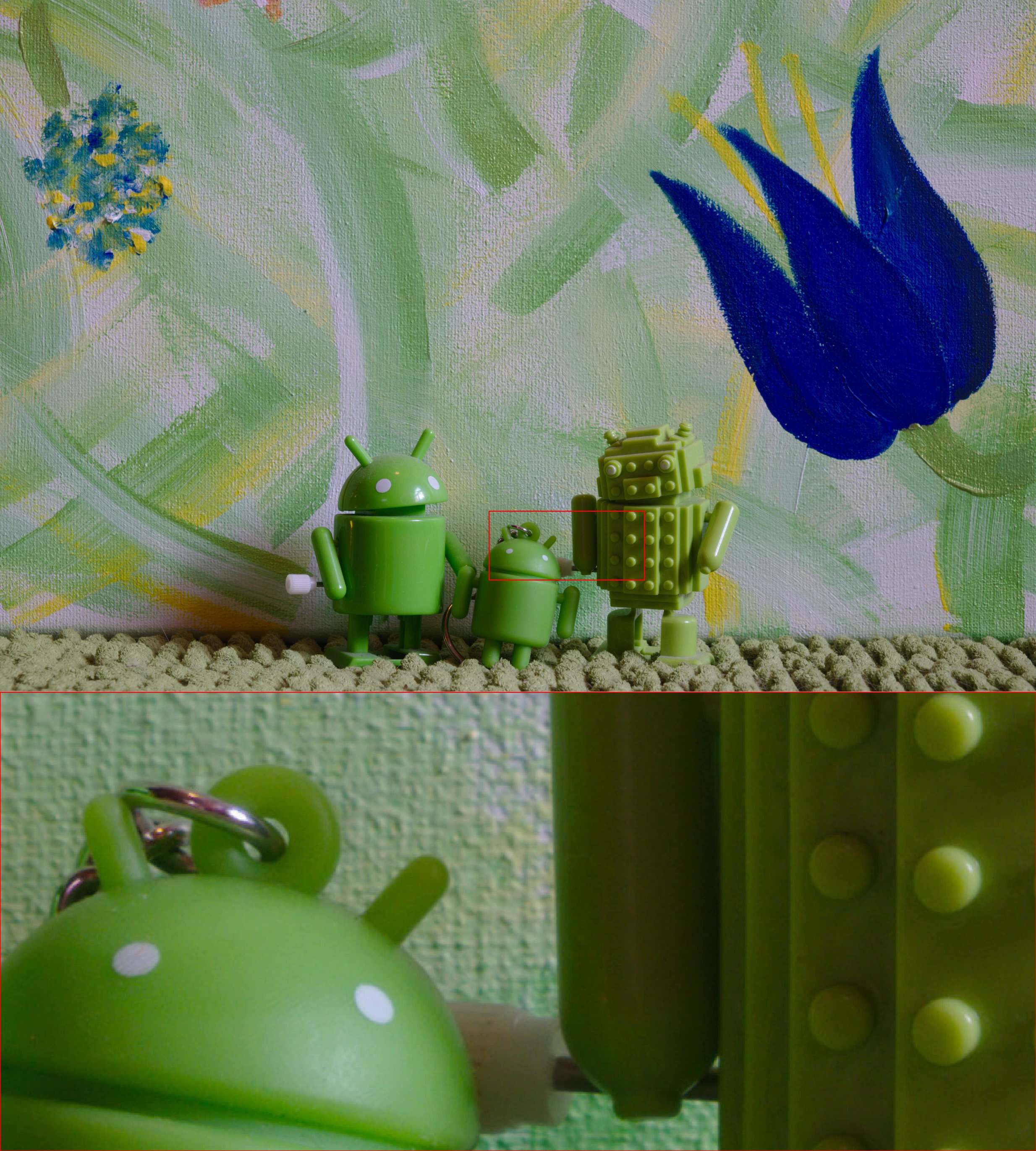}                 \\
        \fontsize{8.5pt}{8.5pt}\selectfont ISO 200
      }\vspace{2mm}                                                                                        \\
    \end{tabular}
  }

  \scalebox{0.95}{
    \begin{tabular}{@{}c@{\hspace{0.3em}}c@{\hspace{0.3em}}c@{}}
      \parbox[t]{0.333\textwidth}{
        \centering
      \includegraphics[width=\linewidth]{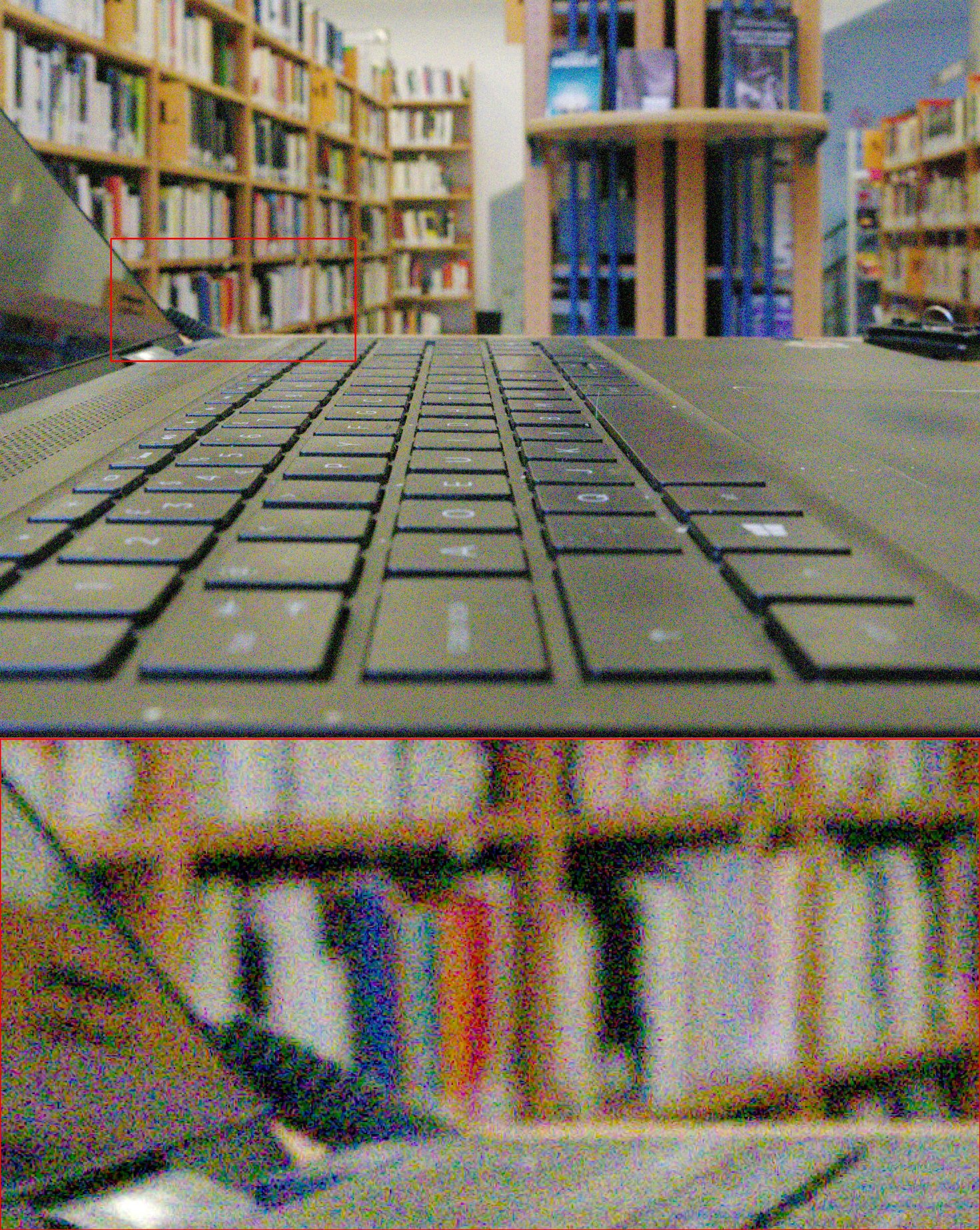}                 \\
        \fontsize{8.5pt}{8.5pt}\selectfont High ISO
      } &
      \parbox[t]{0.333\textwidth}{
        \centering
      \includegraphics[width=\linewidth]{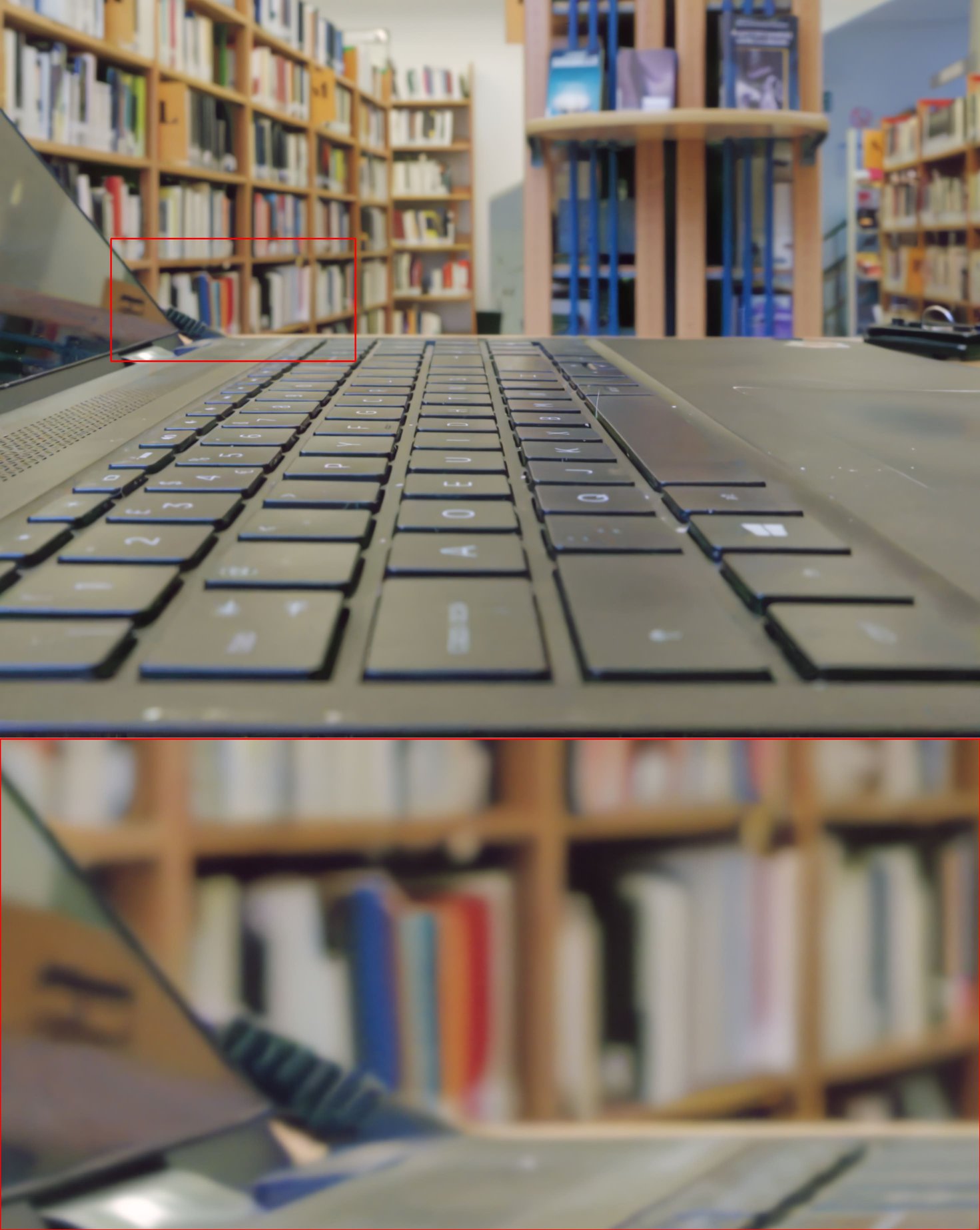} \\
        \fontsize{8.5pt}{8.5pt}\selectfont Denoised High ISO (\bf{Ours})
      } &
      \parbox[t]{0.333\textwidth}{
        \centering
      \includegraphics[width=\linewidth]{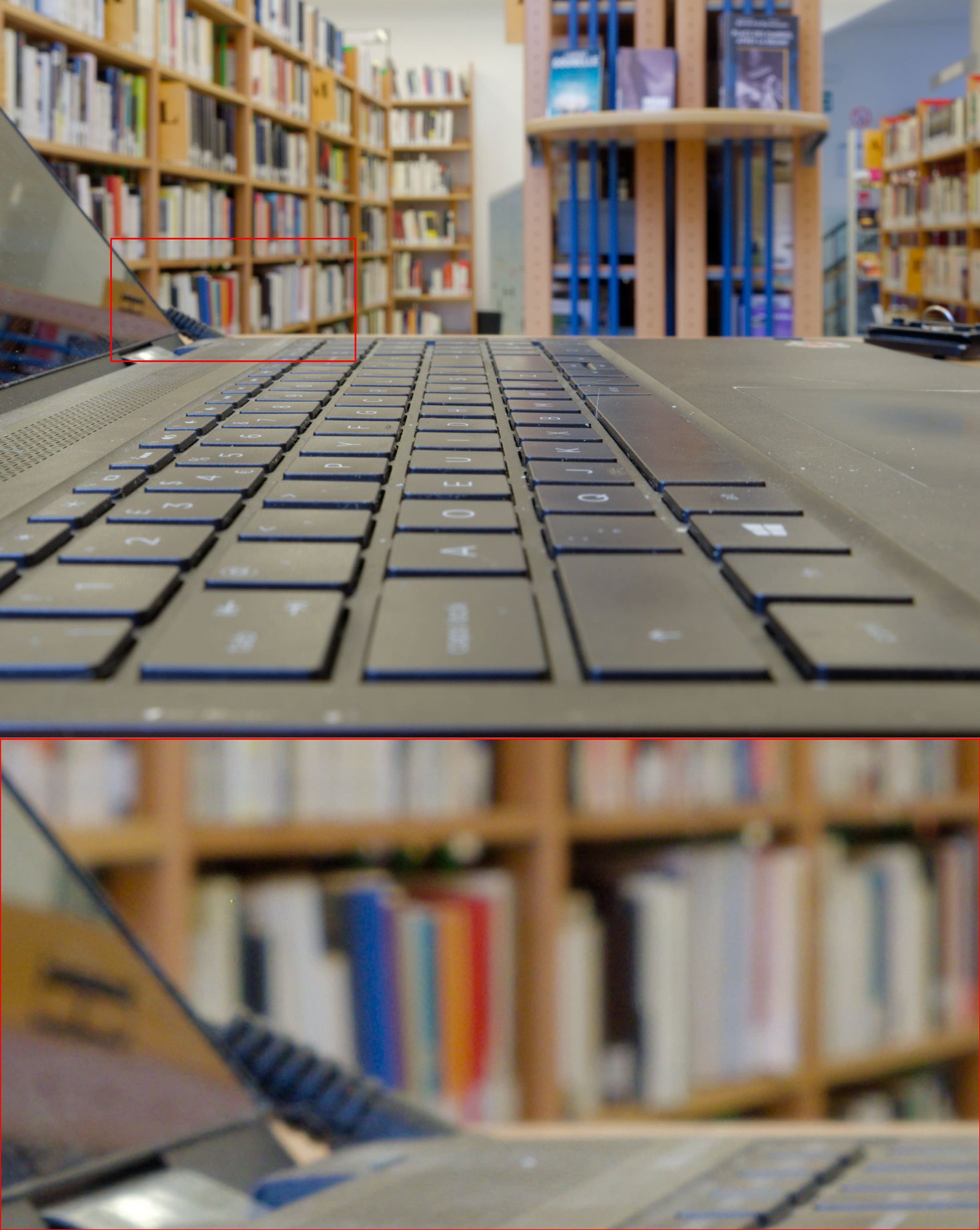}                 \\
        \fontsize{8.5pt}{8.5pt}\selectfont ISO 200
      }   \vspace{2mm}                                                                                              \\
    \end{tabular}
  }
  \vspace{2mm}
  \scalebox{0.95}{
    \begin{tabular}{@{}c@{\hspace{0.3em}}c@{\hspace{0.3em}}c@{}}
      \parbox[t]{0.333\textwidth}{
        \centering
      \includegraphics[width=\linewidth]{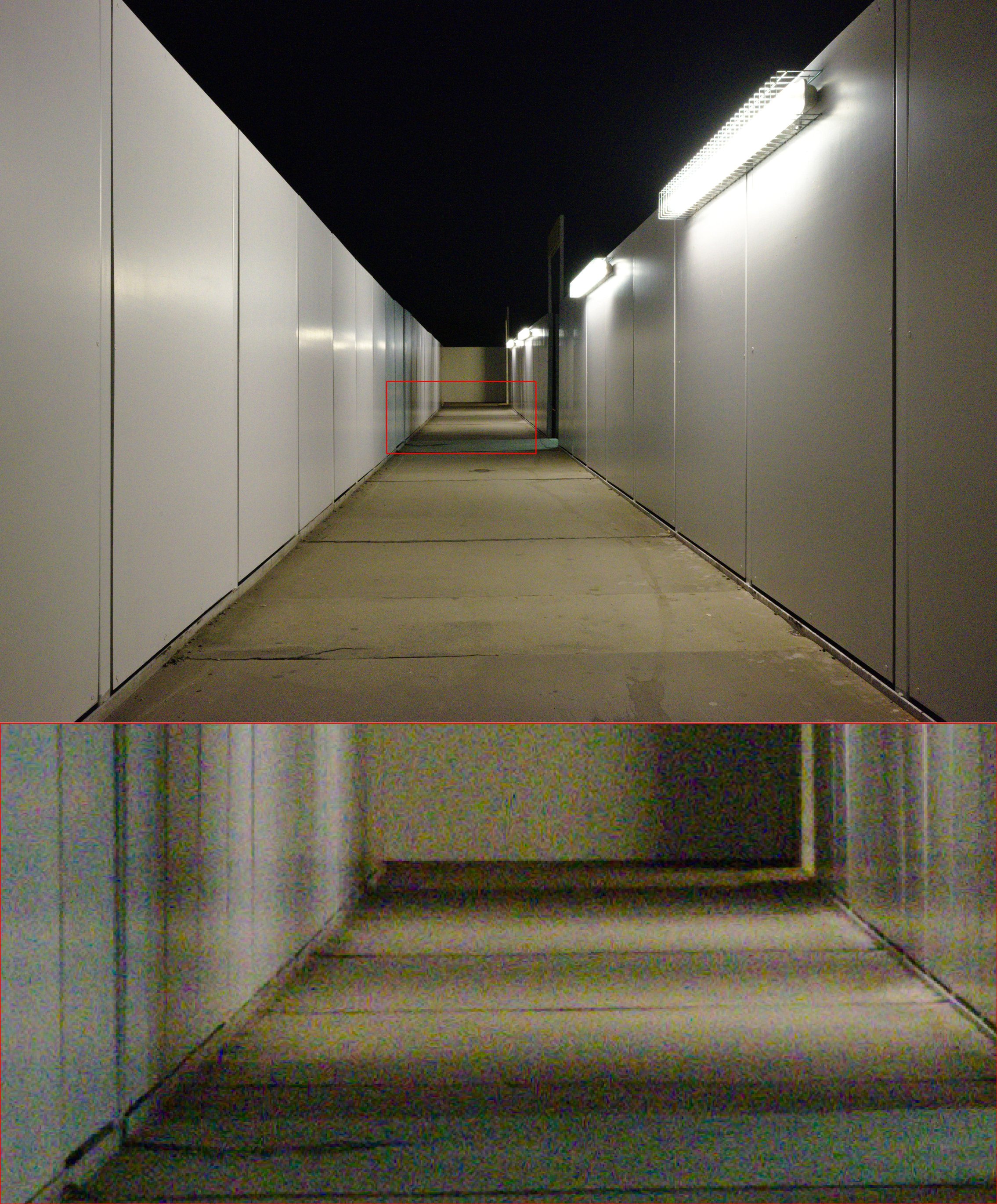}                 \\
        \fontsize{8.5pt}{8.5pt}\selectfont High ISO
      } &
      \parbox[t]{0.333\textwidth}{
        \centering
      \includegraphics[width=\linewidth]{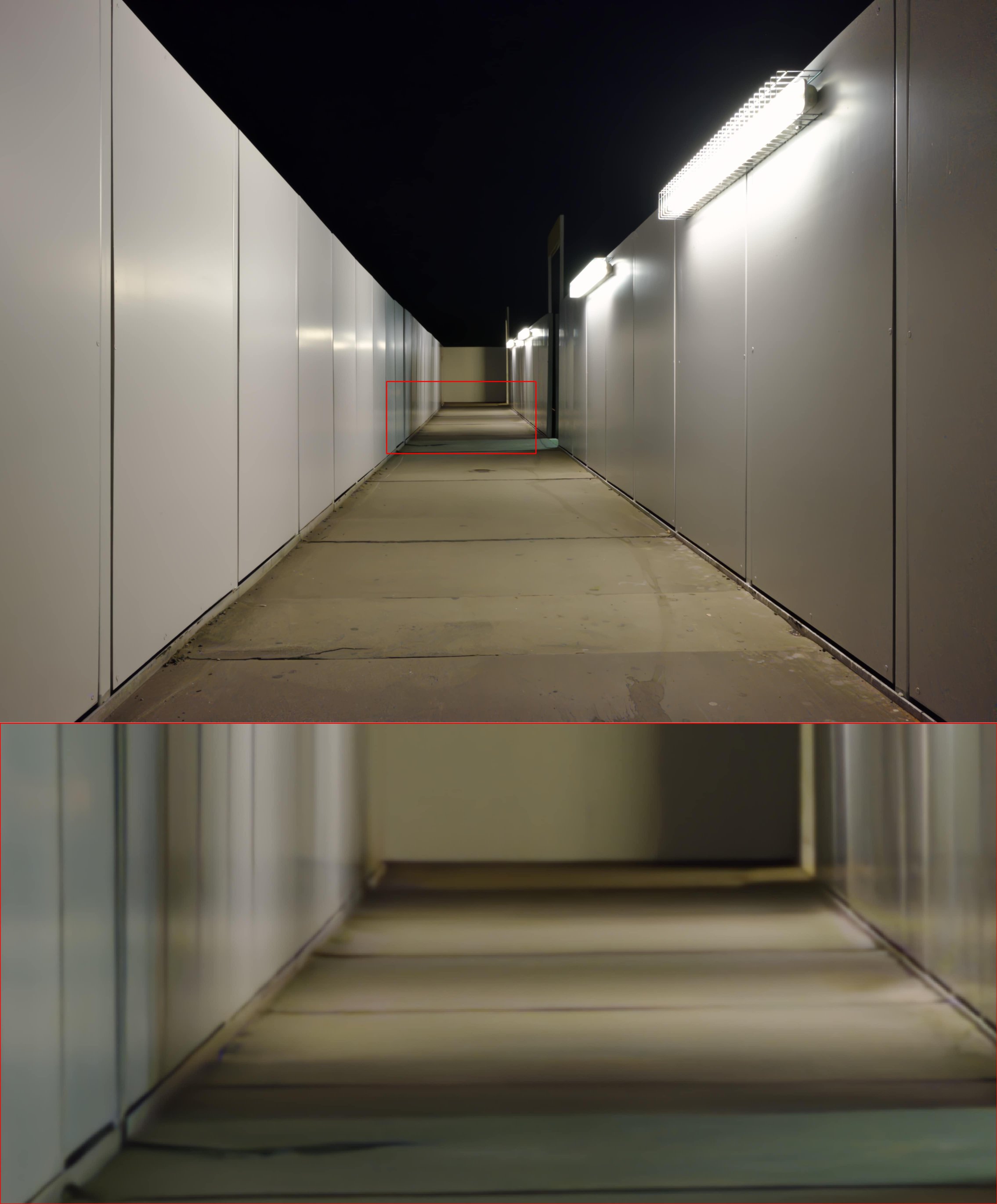} \\
        \fontsize{8.5pt}{8.5pt}\selectfont Denoised High ISO (\bf{Ours})
      } &
      \parbox[t]{0.333\textwidth}{
        \centering
      \includegraphics[width=\linewidth]{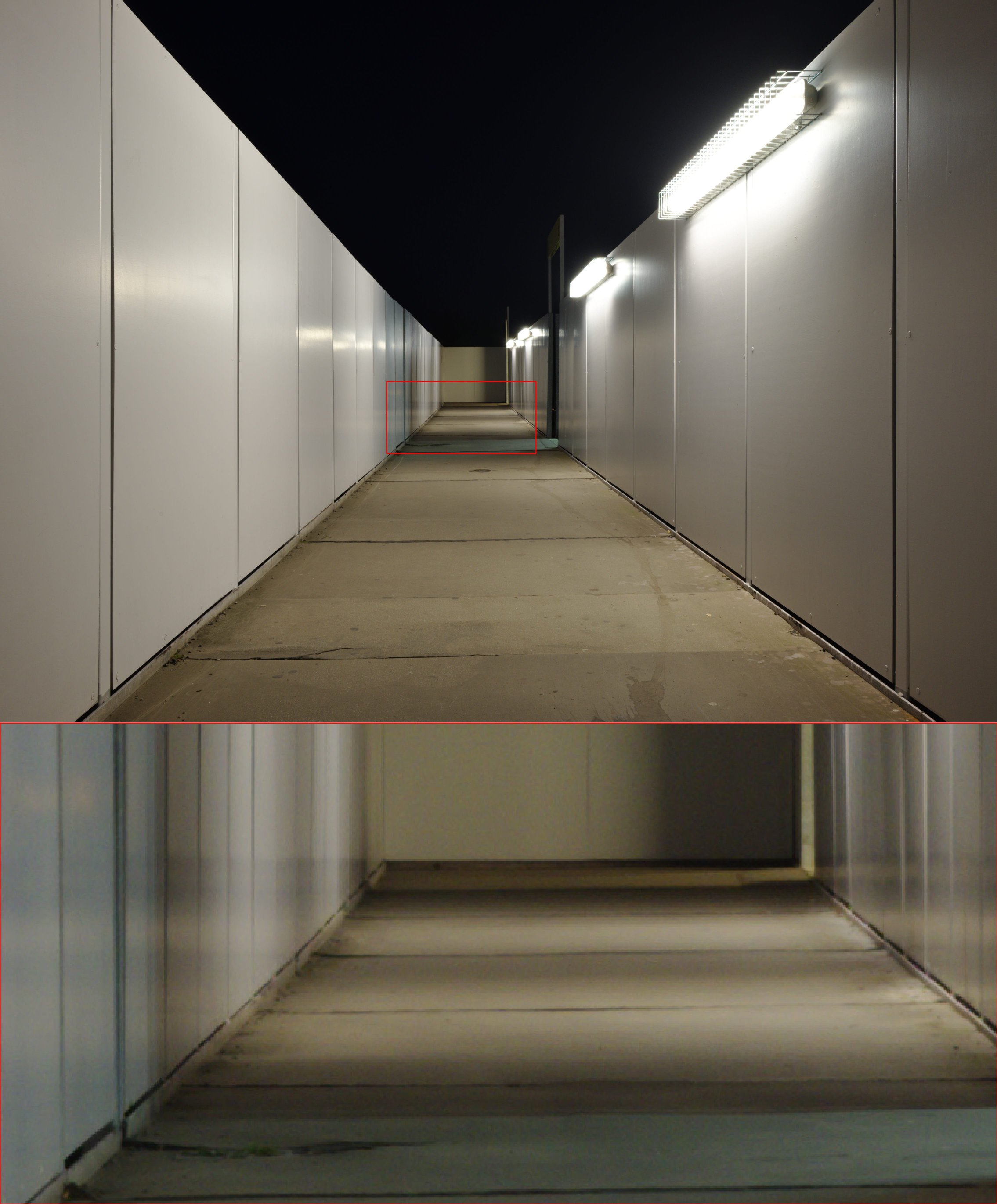}                 \\
        \fontsize{8.5pt}{8.5pt}\selectfont ISO 200
      }                                                                                                        \\
    \end{tabular}
  }
  \caption{
    Qualitative results of our noise translation network with pretrained NAFNet applied to high ISO images from the NIND dataset.  
    Our method effectively removes noise while preserving fine details, producing outputs comparable to low ISO (ISO 200) images.  
    Zoom in for a closer inspection of the visual quality. 
  }
  \label{fig:nind1}
\end{figure*}

\begin{figure*}[t]

  \vspace{-2mm}
  \centering

  \scalebox{0.95}{
    \begin{tabular}{@{}c@{\hspace{0.3em}}c@{\hspace{0.3em}}c@{}}
      \parbox[t]{0.333\textwidth}{
        \centering
      \includegraphics[width=\linewidth]{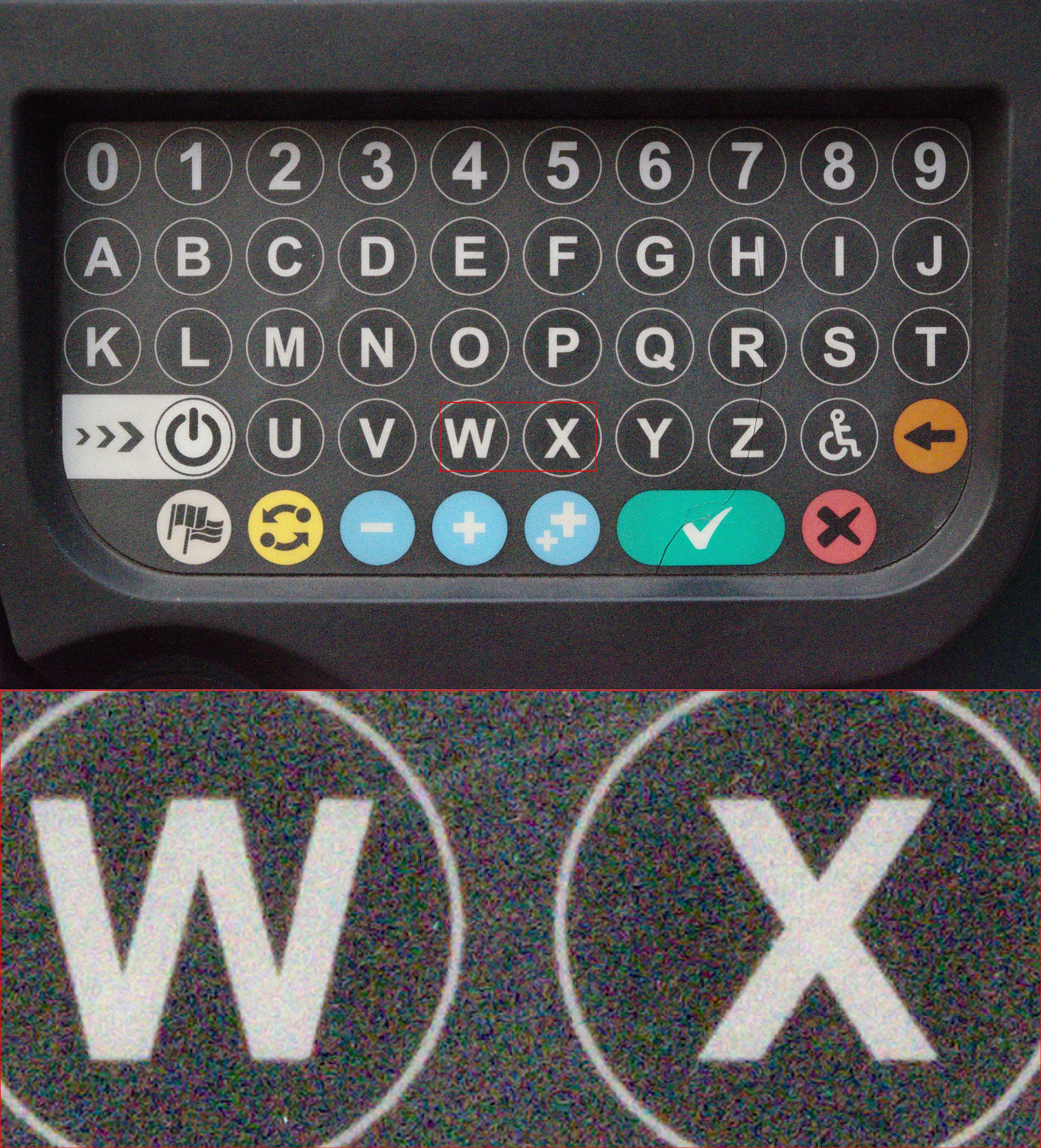}                 \\
        \fontsize{8.5pt}{8.5pt}\selectfont High ISO
      } &
      \parbox[t]{0.333\textwidth}{
        \centering
      \includegraphics[width=\linewidth]{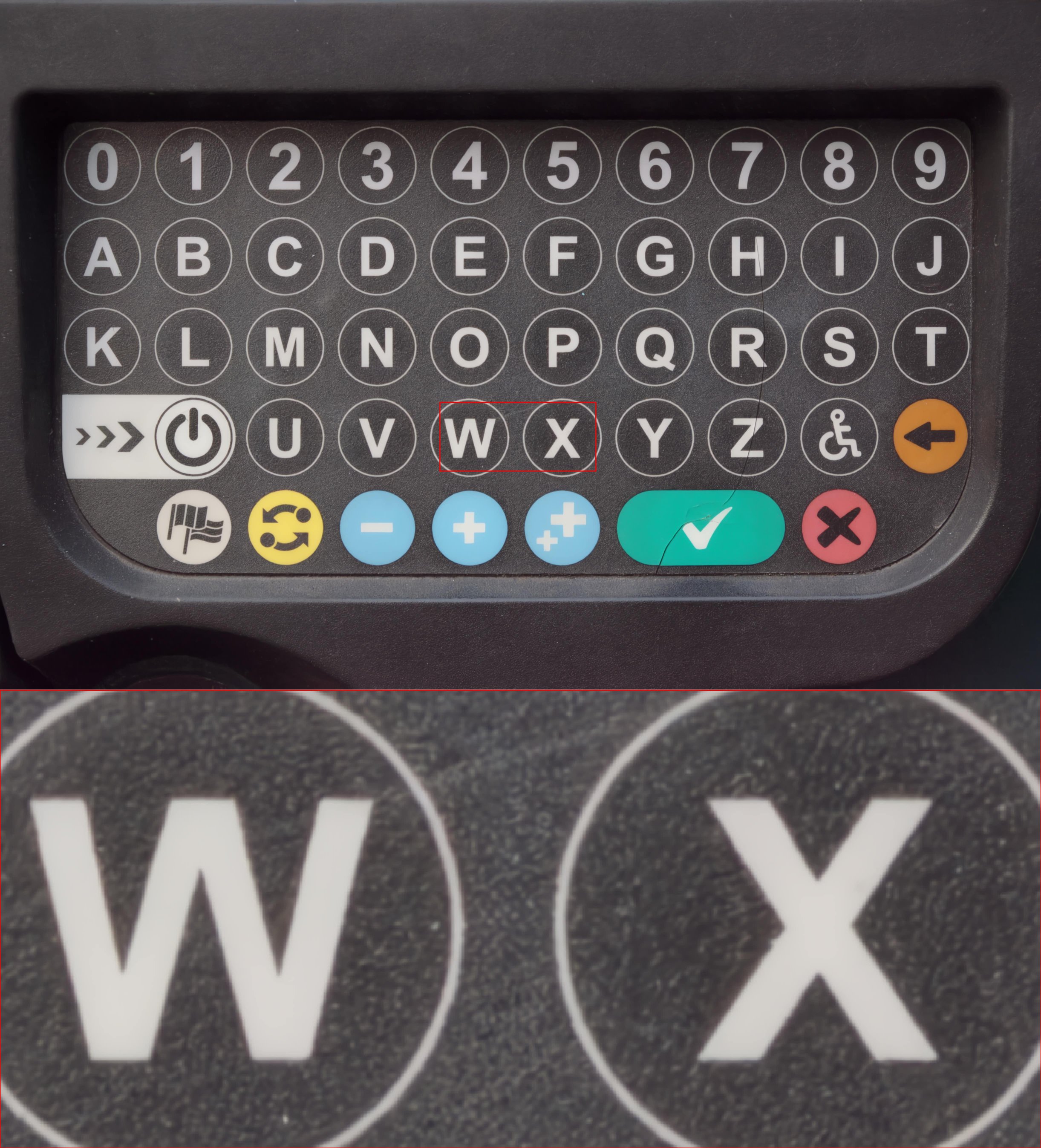} \\
        \fontsize{8.5pt}{8.5pt}\selectfont Denoised High ISO (\bf{Ours})
      } &
      \parbox[t]{0.333\textwidth}{
        \centering
      \includegraphics[width=\linewidth]{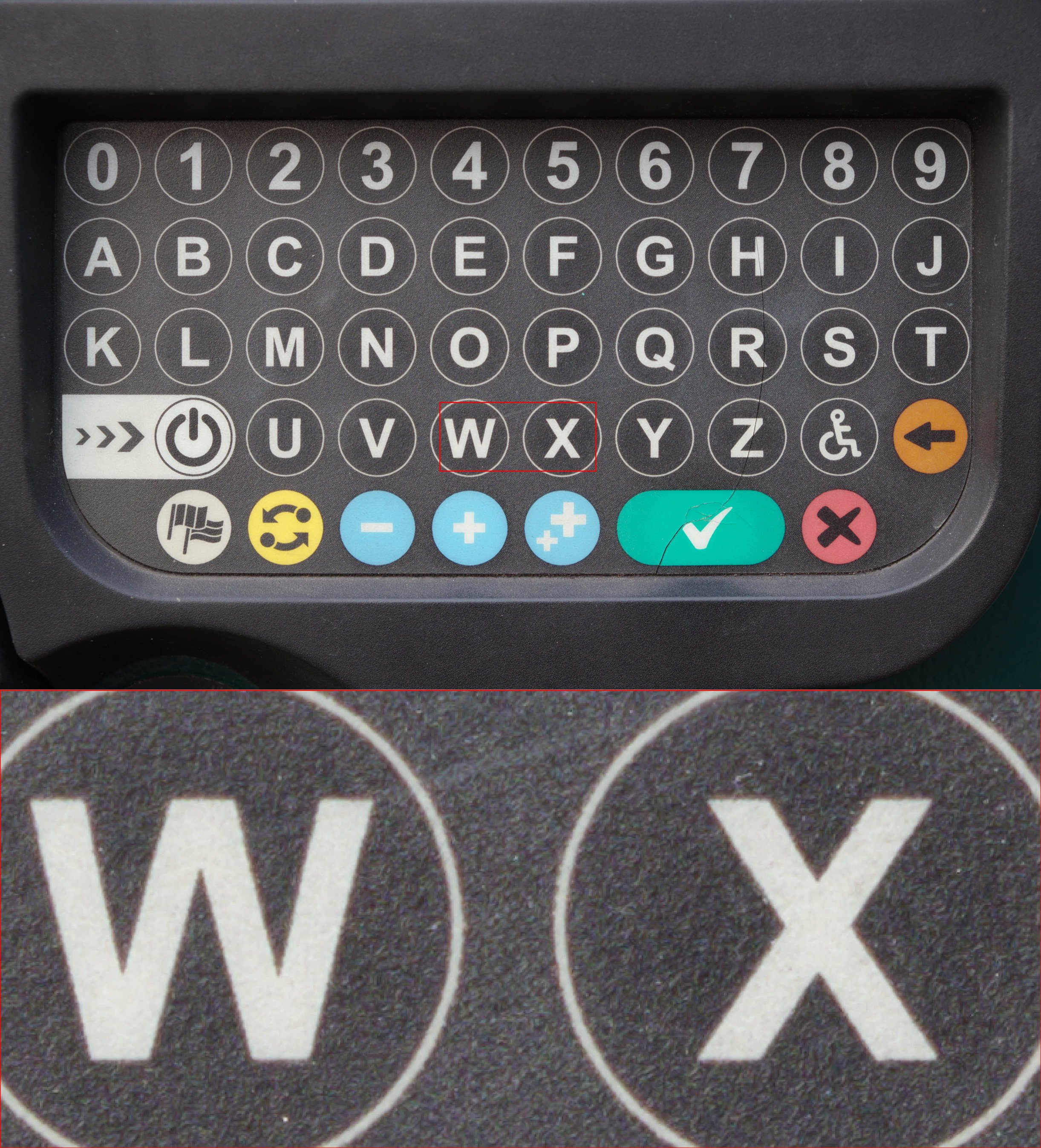}                 \\
        \fontsize{8.5pt}{8.5pt}\selectfont ISO 200
      }  \vspace{0.3em}                                                                                             \\
    \end{tabular}
  }
  \scalebox{0.95}{
    \begin{tabular}{@{}c@{\hspace{0.3em}}c@{\hspace{0.3em}}c@{}}
      \parbox[t]{0.333\textwidth}{
        \centering
      \includegraphics[width=\linewidth]{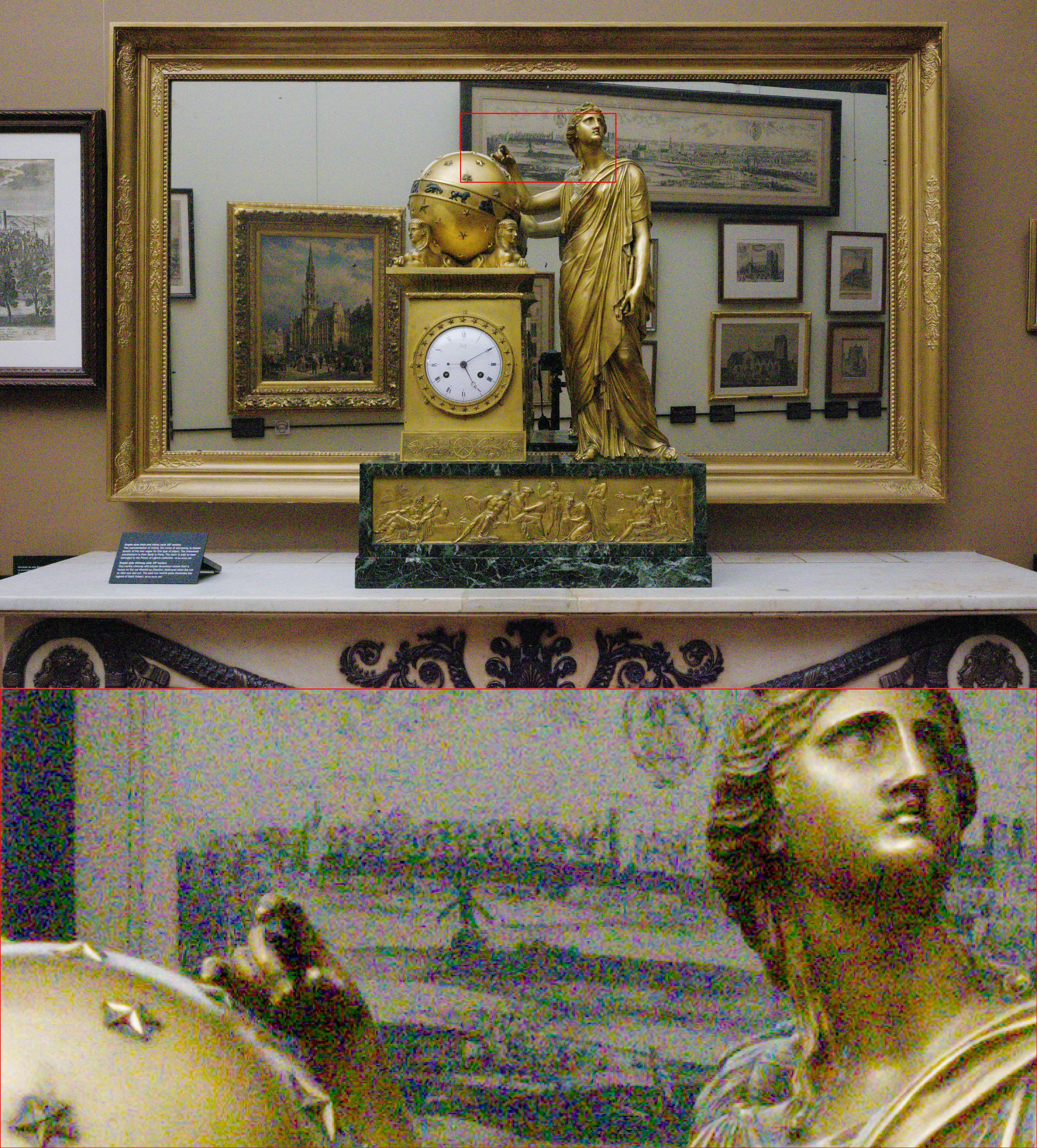}                 \\
        \fontsize{8.5pt}{8.5pt}\selectfont High ISO
      } &
      \parbox[t]{0.333\textwidth}{
        \centering
      \includegraphics[width=\linewidth]{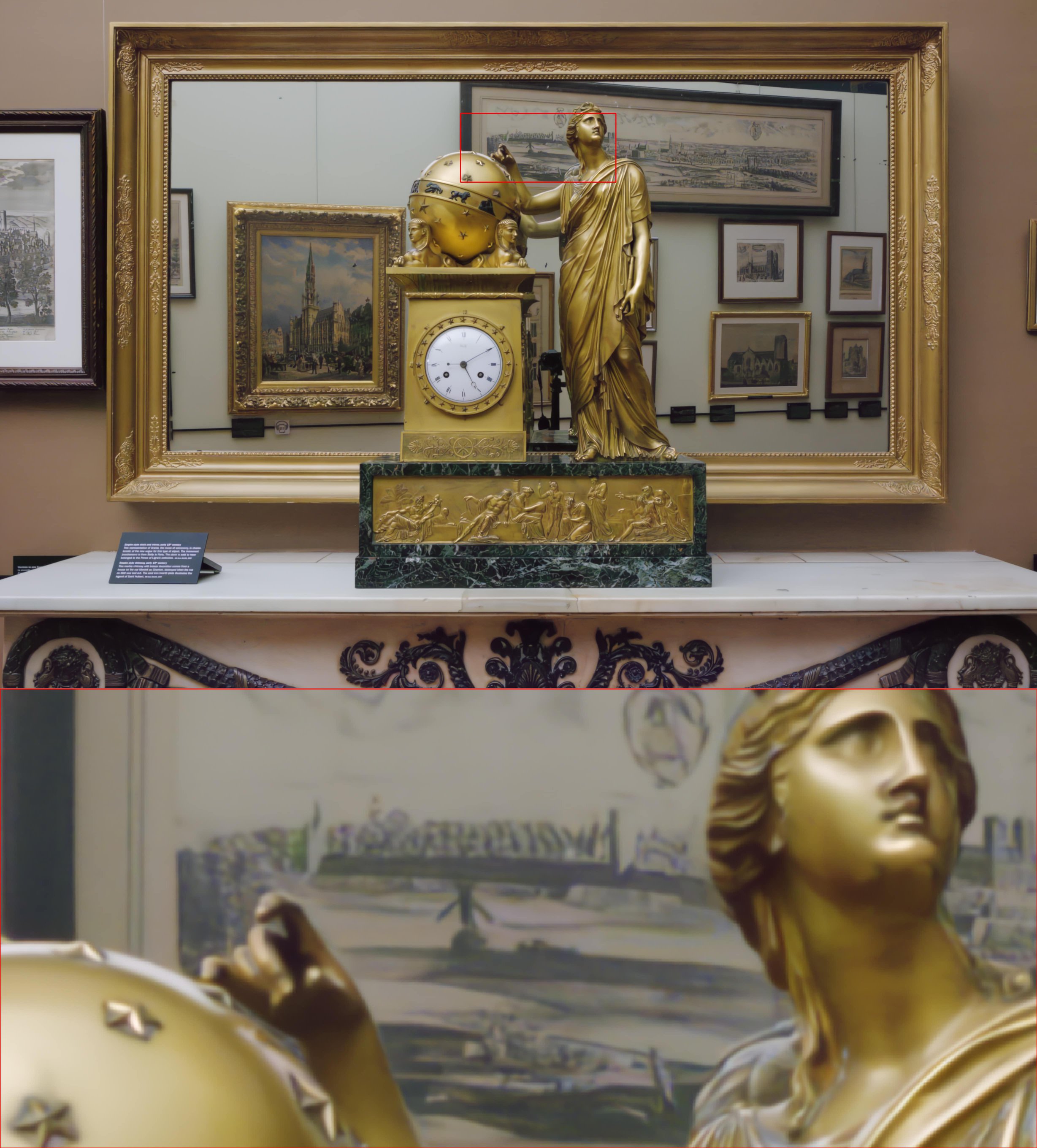} \\
        \fontsize{8.5pt}{8.5pt}\selectfont Denoised High ISO (\bf{Ours})
      } &
      \parbox[t]{0.333\textwidth}{
        \centering
      \includegraphics[width=\linewidth]{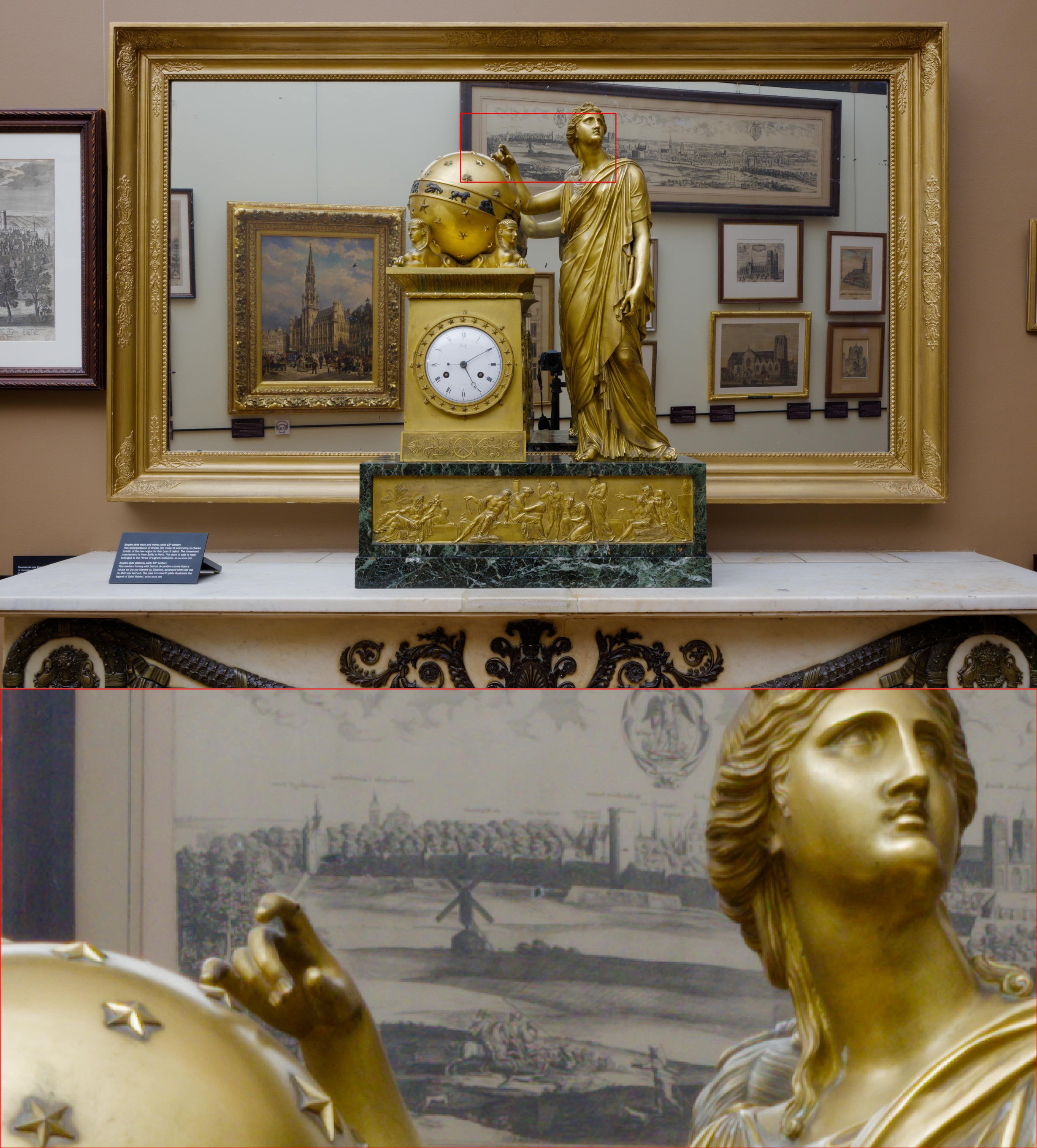}                 \\
        \fontsize{8.5pt}{8.5pt}\selectfont ISO 200
      }  \vspace{0.3em}                                                                                       \\
    \end{tabular}
  }

  \scalebox{0.95}{
    \begin{tabular}{@{}c@{\hspace{0.3em}}c@{\hspace{0.3em}}c@{}}
      \parbox[t]{0.333\textwidth}{
        \centering
      \includegraphics[width=\linewidth]{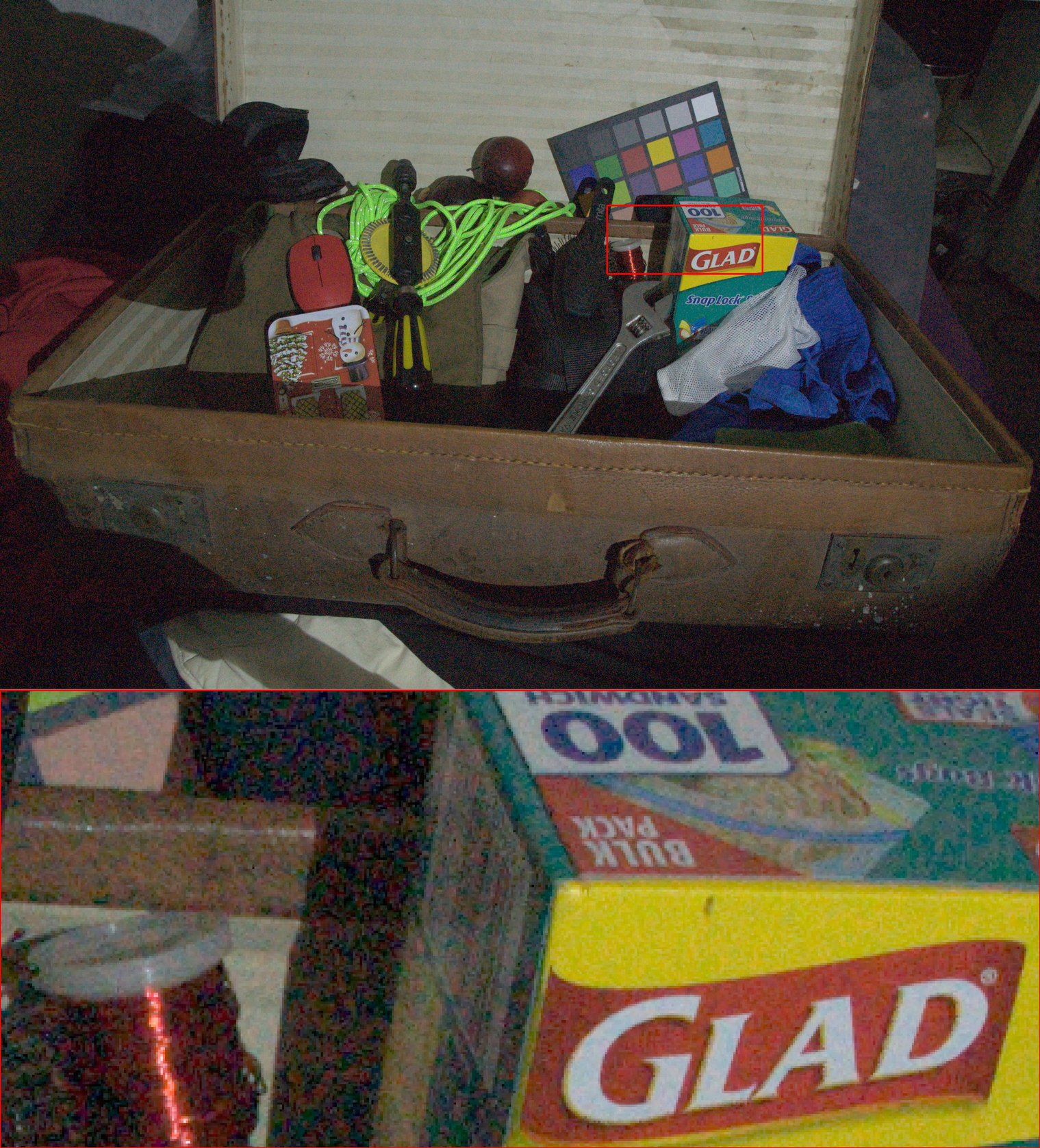}                 \\
        \fontsize{8.5pt}{8.5pt}\selectfont High ISO
      } &
      \parbox[t]{0.333\textwidth}{
        \centering
      \includegraphics[width=\linewidth]{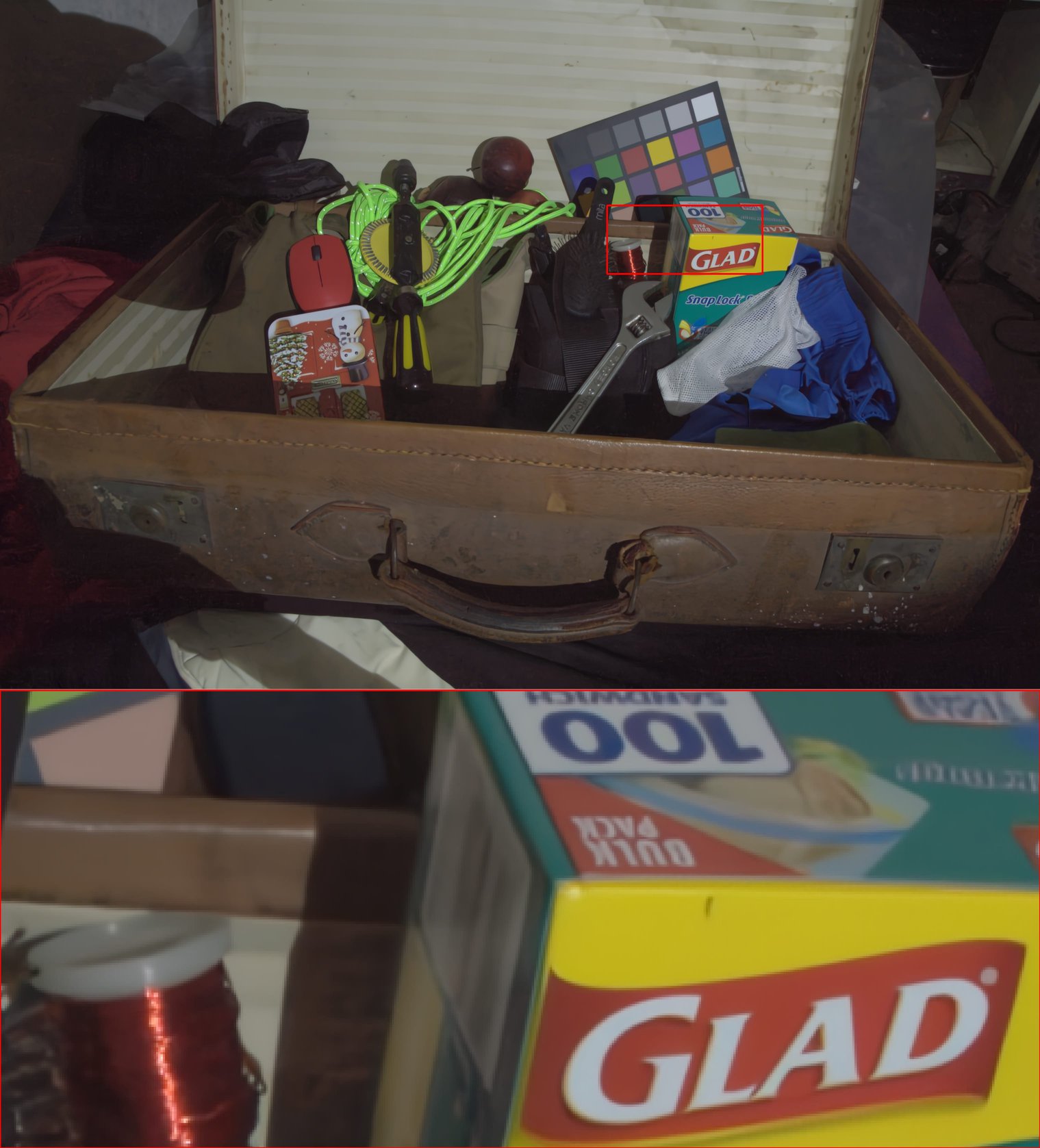} \\
        \fontsize{8.5pt}{8.5pt}\selectfont Denoised High ISO (\bf{Ours})
      } &
      \parbox[t]{0.333\textwidth}{
        \centering
      \includegraphics[width=\linewidth]{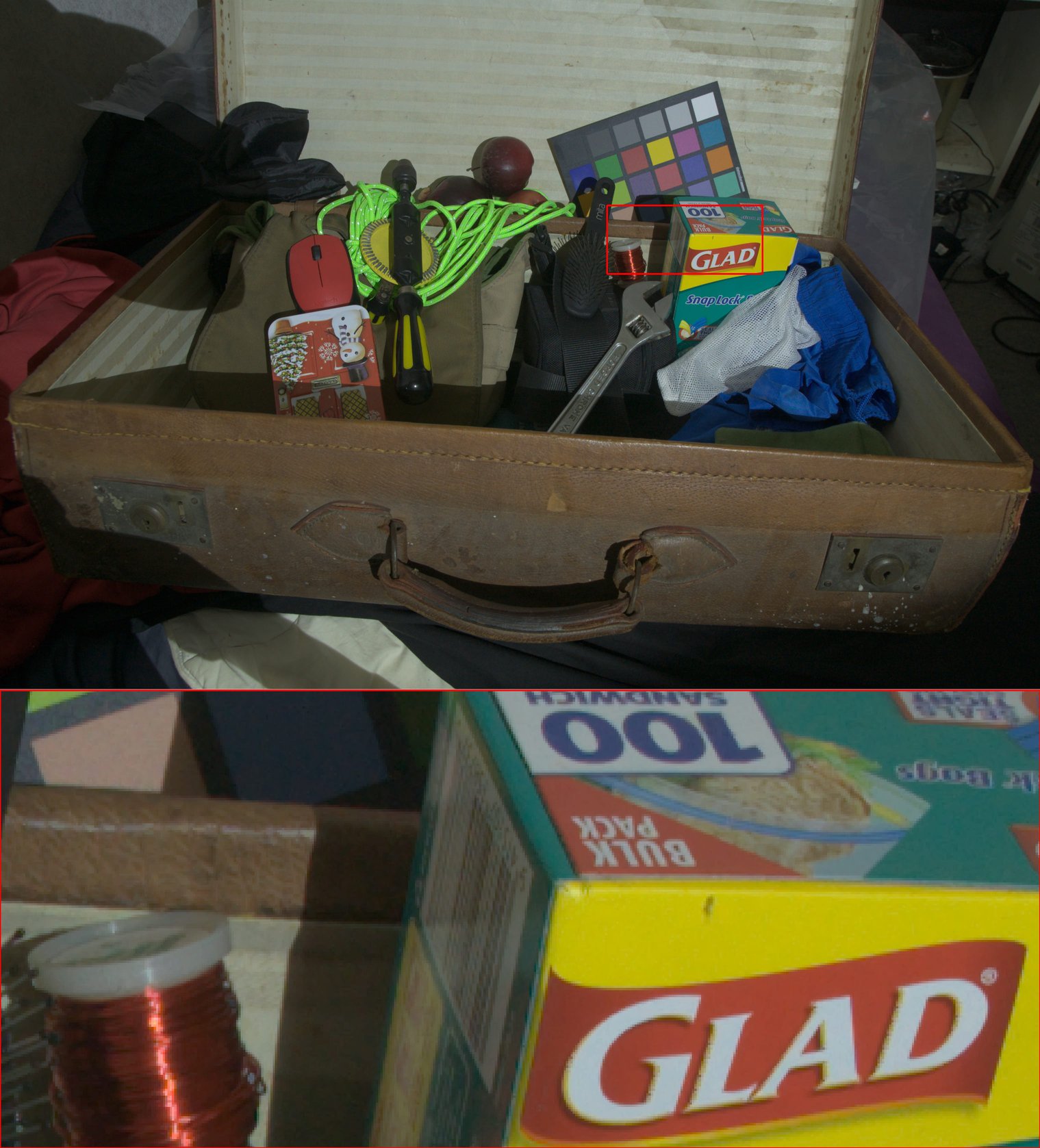}                 \\
        \fontsize{8.5pt}{8.5pt}\selectfont ISO 200
      }  \vspace{0.3em}                                                                                      \\
    \end{tabular}
  }

  \vspace{-2mm}
  \caption{
    Qualitative results of our noise translation network with pretrained NAFNet applied to high ISO images from the NIND dataset. 
    Our method effectively removes noise while preserving fine details, producing outputs comparable to low ISO (ISO 200) images. 
    Zoom in for a closer inspection of the visual quality. 
  }
  \label{fig:nind2}
\end{figure*}

The NIND~\citep{Brummer_2019_CVPR_Workshops} dataset comprises high-resolution images captured at various ISO levels.
High ISO images are noisier due to increased sensor sensitivity, while low ISO images, though cleaner, may require more optimal settings or equipment to capture effectively.
Figure~\ref{fig:nind1} and Figure~\ref{fig:nind2} illustrate the qualitative results of our method applied to high ISO images, demonstrating its ability to effectively remove noise while preserving fine image details.
The denoised outputs are visually clean and free from artifacts, achieving a quality comparable to images captured at low ISO levels.

\subsection{Real-world smartphone image noise and denoising results}
\begin{figure*}[t]
    \centering
    \scalebox{1}{
        \begin{subfigure}[t]{0.195\textwidth}
            \centering
            \includegraphics[width=\linewidth]{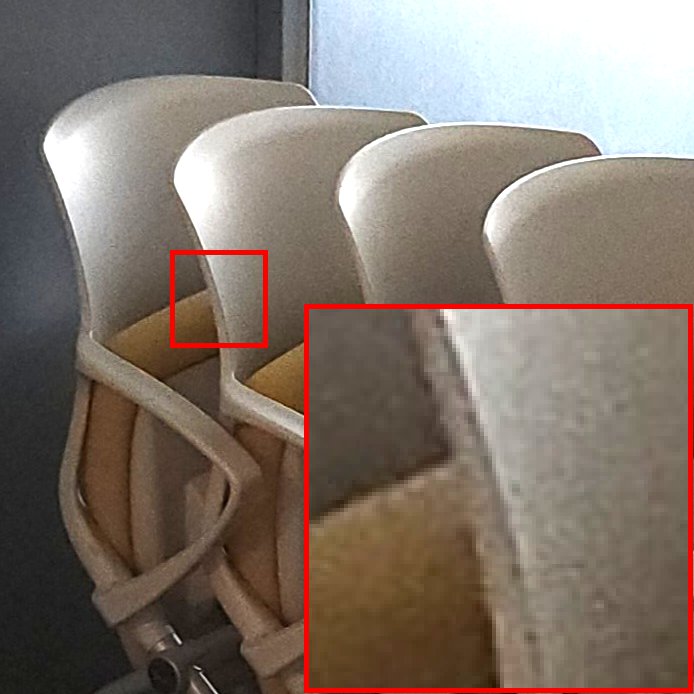}
            \vspace{-5mm}
            \caption*{Noisy Image}
        \end{subfigure}
        \hfill
        \begin{subfigure}[t]{0.195\textwidth}
            \centering
            \includegraphics[width=\linewidth]{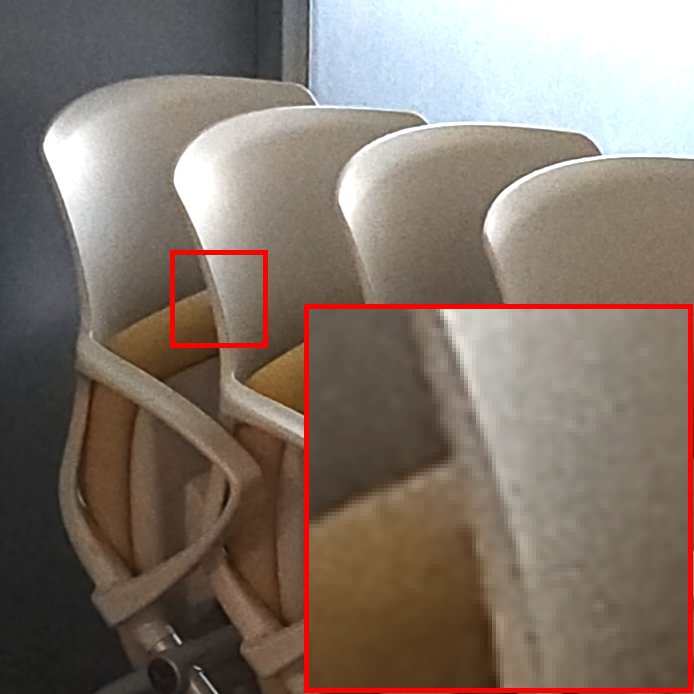}
            \vspace{-5mm}
            \caption{Restormer}
        \end{subfigure}
        \hfill
        \begin{subfigure}[t]{0.195\textwidth}
            \centering
            \includegraphics[width=\linewidth]{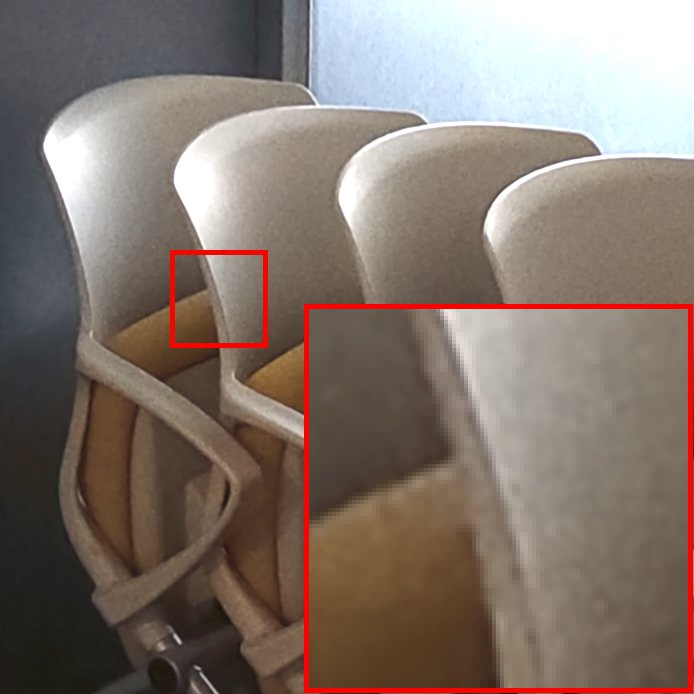}
            \vspace{-5mm}
            \caption{NAFNet}
        \end{subfigure}
        \hfill
        \begin{subfigure}[t]{0.195\textwidth}
            \centering
            \includegraphics[width=\linewidth]{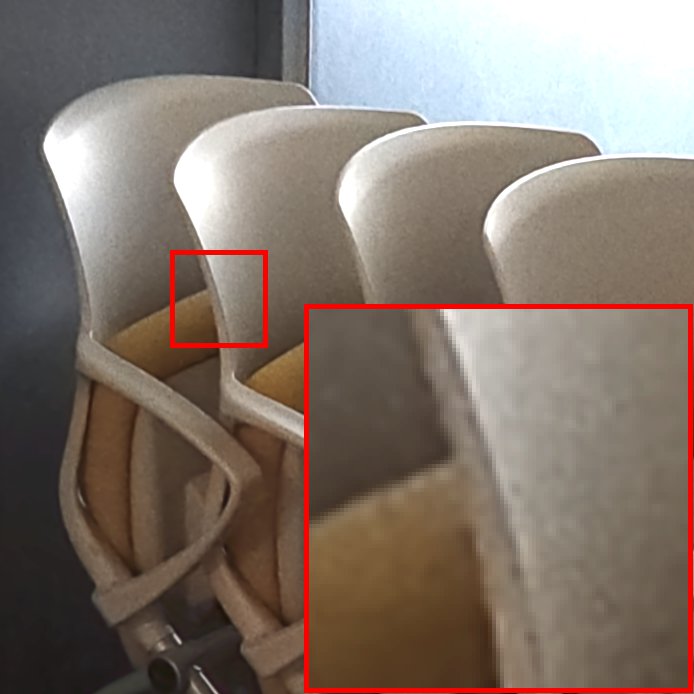}
            \vspace{-5mm}
            \caption{KBNet}
        \end{subfigure}
        \hfill
        \begin{subfigure}[t]{0.195\textwidth}
            \centering
            \includegraphics[width=\linewidth]{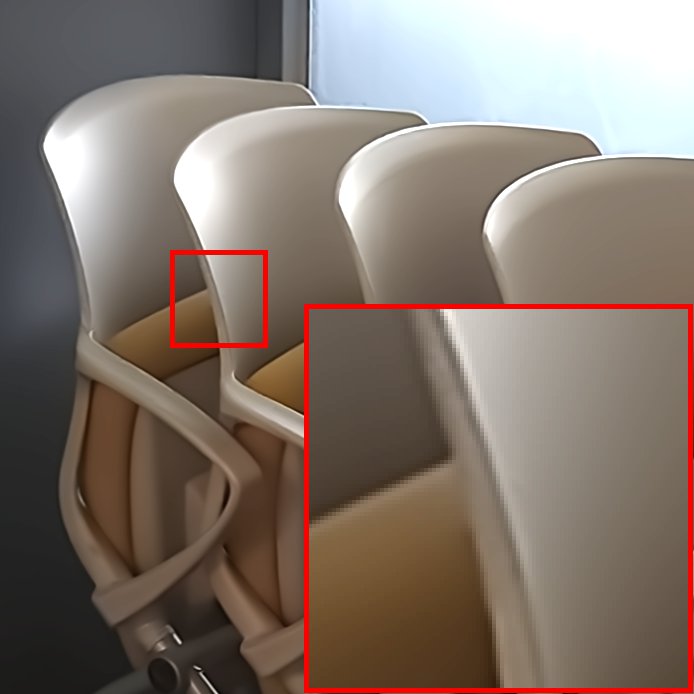}
            \vspace{-5mm}
            \caption{Ours}
        \end{subfigure}
    }

    \scalebox{0.99}{
        \begin{subfigure}[t]{0.195\textwidth}
            \centering
            \includegraphics[width=\linewidth]{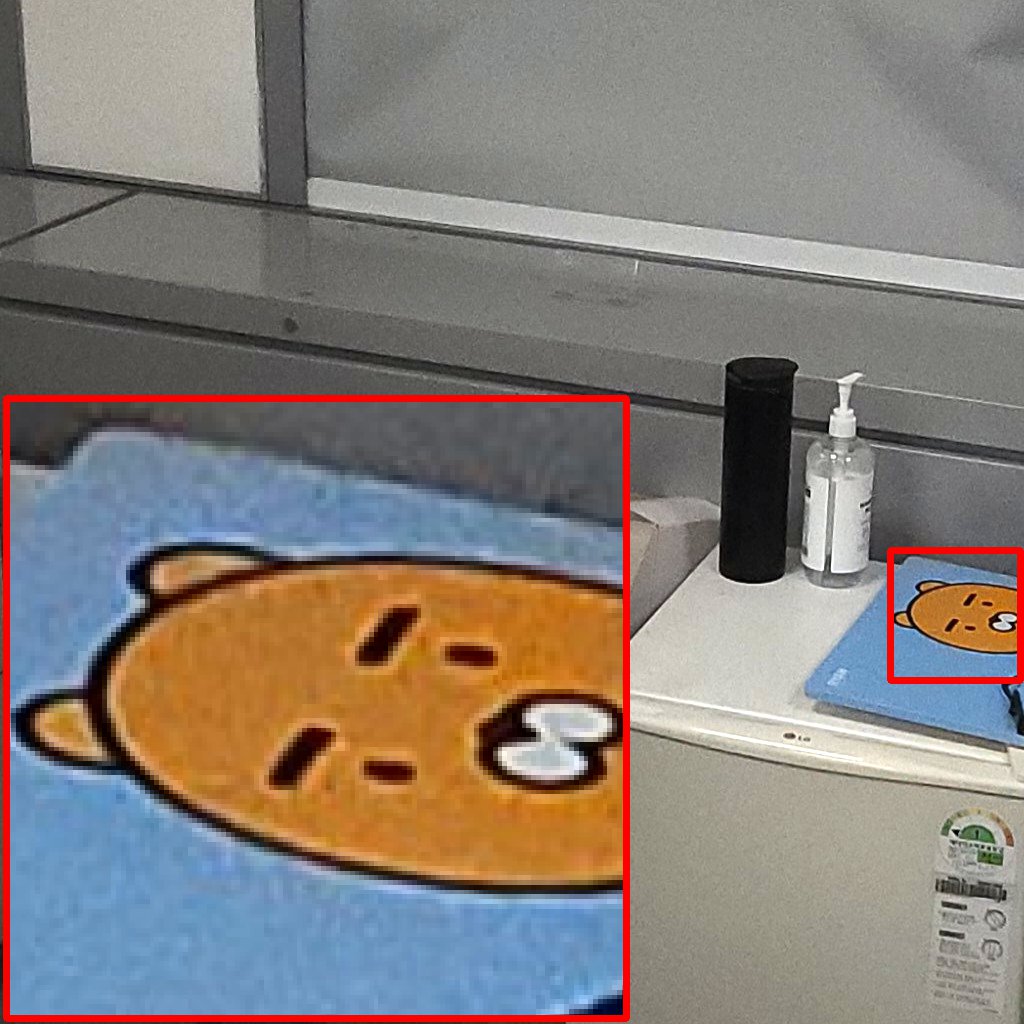}
            \vspace{-5mm}
            \caption*{Noisy Image}
        \end{subfigure}
        \hfill
        \begin{subfigure}[t]{0.195\textwidth}
            \centering
            \includegraphics[width=\linewidth]{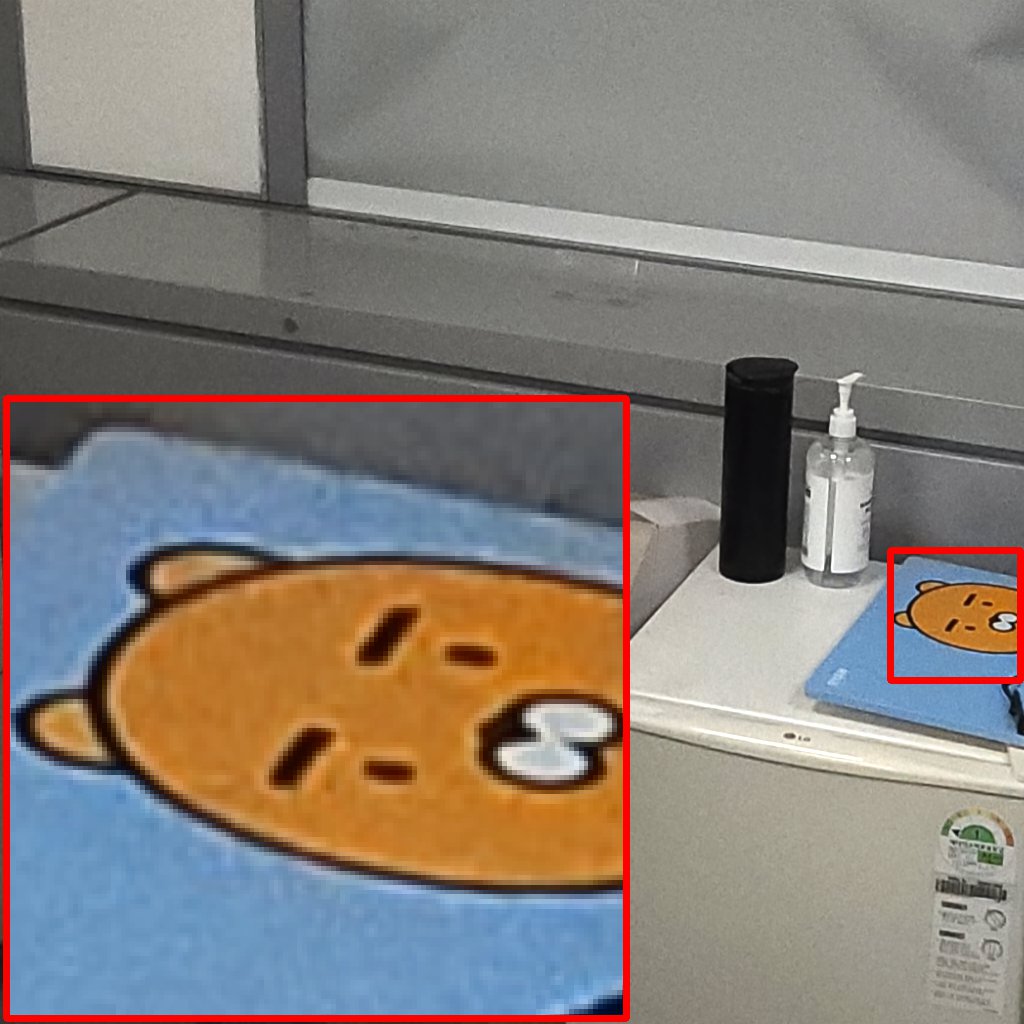}
            \vspace{-5mm}
            \caption{Restormer}
        \end{subfigure}
        \hfill
        \begin{subfigure}[t]{0.195\textwidth}
            \centering
            \includegraphics[width=\linewidth]{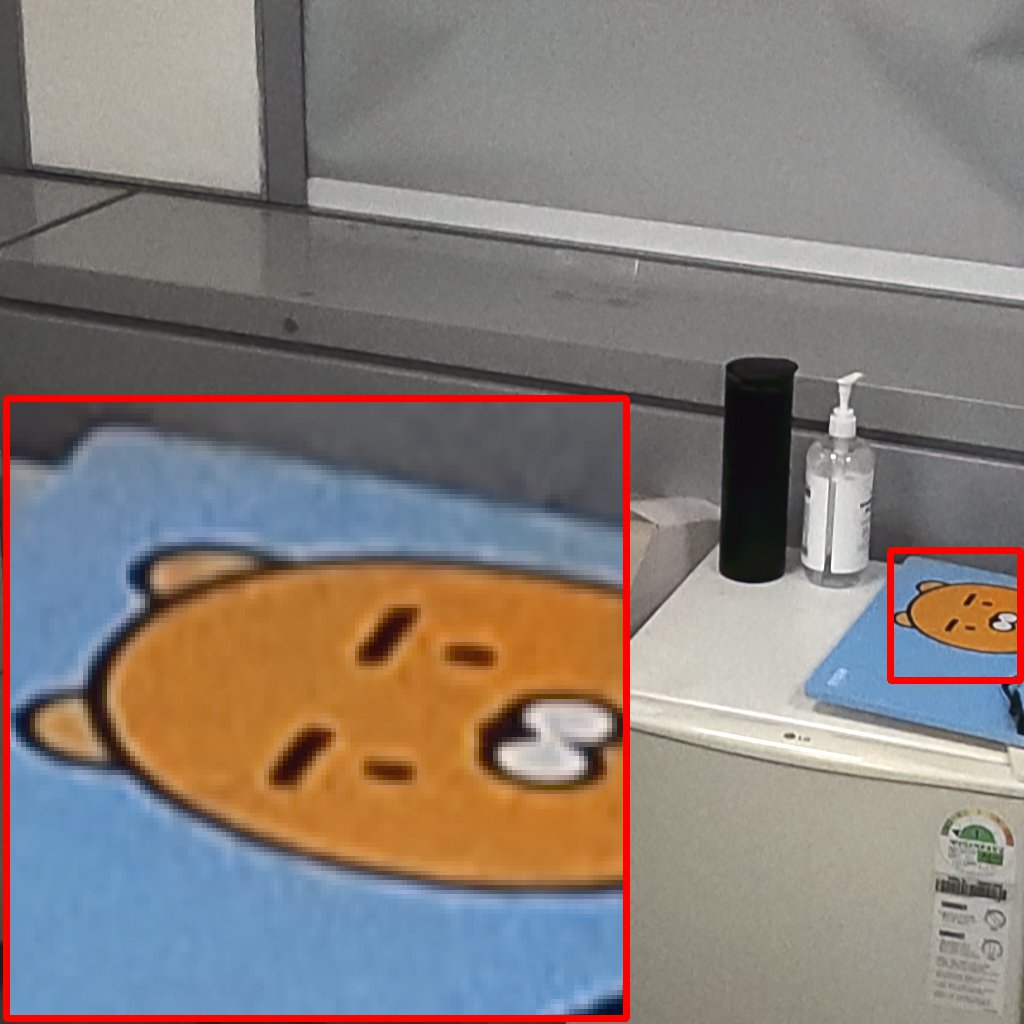}
            \vspace{-5mm}
            \caption{NAFNet}
        \end{subfigure}
        \hfill
        \begin{subfigure}[t]{0.195\textwidth}
            \centering
            \includegraphics[width=\linewidth]{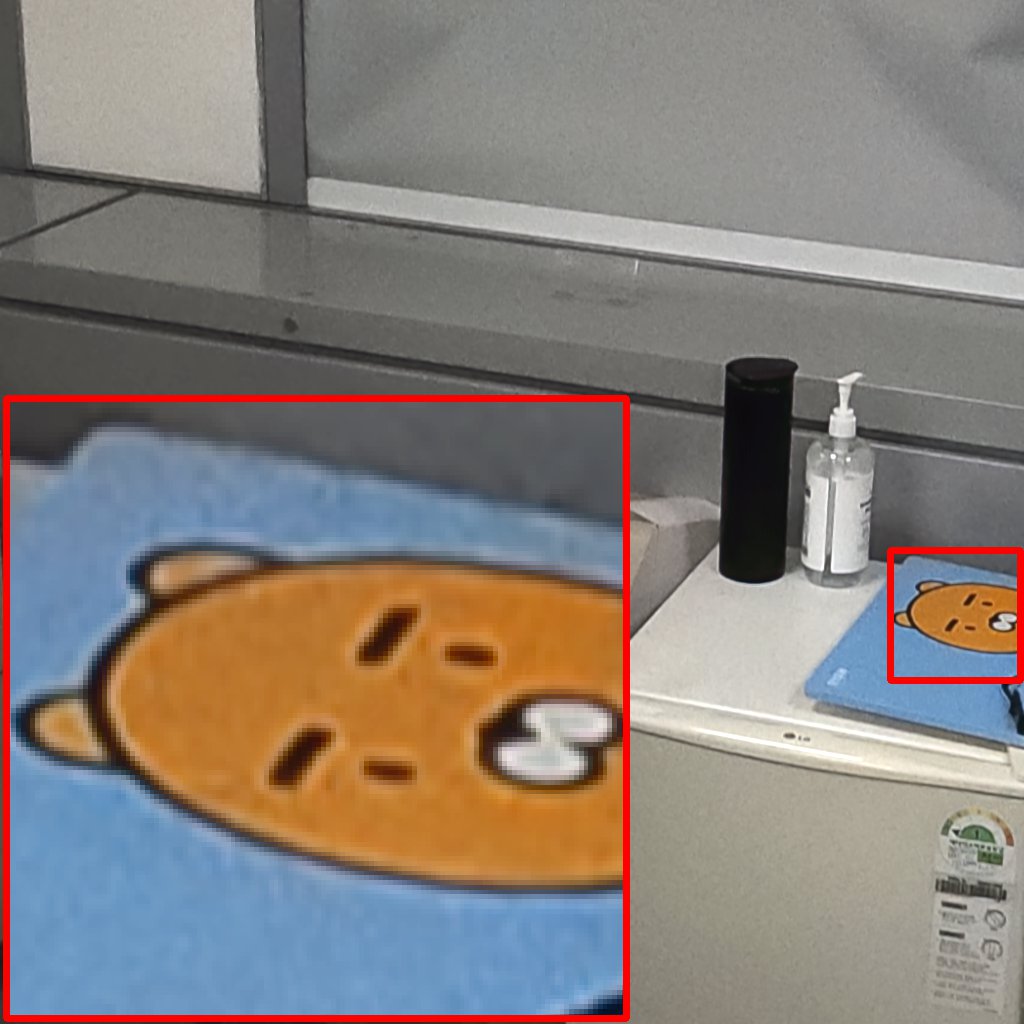}
            \vspace{-5mm}
            \caption{KBNet}
        \end{subfigure}
        \hfill
        \begin{subfigure}[t]{0.195\textwidth}
            \centering
            \includegraphics[width=\linewidth]{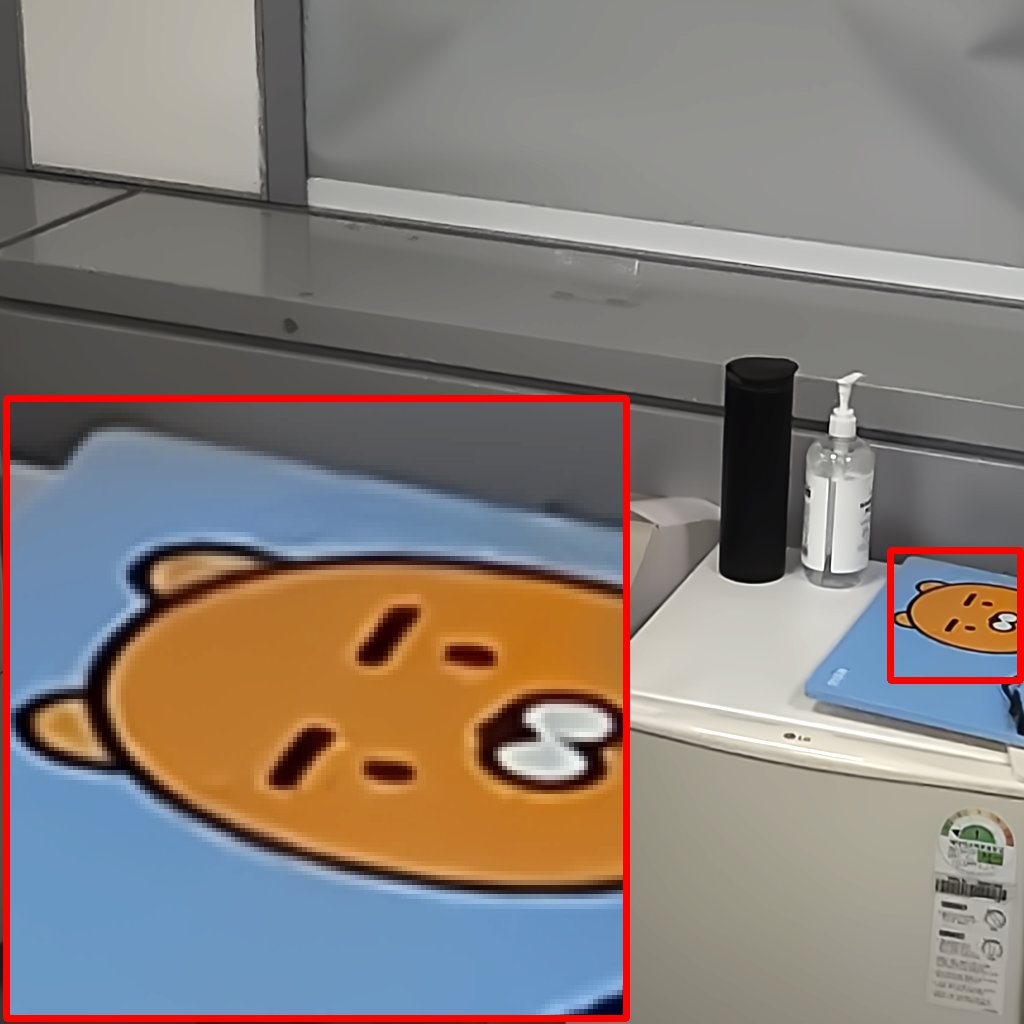}
            \vspace{-5mm}
            \caption{Ours}
        \end{subfigure}
    }
    \caption{
        Comparison between the qualitative results of various denoising networks including ours (noise translation network with pretrained NAFNet) on images captured by Galaxy S22+ smartphone.
        Zoom in for a closer inspection of the visual quality. 
    }
    \label{fig:teaser}
\end{figure*}

Finally, we present the denoised results of images captured using our Galaxy S22+ smartphone, as shown in Figure~\ref{fig:teaser}.

\section{Limitations}
Although our approach provides strong robustness to diverse real-world noise, it may appear to compromise performance on in-distribution (ID) datasets. 
As discussed in Figure 8, this effect is largely due to the pronounced overfitting of conventional denoising models, which tend to reconstruct dataset-specific artifacts rather than genuine image content. 

Moreover, our evaluation primarily relies on PSNR and SSIM computed against the provided ground-truth images. While these metrics are standard practice, they inherently struggle to capture perceptual realism and may not align well with human judgments. 
Although our method consistently improves these metrics, such quantitative measures do not fully represent perceptual quality, which remains subjective and can vary across viewers.
Future work includes exploring alternative evaluation protocols or perceptually aligned metrics that better account for realism, semantic consistency, and human preference.

\clearpage

\end{document}